# Recent Advances in Bianisotropic Boundary Conditions: Theory, Capabilities, Realizations, and Applications

Jordan Budhu and Anthony Grbic

*Abstract*—**In recent years, new functionality and unprecedented wavefront control has been enabled by the introduction of bianisotropic metasurfaces. A bianisotropic metasurface is characterized by an electric response, a magnetic response, and an electro-magnetic/magneto-electric response. In general, these metasurfaces consists of an array of metallic or dielectric particles located within a subwavelength thick host medium, and are approximated and modelled as infinitely-thin, idealized sheet boundaries defined along a surface. An appropriate sheet boundary condition which effectively models the tangential field discontinuity due to the array of magnetoelectric inclusions is the Generalized Sheet Transition Condition or GSTC. Several forms of the GSTC appear in literature. Here, we present each interpretation and show how they are related. Synthesis approaches unique to each form are overviewed. By utilizing the GSTC in metasurface design, new possibilities emerge which are not possible with conventional design techniques incorporating only electric or only magnetic responses. Since the metasurfaces are designed using bianisotropic boundary conditions, they must be realized using particles which contain magnetoelectric responses. This review article discusses the design of metasurfaces using the GSTC, and the bianisotropic particles used to realize GSTC's. Further, it discusses new and recent applications that have emerged due to bianisotropy, and future prospects in metasurface design using bianisotropic boundary conditions. The intent is to provide a comprehensive overview of metasurface design involving bianisotropy and for this review article to serve as a starting point for engineers and scientist that wish to introduce bianisotropy into metasurface design.**

## I. Introduction

BIANISOTROPIC boundaries are surfaces consisting of electric, magnetic, and magneto-electric surface susceptibilities. These boundaries, and their scattering characteristics, have been studied for a number of years. Early studies include those by M. M. Idemen in the late 1980s [1], as well as C. L. Holloway and E. F. Kuester's work in the early 2000s [2]–[12] that revived interest in this topic. However, only in the past few years has the true potential of bianisotropic boundaries been revealed by research community.

In recent years, extreme, reflectionless polarization control [13]–[20], seamless impedance matching between input and output fields [21], [22], as well as wide angle refraction [23]–[28] have been demonstrated using these sheet boundaries. Scientific works have also revealed that arbitrary field transformations can be achieved with bianisotropic boundaries consisting of complex electric, magnetic, and magneto-electric susceptibilities involving loss and gain. In addition, a wide range of local power conserving wavefront transformations have been demonstrated by controlling both the visible (propagating) and invisible spectrum (surface waves) using lossless and passive, bianisotropic sheet boundaries [28]–[40].

Over the past year, nonlocal designs that are passive and lossless have been reported that overcome the local power conservation restriction of earlier designs. These boundaries require only global power conservation to ensure passivity but allow perfect field transformation [32], [37], [38], [41]–[52] or perfect mode conversion [53]–[56]. These passive and lossless designs truly allow unrestricted reciprocal and lossless field transformations. Multiband designs [57]–[59] have also emerged allowing distinct field transformations at different frequencies of operation.

The body of theoretical work and proof of concept experimental demonstrations showing the extreme field manipulation that is possible with bianisotropic sheet boundaries has driven research toward practical realizations and in turn revealed potential applications. Various implementations have been and continue to be proposed. These range from 3D geometries such as spirals or omega particles [21], [60]–[62] to those that can be implemented using planar fabrication approaches [14], [63]–[66]. Planar designs have included cascaded patterned metallic or dielectric claddings [67]–[69] that support zeroth order coupling between the sheets for the manipulation of visible (propagating) electromagnetic spectrum as well as those that support higher order coupling that manipulate both the visible and invisible (evanescent) spectrum. Non-reciprocal particles [70]–[75] and all dielectric bianisotropic particles [76], [77] have also been reported.

In summary, bianisotropic sheet boundaries are ushering in a new generation of ultra-thin electromagnetic devices with revolutionary capabilities. Research in this area opens new

Jordan Budhu and Anthony Grbic are with the Electrical and Computer Engineering Department, University of Michigan, Ann Arbor, MI 48109 USA (e-mail: jbudhu@umich.edu, agrbic@umich.edu).

This work was supported by the Army Research Office under Grant W911NF-19-1-0359, and the Office of Naval Research under Grant N00014-18- 1-2536 opportunities in applications areas that require electromagnetic devices with very small form factors and conformal shapes [78]. The added degrees of freedom afforded by bianisotropic sheet boundaries promise conformal electromagnetic and optical systems that can be seamlessly integrated into various platforms. These include ultra-thin, flat panel antennas [79]–[81] with arbitrary aperture distributions [36], [37], [55], compact mode converters [56], transitions and couplers, conformal cloaking membranes [82]–[88], as well as ultra-thin cameras, detectors, and high-resolution 3D holographic displays.

This review article will present the bianisotropic boundary conditions, several synthesis methods used to utilize them in metasurface design, and several magnetoelectric particle options that can be used to realize the metasurfaces designed from these boundaries. Furthermore, this review article will chronicle recent work in the use of bianisotropic boundaries in electromagnetic design. The technological advancements will be highlighted, and future prospects discussed.

The article begins with section II which derives the generalized sheet transition conditions (GSTC). Section III outlines various synthesis methods used to design bianisotropic metasurfaces. Section IV provides designs for magnetoelectric particles used to realize bianisotropic metasurfaces. Section V then provides a chronology of scientific works on bianisotropic metasurfaces taken from literature. Section VI presents prospects in the field of bianisotropic metasurfaces. Finally, in Section VII the paper is concluded. Note, an $e^{j\omega t}$ time convention is assumed and suppressed throughout the paper.

## II. Bianisotropic Boundary Conditions in Metasurface Design

### A. Generalized Sheet Transition Conditions (GSTC)

Consider an infinite planar metasurface separating two dielectric half spaces of intrinsic impedance $\eta_1$ and $\eta_2$ (see Fig. 1). The normal to the metasurface is denoted by the unit vector $\hat{z}$ and therefore the metasurface spans the $xy$-plane at $z$=0. The metasurface consists of a periodic arrangement of polarizable bianisotropic particles separated by a subwavelength period. Assuming the metasurface can be homogenized, we will derive a bianisotropic sheet boundary condition which models the metasurface as a sheet of electric and magnetic polarization currents and relates these to the tangential field discontinuities. To begin, Maxwell's equations are written with their transverse and normal components separated [89]

$$\begin{aligned}
&\nabla_t \times \vec{E}_t = -\hat{z}K_z - j\omega\mu_0\hat{z}H_z \\
&\nabla_t \times \hat{z}E_z + \hat{z}\times\frac{\partial}{\partial z}\vec{E}_t = -\vec{K}_t - j\omega\mu_0\vec{H}_t \\
&\nabla_t \times \vec{H}_t = \hat{z}J_z + j\omega\varepsilon_0\hat{z}E_z \\
&\nabla_t \times \hat{z}H_z + \hat{z}\times\frac{\partial}{\partial z}\vec{H}_t = \vec{J}_t + j\omega\varepsilon_0\vec{E}_t
\end{aligned} \tag{1}$$

where $\vec{E}_t$ and $\vec{H}_t$ are the transverse components of the electric and magnetic field, $\vec{J}_t$ and $\vec{K}_t$ are the tangential electric and magnetic surface current densities, $\omega$ is the angular frequency of the excitation, and $\mu_0$ and $\epsilon_0$ are the permeability and permittivity of free space. The operator $\nabla_t = (\frac{\partial}{\partial x}\hat{x} + \frac{\partial}{\partial y}\hat{y})$.

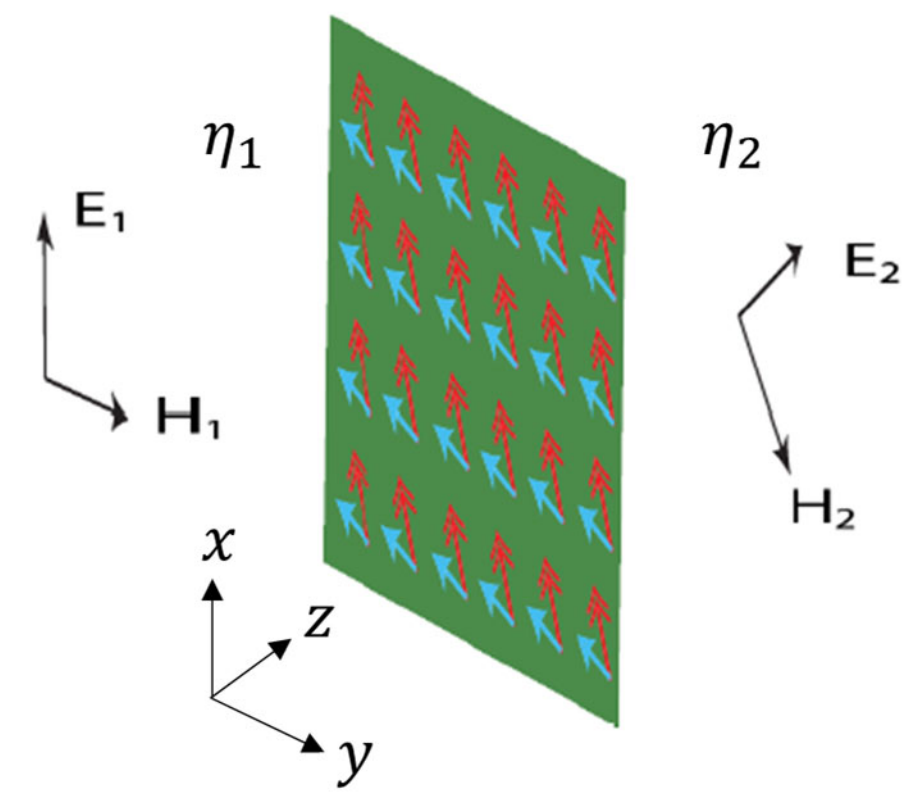

Fig. 1. Metasurface geometry. The metasurface consists of a periodic arrangement of polarizable magnetoelectric particles separated by a subwavelength period.

Enforcing (1) along a sheet boundary at $z$=0, which supports the surface current densities

$$\begin{aligned}
\vec{J} &= \delta(z)\left(\vec{J}_{st} + \hat{z}J_{sz}\right) \\
\vec{K} &= \delta(z)\left(\vec{K}_{st} + \hat{z}K_{sz}\right)
\end{aligned}, \tag{2}$$

yields

$$\begin{aligned}
\Delta\vec{E}\times\hat{z} &= \vec{K}_{st} - \nabla_t\left(\frac{J_{sz}}{j\omega\varepsilon_0}\right)\times\hat{z} \\
\hat{z}\times\Delta\vec{H} &= \vec{J}_{st} - \hat{z}\times\nabla_t\left(\frac{K_{sz}}{j\omega\mu_0}\right)
\end{aligned} \tag{3}$$

where

$$\begin{aligned}
\Delta\vec{E} &= \vec{E}\Big|_{z=0^-}^{0^+} = \vec{E}\left(x,y,z=0^+\right) - \vec{E}\left(x,y,z=0^-\right) \\
\Delta\vec{H} &= \vec{H}\Big|_{z=0^-}^{0^+} = \vec{H}\left(x,y,z=0^+\right) - \vec{H}\left(x,y,z=0^-\right)
\end{aligned} \tag{4}$$

are the tangential field discontinuities in the electric and magnetic fields across the surface. Next, the surface currents are related to the surface polarization density and surface magnetization as

$$\begin{aligned}
\vec{J} &= \delta(z)\left(\vec{J}_{st} + \hat{z}J_{sz}\right) = j\omega\vec{P}_s\delta(z) \\
&= j\omega\left(\vec{P}_{st} + \hat{z}P_{sz}\right)\delta(z) \\
\vec{K} &= \delta(z)\left(\vec{K}_{st} + \hat{z}K_{sz}\right) = j\omega\mu_0\vec{M}_s\delta(z) \\
&= j\omega\mu_0\left(\vec{M}_{st} + \hat{z}M_{sz}\right)\delta(z)
\end{aligned} \tag{5}$$

Substitution of (5) into (3) gives

$$\begin{aligned}
\hat{z}\times\Delta\vec{E} &= -j\omega\mu_0\vec{M}_{st} - \hat{z}\times\nabla_t\left(\frac{P_{sz}}{\varepsilon_0}\right) \\
\hat{z}\times\Delta\vec{H} &= j\omega\vec{P}_{st} - \hat{z}\times\nabla_t M_{sz}
\end{aligned} \tag{6}$$

The result (6) has been derived by Ideman in [1]. The boundary condition in (6) links the discontinuities in the macroscopic averaged total fields (the incident field plus the fields due to the array of polarizable particles) to the surface electric polarization and magnetization densities. These densities are a consequence of averaging a distribution of electric and magnetic dipoles [90]–[92]

$$\vec{P}_s = N\left\langle \vec{p}_{ee} \right\rangle + N\left\langle \vec{p}_{em} \right\rangle$$
$$\vec{M}_s = N\left\langle \vec{p}_{mm} \right\rangle + N\left\langle \vec{p}_{me} \right\rangle \tag{7}$$

In (7), the units of $\vec{P}_s$ and $\vec{M}_s$ are Coulomb/meter and Amp, $N$ is the number scatterers per unit area, and $\langle \vec{p} \rangle$ represents the dipole moment averaged over a unit cell area of $S$

$$\left\langle \vec{p} \right\rangle = \frac{1}{S}\int \vec{p} ds \tag{8}$$

The dipole moment associated with each particle is proportional to the local field acting on the particle and its polarizability tensor

$$\begin{bmatrix} \vec{p}_e \\ \vec{p}_m \end{bmatrix} = \begin{bmatrix} \vec{p}_{ee} + \vec{p}_{em} \\ \vec{p}_{me} + \vec{p}_{mm} \end{bmatrix} = \begin{bmatrix} \overline{\overline{\alpha}}_{ee} & \overline{\overline{\alpha}}_{em} \\ \overline{\overline{\alpha}}_{me} & \overline{\overline{\alpha}}_{mm} \end{bmatrix} \begin{bmatrix} \vec{E}_{loc} \\ \vec{H}_{loc} \end{bmatrix} \tag{9}$$

Each of the polarizability tensors (the electric $\bar{\bar{\alpha}}_{ee}$, the magnetic $\bar{\bar{\alpha}}_{mm}$, the magnetoelectric $\bar{\bar{\alpha}}_{me}$, and the electromagnetic $\bar{\bar{\alpha}}_{em}$ polarizability tensor) in (9) are of dimension $3 \times 3$. For example, the electric polarizability tensor is

$$\overline{\overline{\alpha}}_{ee} = \begin{bmatrix} \alpha_{ee}^{xx} & \alpha_{ee}^{xy} & \alpha_{ee}^{xz} \\ \alpha_{ee}^{yx} & \alpha_{ee}^{yy} & \alpha_{ee}^{yz} \\ \alpha_{ee}^{zx} & \alpha_{ee}^{zy} & \alpha_{ee}^{zz} \end{bmatrix} \tag{10}$$

where the notation $\alpha_{ab}^{uv}$ denotes the $u$ component of the dipole response $a$ due to the $v$ component of the excitation field $b$ (Note, $a, b = e, m$ and $u, v = x, y, z$). Substituting the results of (7)-(10) back into (6) gives

$$\begin{aligned} \hat{z} \times \Delta\vec{E} &= -j\omega\mu_0 N\left\langle \overline{\overline{\alpha}}_{mm} \right\rangle \bullet \vec{H}_{t,loc} - j\omega\mu_0 N\left\langle \overline{\overline{\alpha}}_{me} \right\rangle \bullet \vec{E}_{t,loc} \\ &\quad -\left(\frac{1}{\varepsilon_0}\right)\hat{z} \times \nabla_t \left[ N\left\langle \overline{\overline{\alpha}}_{ee} \right\rangle \bullet \hat{z}E_{z,loc} + N\left\langle \overline{\overline{\alpha}}_{em} \right\rangle \bullet \hat{z}H_{z,loc} \right] \\ \hat{z} \times \Delta\vec{H} &= j\omega N\left\langle \overline{\overline{\alpha}}_{ee} \right\rangle \bullet \vec{E}_{t,loc} + j\omega N\left\langle \overline{\overline{\alpha}}_{em} \right\rangle \bullet \vec{H}_{t,loc} \\ &\quad -\hat{z} \times \nabla_t \left[ N\left\langle \overline{\overline{\alpha}}_{mm} \right\rangle \bullet \hat{z}H_{z,loc} + N\left\langle \overline{\overline{\alpha}}_{me} \right\rangle \bullet \hat{z}E_{z,loc} \right] \end{aligned} \tag{11}$$

where the notation $\vec{E}_{t,loc} = \left[E_{x,loc}, E_{y,loc}, 0\right]^T$, $\vec{H}_{t,loc} = \left[H_{x,loc}, H_{y,loc}, 0\right]^T$, $\hat{z}E_{z,loc} = \left[0,0,E_{z,loc}\right]^T$, and $\hat{z}H_{z,loc} = \left[0,0,H_{z,loc}\right]^T$. Thus, to make use of (6) for an array of polarizable magnetoelectric particles, the local field $\vec{E}_{loc}, \vec{H}_{loc}$ acting on each particle must be obtained. The local field cannot be the macroscopic averaged total fields since these fields are discontinuous in the plane of the particle by (6). The local field must be continuous and well defined. The local fields polarizing the particle are defined to be the incident field plus the field due to the array of particles excluding the particle of interest. Kuester et. al in [2], find the local field by assuming that the array of particles can be modeled as a sheet of continuous electric and magnetic polarization density distributions (obtained by averaging the dipole moments of the particles as in (7)) from which the polarization and magnetization in a small disk of radius $R$ surrounding the particle of interest has been removed. The fields due to this punctured sheet are calculated by finding the fields due to the entire sheet of polarization densities (without the hole removed) and subtracting from it the fields (averaged over the disk area) due to a uniformly polarized and magnetized disk of radius $R$. The result is a continuous and well defined local field which is a function of the macroscopic averaged field allowing(7)-(9) to be written as

$$\begin{bmatrix} \vec{P}_s \\ \vec{M}_s \end{bmatrix} = \begin{bmatrix} \varepsilon_0 \overline{\overline{\chi}}_{ee} & \sqrt{\varepsilon_0\mu_0}\, \overline{\overline{\chi}}_{em} \\ \sqrt{\varepsilon_0/\mu_0}\, \overline{\overline{\chi}}_{me} & \overline{\overline{\chi}}_{mm} \end{bmatrix} \begin{bmatrix} \vec{E}_{av} \\ \vec{H}_{av} \end{bmatrix} \tag{12}$$

where $\bar{\bar{\chi}}_{ab}$ are the surface susceptibilities, have units of meters, and can be interpreted as the actual polarizability multiplied by a correction factor called the interaction constant $\beta$ accounting for the effects of the array on the local field ($c\bar{\bar{\chi}}_{ab} = N\langle \bar{\bar{\alpha}}_{ab} \rangle \bar{\bar{\beta}}$, where $c$ is the corresponding free space material constant coefficient of (12)). Note, Kuester, in [2], calls the surface susceptibilities ($\bar{\bar{\chi}}_{ab}$) effective polarizability densities ($\bar{\bar{\alpha}}_{ab}^{eff}$), however, this notation causes confusion as the quantities are not polarizabilities relating dipole moments to fields but rather quantities relating surface densities to fields. He later switched notation to the correct one of surface susceptibilities ([9], Eq. (1) for example). In (12), the free space material constants ensure (6) is dimensionally correct. Kuester derives the interaction constants in [2] at least for the electric and magnetic susceptibilities. Note, the averaged fields in (12) are defined as

$$\begin{aligned} \vec{E}_{av} &= \frac{1}{2}\left[ \vec{E}\Big|_{z=0^+} + \vec{E}\Big|_{z=0^-} \right] \\ \vec{H}_{av} &= \frac{1}{2}\left[ \vec{H}\Big|_{z=0^+} + \vec{H}\Big|_{z=0^-} \right] \end{aligned} \tag{13}$$

Using (12) gives the result derived by Kuester (although we have generalized his result to include magnetoelectric susceptibilities as Kuester originally did not)

$$\begin{aligned} \hat{z} \times \Delta\vec{E} &= -j\omega\mu_0 \overline{\overline{\chi}}_{mm} \bullet \vec{H}_{t,av} - j\omega\sqrt{\varepsilon_0\mu_0}\, \overline{\overline{\chi}}_{me} \bullet \vec{E}_{t,av} \\ &\quad -\hat{z} \times \nabla_t \left[ \overline{\overline{\chi}}_{ee} \bullet \hat{z}E_{z,av} + \sqrt{\mu_0/\varepsilon_0}\, \overline{\overline{\chi}}_{em} \bullet \hat{z}H_{z,av} \right] \\ \hat{z} \times \Delta\vec{H} &= j\omega\varepsilon_0 \overline{\overline{\chi}}_{ee} \bullet \vec{E}_{t,av} + j\omega\sqrt{\varepsilon_0\mu_0}\, \overline{\overline{\chi}}_{em} \bullet \vec{H}_{t,av} \\ &\quad -\hat{z} \times \nabla_t \left[ \overline{\overline{\chi}}_{mm} \bullet \hat{z}H_{z,av} + \sqrt{\varepsilon_0/\mu_0}\, \overline{\overline{\chi}}_{me} \bullet \hat{z}E_{z,av} \right] \end{aligned} \tag{14}$$

It should be noted that (14) appeared in [9] incorrectly without the free space material constants. The susceptibilities are a more natural parameter to use as they easily generalize to nonlinear (higher order $\chi^{(\ )}$) and time-modulated cases. The result (14) is known as the Generalized Sheet Transition Condition or GSTC. In essence a GSTC connects the tangential field discontinuities across a surface to the averaged fields of both sides of the surface through material (surface) proportionality constants. These proportionality constants are sometimes referred to as effective polarization densities as in [2], or surface susceptibilities as in (14), or even surface impedances as in (22). These different nomenclatures all come from the same model with their own interpretations. We discuss these models, their origins and interpretations, and the synthesis techniques built from them next.

### B. Particle Polarizability Model

In [20], Niemi derives the interaction constants for all components including the magnetoelectric polarizabilities. The interaction constant is derived in the same way as Kuester, namely, by finding the fields of a sheet of continuous polarization densities with a hole at the center and adding to it the incident field [20], [93]

$$\vec{E}_{loc} = \vec{E}_{inc} + \overline{\overline{\beta}}_e \bullet \vec{p}_e$$
$$\vec{H}_{loc} = \vec{H}_{inc} + \overline{\overline{\beta}}_m \bullet \vec{p}_m \tag{15}$$

The interaction constants $\overline{\overline{\beta}}$, which model the effect of the array on the local field, are found as

$$\overline{\overline{\beta}}_e = -\mathrm{Re}\left\{ \frac{j\omega\eta_0}{4S}\left(1-\frac{1}{jkR}\right)e^{-jkR} \right\}\overline{\overline{I}}_t + j\left(\frac{\eta_0\varepsilon_0\mu_0\omega^3}{6\pi} - \frac{\eta_0\omega}{2S}\right)\overline{\overline{I}}_t$$
$$\overline{\overline{\beta}}_m = \frac{\overline{\overline{\beta}}_e}{\eta_0^2} \tag{16}$$

Note in (16), Re[ ] denotes the real part operator, $k$ is the free space wavenumber, and $\overline{\overline{I}}_t = \overline{\overline{I}} - \hat{z}\hat{z}$ is the transverse unit dyadic. The interaction constants are a function of the chosen disk radius $R$. In (16), $R = a/1.438$ ($S = a^2$ is the unit cell area) and hence the hole only encompasses a single dipole. The interaction constants were derived for normal incidence but are valid also for small oblique angles of incidence [93]. From (9), (15), and (16), one could construct effective polarizabilities and form a matrix equation similar to (12) only now relating the dipole moments to the incident fields as

$$\begin{bmatrix} \vec{p}_e \\ \vec{p}_m \end{bmatrix} = \begin{bmatrix} \hat{\overline{\overline{\alpha}}}_{ee} & \hat{\overline{\overline{\alpha}}}_{em} \\ \hat{\overline{\overline{\alpha}}}_{me} & \hat{\overline{\overline{\alpha}}}_{mm} \end{bmatrix} \begin{bmatrix} \vec{E}_{inc} \\ \vec{H}_{inc} \end{bmatrix} \tag{17}$$

Relating the dipole moments to the incident field rather than the average fields as in (12) allows for a direct synthesis formulation for stipulated incident, reflected, and transmitted fields. In this case, the effective polarizability densities take a more complex form and are related to the individual particle polarizabilities as

$$\hat{\overline{\overline{\alpha}}}_{ee} = \left(\overline{\overline{I}}_t - \overline{\overline{\alpha}}_{ee}\bullet\overline{\overline{\beta}}_e - \overline{\overline{\alpha}}_{em}\bullet\overline{\overline{\beta}}_m\bullet\left(\overline{\overline{I}}_t - \overline{\overline{\alpha}}_{mm}\bullet\overline{\overline{\beta}}_m\right)^{-1}\bullet\overline{\overline{\alpha}}_{me}\bullet\overline{\overline{\beta}}_e\right)^{-1} \bullet\left(\overline{\overline{\alpha}}_{ee} + \overline{\overline{\alpha}}_{em}\bullet\overline{\overline{\beta}}_m\bullet\left(\overline{\overline{I}}_t - \overline{\overline{\alpha}}_{mm}\bullet\overline{\overline{\beta}}_m\right)^{-1}\bullet\overline{\overline{\alpha}}_{me}\right)$$
$$\hat{\overline{\overline{\alpha}}}_{em} = \left(\overline{\overline{I}}_t - \overline{\overline{\alpha}}_{ee}\bullet\overline{\overline{\beta}}_e - \overline{\overline{\alpha}}_{em}\bullet\overline{\overline{\beta}}_m\bullet\left(\overline{\overline{I}}_t - \overline{\overline{\alpha}}_{mm}\bullet\overline{\overline{\beta}}_m\right)^{-1}\bullet\overline{\overline{\alpha}}_{me}\bullet\overline{\overline{\beta}}_e\right)^{-1} \bullet\left(\overline{\overline{\alpha}}_{em} + \overline{\overline{\alpha}}_{em}\bullet\overline{\overline{\beta}}_m\bullet\left(\overline{\overline{I}}_t - \overline{\overline{\alpha}}_{mm}\bullet\overline{\overline{\beta}}_m\right)^{-1}\bullet\overline{\overline{\alpha}}_{mm}\right)$$
$$\hat{\overline{\overline{\alpha}}}_{me} = \left(\overline{\overline{I}}_t - \overline{\overline{\alpha}}_{mm}\bullet\overline{\overline{\beta}}_m - \overline{\overline{\alpha}}_{me}\bullet\overline{\overline{\beta}}_e\bullet\left(\overline{\overline{I}}_t - \overline{\overline{\alpha}}_{ee}\bullet\overline{\overline{\beta}}_e\right)^{-1}\bullet\overline{\overline{\alpha}}_{em}\bullet\overline{\overline{\beta}}_m\right)^{-1} \bullet\left(\overline{\overline{\alpha}}_{me} + \overline{\overline{\alpha}}_{me}\bullet\overline{\overline{\beta}}_e\bullet\left(\overline{\overline{I}}_t - \overline{\overline{\alpha}}_{ee}\bullet\overline{\overline{\beta}}_e\right)^{-1}\bullet\overline{\overline{\alpha}}_{ee}\right) \tag{18}$$
$$\hat{\overline{\overline{\alpha}}}_{mm} = \left(\overline{\overline{I}}_t - \overline{\overline{\alpha}}_{mm}\bullet\overline{\overline{\beta}}_m - \overline{\overline{\alpha}}_{me}\bullet\overline{\overline{\beta}}_e\bullet\left(\overline{\overline{I}}_t - \overline{\overline{\alpha}}_{ee}\bullet\overline{\overline{\beta}}_e\right)^{-1}\bullet\overline{\overline{\alpha}}_{em}\bullet\overline{\overline{\beta}}_m\right)^{-1} \bullet\left(\overline{\overline{\alpha}}_{mm} + \overline{\overline{\alpha}}_{me}\bullet\overline{\overline{\beta}}_e\bullet\left(\overline{\overline{I}}_t - \overline{\overline{\alpha}}_{ee}\bullet\overline{\overline{\beta}}_e\right)^{-1}\bullet\overline{\overline{\alpha}}_{em}\right)$$

To design an array, the reflected and transmitted fields in terms of the dipole moments must be obtained. The induced dipole moments in (17) correspond to averaged electric and magnetic current sheets with the surface averaged current densities $\vec{J}_s = j\omega\vec{p}_e/S$ and $\vec{M}_s = j\omega\vec{p}_m/S$. The reflected field can be found as the field radiated by these currents [20]

$$\vec{E}_{ref} = -\frac{j\omega}{2S}\left[\eta_0\vec{p}_e - \hat{z}\times\vec{p}_m\right] \tag{19}$$

Similarly, the field transmitted can be found as [20]

$$\vec{E}_{trans} = \vec{E}_{inc} - \frac{j\omega}{2S}\left[\eta_0\vec{p}_e + \hat{z}\times\vec{p}_m\right] \tag{20}$$

By defining the reflected, transmitted, and incident fields and making use of (17), one can synthesize the needed polarizabilities (see section III.E). Then appropriate particle designs can be utilized to realize the necessary polarizabilities by relating their polarizabilities to the effective polarizability densities as in (18). For particle designs and their polarizabilities, see section IV.

### *C. Sheet Impedance Model*

The GSTC can be recast into a form involving sheet impedances and admittances. By ignoring normal surface polarization and magnetization densities, $P_{sz} = M_{sz} = 0$, the GSTC in (14) can be written as

$$\hat{z}\times\Delta\vec{E} = -j\omega\mu_0\overline{\overline{\chi}}_{mm}\bullet\vec{H}_{t,av} - j\omega\sqrt{\varepsilon_0\mu_0}\,\overline{\overline{\chi}}_{me}\bullet\vec{E}_{t,av}$$
$$\hat{z}\times\Delta\vec{H} = j\omega\varepsilon_0\overline{\overline{\chi}}_{ee}\bullet\vec{E}_{t,av} + j\omega\sqrt{\varepsilon_0\mu_0}\,\overline{\overline{\chi}}_{em}\bullet\vec{H}_{t,av} \tag{21}$$

By defining an electric sheet admittance ($\overline{\overline{Y}} = j\omega\epsilon_0\overline{\overline{\chi}}_{ee}$), a magnetic surface impedance ($\overline{\overline{Z}} = j\omega\mu_0\overline{\overline{\chi}}_{mm}$), and dimensionless electromagnetic coupling ($\overline{\overline{\chi}} = j\omega\sqrt{\epsilon_0\mu_0}\overline{\overline{\chi}}_{em}$) and magnetoelectric coupling ($\overline{\overline{\gamma}} = j\omega\sqrt{\epsilon_0\mu_0}\overline{\overline{\chi}}_{me}$) tensors, we can relate the surface current densities established on the metasurface to the average tangential electric and magnetic field

$$\begin{bmatrix} \vec{J}_s \\ \vec{K}_s \end{bmatrix} = \begin{bmatrix} \overline{\overline{Y}} & \overline{\overline{\chi}} \\ \overline{\overline{\gamma}} & \overline{\overline{Z}} \end{bmatrix} \begin{bmatrix} \vec{E}_{t,av} \\ \vec{H}_{t,av} \end{bmatrix} \tag{22}$$

This form of bianisotropic boundary conditions is known as the impedance boundary condition (IBC). The tensors $\overline{\overline{Y}}, \overline{\overline{Z}}, \overline{\overline{\chi}}$, and $\overline{\overline{\gamma}}$ are collectively known as the constituent surface parameters. Note, (22) can also be expressed as [91]

$$\hat{z}\times\vec{E}_{av} = \overline{\overline{Z}}_{ee}\bullet\left[\hat{z}\times\Delta\vec{H}\right] + \overline{\overline{K}}_{me}\bullet\left[\hat{z}\times\Delta\vec{E}\right]$$
$$\hat{z}\times\vec{H}_{av} = \overline{\overline{K}}_{em}\bullet\left[\hat{z}\times\Delta\vec{H}\right] + \overline{\overline{Y}}_{mm}\bullet\left[\hat{z}\times\Delta\vec{E}\right] \tag{23}$$

where $\overline{\overline{Z}}_{ee}$, $\overline{\overline{K}}_{me}$, $\overline{\overline{K}}_{em}$, and $\overline{\overline{Y}}_{mm}$ are the related to the inverses of the tensors appearing in (22). Each tensor in (22) and (23) are of dimension $2\times 2$, for example, the electric surface impedance tensor is

$$\overline{\overline{Z}}_{ee} = \begin{bmatrix} Z_{ee}^{xx} & Z_{ee}^{xy} \\ Z_{ee}^{yx} & Z_{ee}^{yy} \end{bmatrix} \tag{24}$$

and hence the tangential average fields appearing in (22) are of dimension $2\times 1$. The IBC has also been extended to include the normal constituent surface parameters [94]. In this case, each vector in (22) contains three components and each tensor contains 9 components. These 3-dimensional constituent surface parameter tensors are expressed as equivalent spatially dispersive 2-dimensional constituent surface parameters [95]. The spatially dispersive 2-dimensional constituent surface parameters can be used to mimic non-reciprocal phenomena from ordinary reciprocal materials. The inclusion of nor-mal constituent surface parameters can be exploited to significantly expand the scope of the electromagnetic phenomena that can be engineered with reciprocal materials [89], [94], [96]. However, in most cases, modelling of only the in-plane surface material parameters is sufficient as argued in [91] since by uniqueness theorems, the tangential field components are sufficient to define the full field vectors. By relating the desired scattering parameters to the constituent surface parameters, one can obtain the description of the required constituent surface parameters (see section III.B and III.C). Alternatively, one can construct an integral equation around the IBC and synthesize the constituent surface parameters needed to achieve a desired field transformation (see section III.D). Then, printed circuit techniques [97] can be used in conjunction with multi-sheet realizations of bianisotropic particles (see Section IV.C) to realize the constituent surface parameters.

### D. *Susceptibility Model*

Omitting normal polarization densities, (14) can be written as

$$\begin{aligned}\hat{z}\times\Delta\vec{E} &= -j\omega\mu_0\bar{\bar{\chi}}_{mm}\bullet\vec{H}_{t,av} - j\omega\sqrt{\varepsilon_0\mu_0}\,\bar{\bar{\chi}}_{me}\bullet\vec{E}_{t,av}\\ \hat{z}\times\Delta\vec{H} &= j\omega\varepsilon_0\bar{\bar{\chi}}_{ee}\bullet\vec{E}_{t,av} + j\omega\sqrt{\varepsilon_0\mu_0}\,\bar{\bar{\chi}}_{em}\bullet\vec{H}_{t,av}\end{aligned} \tag{25}$$

The tensors in (25) are the electric, $\bar{\bar{\chi}}_{ee}$, the magnetic, $\bar{\bar{\chi}}_{mm}$,, the electromagnetic, $\bar{\bar{\chi}}_{em}$, and the magnetoelectric, $\bar{\bar{\chi}}_{me}$, surface susceptibility tensors. Each of these tensors are of dimension $3\times3$. For example, the electric susceptibility tensor is

$$\bar{\bar{\chi}}_{ee} = \begin{bmatrix}\chi_{ee}^{xx} & \chi_{ee}^{xy} & \chi_{ee}^{xz}\\ \chi_{ee}^{yx} & \chi_{ee}^{yy} & \chi_{ee}^{yz}\\ \chi_{ee}^{zx} & \chi_{ee}^{zy} & \chi_{ee}^{zz}\end{bmatrix} \tag{26}$$

Thus, knowing the tangential electric and magnetic field on both sides of the surface, use of (25) allows one to solve for the surface susceptibilities required to achieve the field transformation (see section III.A). Then, polarizable particles which realize the susceptibilities can be obtained by relating the susceptibilities to the effective polarizability densities as [91]

$$\begin{aligned}
\bar{\bar{\chi}}_{ee} &= \frac{1}{\varepsilon_0}\bar{\bar{C}}_p^{-1}\bullet\left[\hat{\bar{\bar{\alpha}}}_{ee} + \frac{j\omega}{2\eta_0}\hat{\bar{\bar{\alpha}}}_{em}\bullet\left(S\bar{\bar{I}}_t - \hat{\bar{\bar{\alpha}}}_{mm}\frac{j\omega}{2\eta_0}\right)^{-1}\bullet\hat{\bar{\bar{\alpha}}}_{me}\right]\\
\bar{\bar{\chi}}_{em} &= \frac{1}{\sqrt{\varepsilon_0\mu_0}}\bar{\bar{C}}_p^{-1}\bullet\left[\hat{\bar{\bar{\alpha}}}_{em} + \frac{j\omega}{2\eta_0}\hat{\bar{\bar{\alpha}}}_{em}\bullet\left(S\bar{\bar{I}}_t - \hat{\bar{\bar{\alpha}}}_{mm}\frac{j\omega}{2\eta_0}\right)^{-1}\bullet\hat{\bar{\bar{\alpha}}}_{mm}\right]\\
\bar{\bar{\chi}}_{me} &= \sqrt{\frac{\mu_0}{\varepsilon_0}}\bar{\bar{C}}_m^{-1}\bullet\left[\hat{\bar{\bar{\alpha}}}_{me} + \frac{j\omega\eta_0}{2}\hat{\bar{\bar{\alpha}}}_{me}\bullet\left(S\bar{\bar{I}}_t - \hat{\bar{\bar{\alpha}}}_{ee}\frac{j\omega\eta_0}{2}\right)^{-1}\bullet\hat{\bar{\bar{\alpha}}}_{ee}\right]\\
\bar{\bar{\chi}}_{mm} &= \bar{\bar{C}}_m^{-1}\bullet\left[\hat{\bar{\bar{\alpha}}}_{mm} + \frac{j\omega\eta_0}{2}\hat{\bar{\bar{\alpha}}}_{me}\bullet\left(S\bar{\bar{I}}_t - \hat{\bar{\bar{\alpha}}}_{ee}\frac{j\omega\eta_0}{2}\right)^{-1}\bullet\hat{\bar{\bar{\alpha}}}_{em}\right]\\
\bar{\bar{C}}_p &= S\bar{\bar{I}}_t - \hat{\bar{\bar{\alpha}}}_{ee}\frac{j\omega\eta_0}{2} + \frac{\omega^2}{4}\hat{\bar{\bar{\alpha}}}_{em}\bullet\left(S\bar{\bar{I}}_t - \hat{\bar{\bar{\alpha}}}_{mm}\frac{j\omega}{2\eta_0}\right)^{-1}\bullet\hat{\bar{\bar{\alpha}}}_{me}\\
\bar{\bar{C}}_m &= S\bar{\bar{I}}_t - \hat{\bar{\bar{\alpha}}}_{mm}\frac{j\omega}{2\eta_0} + \frac{\omega^2}{4}\hat{\bar{\bar{\alpha}}}_{me}\bullet\left(S\bar{\bar{I}}_t - \hat{\bar{\bar{\alpha}}}_{ee}\frac{j\omega\eta_0}{2}\right)^{-1}\bullet\hat{\bar{\bar{\alpha}}}_{em}
\end{aligned} \tag{27}$$

and then relating the effective polarizability densities to the individual particle polarizabilities as in (18). For particle designs and their polarizabilities, see section IV.

## III. Metasurface Synthesis Methods

### A. *Synthesis using the Polarizability Model: Reflection and Transmission to Effective Polarizabilities*

The polarizabilities of the particles necessary to achieve a particular field transformation can be synthesized directly [20] starting from the polarizability model. Assuming plane wave incidence, (15)-(17) can be combined and written as

$$\begin{aligned}
\vec{E}_r &= -\frac{j\omega}{2S}\left[\left(\eta_0\hat{\bar{\bar{\alpha}}}_{ee} - \hat{\bar{\bar{\alpha}}}_{em}\bullet\bar{\bar{J}}_t\right) - \hat{z}\times\left(\hat{\bar{\bar{\alpha}}}_{me} - \frac{1}{\eta_0}\hat{\bar{\bar{\alpha}}}_{mm}\bullet\bar{\bar{J}}_t\right)\right]\bullet\vec{E}_{inc}\\
\vec{E}_t &= \left(1 - \frac{j\omega}{2S}\left[\left(\eta_0\hat{\bar{\bar{\alpha}}}_{ee} - \hat{\bar{\bar{\alpha}}}_{em}\bullet\bar{\bar{J}}_t\right) + \hat{z}\times\left(\hat{\bar{\bar{\alpha}}}_{me} - \frac{1}{\eta_0}\hat{\bar{\bar{\alpha}}}_{mm}\bullet\bar{\bar{J}}_t\right)\right]\right)\bullet\vec{E}_{inc}
\end{aligned} \tag{28}$$

where $\bar{\bar{J}}_t = \hat{z}\times\bar{\bar{I}}$ and $\bar{\bar{I}}$ is the $2\times2$ unit dyadic. The metasurface is synthesized by stipulating the incident, reflected, and transmitted fields and solving (28) for the required polarizabilities. Then particles can be chosen according to (18). Several uses of this synthesis approach will also be highlighted in section V (see Table II).

### B. *Synthesis using the Sheet Impedance Model: S-parameter to Constituent Surface Parameters*

When the metasurface shown in Fig. 1 is illuminated by an incident plane wave, scattering parameters (S-parameters) can be defined as the ratio of the scattered electric field into region $n$ to the incident electric field from region $m$

$$\bar{\bar{S}}_{nm} = \begin{bmatrix}S_{nm}^{xx} & S_{nm}^{xy}\\ S_{nm}^{yx} & S_{nm}^{yy}\end{bmatrix} \tag{29}$$

Treating all possible excitation/response combinations, (29) is generalized to

$$\bar{\bar{S}} = \begin{bmatrix}\bar{\bar{S}}_{11} & \bar{\bar{S}}_{12}\\ \bar{\bar{S}}_{21} & \bar{\bar{S}}_{22}\end{bmatrix} \tag{30}$$

The S-parameters $\bar{\bar{S}}_{11}$ and $\bar{\bar{S}}_{22}$ are the reflection coefficients when viewed from regions 1 and 2, respectively. Similarly, the S-parameters $\bar{\bar{S}}_{21}$ and $\bar{\bar{S}}_{12}$ are the transmission coefficients when viewed from regions 1 and 2, respectively. The S-parameters in (30) can be related to the constituent surface parameters of the IBC in (22) from [15], [65]

$$\begin{bmatrix}\bar{\bar{S}}_{11} & \bar{\bar{S}}_{12}\\ \bar{\bar{S}}_{21} & \bar{\bar{S}}_{22}\end{bmatrix} = \begin{bmatrix}\frac{\bar{\bar{Y}}}{2} - \frac{\bar{\bar{\chi}}\bar{\bar{n}}}{2\eta_1} + \frac{\bar{\bar{I}}}{\eta_1} & \frac{\bar{\bar{Y}}}{2} - \frac{\bar{\bar{\chi}}\bar{\bar{n}}}{2\eta_2} + \frac{\bar{\bar{I}}}{\eta_2}\\ -\frac{\bar{\bar{Z}}\bar{\bar{n}}}{2\eta_1} + \frac{\bar{\bar{\gamma}}}{2} - \bar{\bar{n}} & \frac{\bar{\bar{Z}}\bar{\bar{n}}}{2\eta_2} + \frac{\bar{\bar{\gamma}}}{2} + \bar{\bar{n}}\end{bmatrix}^{-1} \bullet \begin{bmatrix}-\frac{\bar{\bar{Y}}}{2} - \frac{\bar{\bar{\chi}}\bar{\bar{n}}}{2\eta_1} + \frac{\bar{\bar{I}}}{\eta_1} & -\frac{\bar{\bar{Y}}}{2} + \frac{\bar{\bar{\chi}}\bar{\bar{n}}}{2\eta_2} + \frac{\bar{\bar{I}}}{\eta_2}\\ -\frac{\bar{\bar{Z}}\bar{\bar{n}}}{2\eta_1} - \frac{\bar{\bar{\gamma}}}{2} + \bar{\bar{n}} & \frac{\bar{\bar{Z}}\bar{\bar{n}}}{2\eta_2} - \frac{\bar{\bar{\gamma}}}{2} - \bar{\bar{n}}\end{bmatrix} \tag{31}$$

where

$$\overline{\overline{I}}=\begin{bmatrix}1 & 0\\ 0 & 1\end{bmatrix},\quad \overline{\overline{n}}=\begin{bmatrix}0 & -1\\ 1 & 0\end{bmatrix}$$

Note, the notation $\bar{\bar{a}}\bar{\bar{b}}$ juxtapositioning two tensors means to multiply the matrix representation of the tensors using the usual rules of matrix multiplication. Inversely, the constituent surface parameters can be defined in terms of the S-parameters as [15], [65]

$$\begin{bmatrix}\overline{\overline{Y}} & \overline{\overline{\chi}}\\ \overline{\overline{\gamma}} & \overline{\overline{Z}}\end{bmatrix}=2\begin{bmatrix}\frac{\overline{\overline{I}}}{\eta_1}-\frac{\overline{\overline{S}}_{11}}{\eta_1}-\frac{\overline{\overline{S}}_{21}}{\eta_2} & \frac{\overline{\overline{I}}}{\eta_2}-\frac{\overline{\overline{S}}_{12}}{\eta_1}-\frac{\overline{\overline{S}}_{22}}{\eta_2}\\ \overline{\overline{n}}+\overline{\overline{n}}\overline{\overline{S}}_{11}-\overline{\overline{n}}\overline{\overline{S}}_{21} & -\overline{\overline{n}}+\overline{\overline{n}}\overline{\overline{S}}_{12}-\overline{\overline{n}}\overline{\overline{S}}_{22}\end{bmatrix} \bullet \begin{bmatrix}\overline{\overline{I}}+\overline{\overline{S}}_{11}+\overline{\overline{S}}_{21} & \overline{\overline{I}}+\overline{\overline{S}}_{12}+\overline{\overline{S}}_{22}\\ \frac{\overline{\overline{n}}}{\eta_1}-\frac{\overline{\overline{n}}\overline{\overline{S}}_{11}}{\eta_1}+\frac{\overline{\overline{n}}\overline{\overline{S}}_{21}}{\eta_2} & -\frac{\overline{\overline{n}}}{\eta_2}-\frac{\overline{\overline{n}}\overline{\overline{S}}_{12}}{\eta_1}+\frac{\overline{\overline{n}}\overline{\overline{S}}_{22}}{\eta_2}\end{bmatrix}^{-1} \tag{32}$$

Using (32) one can synthesize a metasurface's constituent surface parameters, and using (31), one can analyze a given metasurfaces response to plane wave fields. Network parameter matrix representations which can be cascaded allow multi-layer metasurfaces made from stacks of bianisotropic sheets and dielectric spacers to be analyzed or synthesized (see section IV.C). Examples of metasurfaces synthesized using this approach will be presented in section V (see Table II).

### *C. Synthesis using the Sheet Impedance Model: Wave Matrices to Constituent Surface Parameters*

Another approach to synthesize the constituent surface parameters are to relate them to wave matrices [66]. Wave matrices relate the forward and backward propagating fields in region 1 to those in region 2 (see Fig. 1) in the following manner (+ means forward travelling modes, from region 1 to region 2, and – means backward travelling modes, from region 2 to region 1)

$$\begin{bmatrix}\vec{E}_1^{+}\\ \vec{E}_1^{-}\end{bmatrix}=\begin{bmatrix}\overline{\overline{M}}_{11} & \overline{\overline{M}}_{12}\\ \overline{\overline{M}}_{21} & \overline{\overline{M}}_{22}\end{bmatrix}\begin{bmatrix}\vec{E}_2^{+}\\ \vec{E}_2^{-}\end{bmatrix} \tag{33}$$

The relationship between the wave matrices and the scattering parameters is

$$\begin{bmatrix}\overline{\overline{M}}_{11} & \overline{\overline{M}}_{12}\\ \overline{\overline{M}}_{21} & \overline{\overline{M}}_{22}\end{bmatrix}=\begin{bmatrix}\overline{\overline{I}} & \overline{\overline{0}}\\ \overline{\overline{S}}_{11} & \overline{\overline{S}}_{12}\end{bmatrix}\begin{bmatrix}\overline{\overline{S}}_{21} & \overline{\overline{S}}_{22}\\ \overline{\overline{0}} & \overline{\overline{I}}\end{bmatrix}^{-1} \tag{34}$$

The wave matrix for a bianisotropic sheet consisting of the constituent surface parameters $\bar{\bar{Y}}, \bar{\bar{Z}}, \bar{\bar{\gamma}}$, and $\bar{\bar{\chi}}$ is

$$\begin{bmatrix}\overline{\overline{M}}_{11} & \overline{\overline{M}}_{12}\\ \overline{\overline{M}}_{21} & \overline{\overline{M}}_{22}\end{bmatrix}=\begin{bmatrix}\frac{\overline{\overline{Y}}}{2}+\frac{\overline{\overline{\chi}}\overline{\overline{n}}}{2\eta_1}-\frac{\overline{\overline{I}}}{\eta_1} & \frac{\overline{\overline{Y}}}{2}-\frac{\overline{\overline{\chi}}\overline{\overline{n}}}{2\eta_1}+\frac{\overline{\overline{I}}}{\eta_1}\\ \frac{\overline{\overline{Z}}\overline{\overline{n}}}{2\eta_1}+\frac{\overline{\overline{\gamma}}}{2}-\overline{\overline{n}} & -\frac{\overline{\overline{Z}}\overline{\overline{n}}}{2\eta_1}+\frac{\overline{\overline{\gamma}}}{2}-\overline{\overline{n}}\end{bmatrix}^{-1} \bullet \begin{bmatrix}-\frac{\overline{\overline{Y}}}{2}-\frac{\overline{\overline{\chi}}\overline{\overline{n}}}{2\eta_2}-\frac{\overline{\overline{I}}}{\eta_2} & -\frac{\overline{\overline{Y}}}{2}+\frac{\overline{\overline{\chi}}\overline{\overline{n}}}{2\eta_2}+\frac{\overline{\overline{I}}}{\eta_2}\\ -\frac{\overline{\overline{Z}}\overline{\overline{n}}}{2\eta_2}-\frac{\overline{\overline{\gamma}}}{2}-\overline{\overline{n}} & \frac{\overline{\overline{Z}}\overline{\overline{n}}}{2\eta_2}-\frac{\overline{\overline{\gamma}}}{2}-\overline{\overline{n}}\end{bmatrix} \tag{35}$$

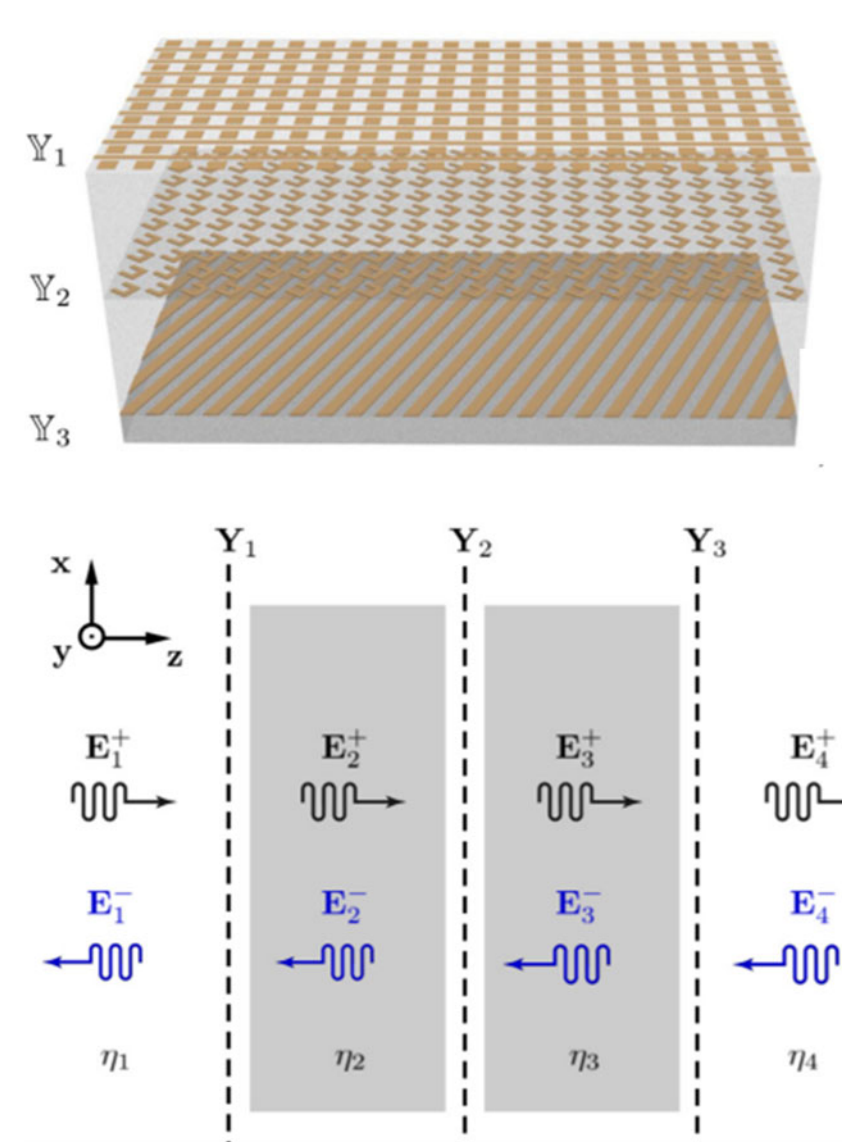

Fig. 2. A metasurface consisting of three cascaded electric sheet admittances designed to realize a stipulated S matrix (reprinted with permission from [66]).

By relating (34) and (35), the constituent surface parameters can be obtained in terms of the desired S-parameters.

The real power of the wave matrix approach, however, lies in its ability to easily model cascades of sheets and dielectric spacers. For example, a common way to realize a bianisotropic boundary is the three-sheet or four-sheet method (which allows arbitrary polarization conversions), as shown in see section IV.C. The three-sheet realization of a bianisotropic boundary consists of a stack of three electric sheet admittances each separated by dielectric spacers, as shown in Fig. 2. The wave matrix associated with the cascaded metasurface in Fig. 2 is,

$$\begin{aligned}\begin{bmatrix}\overline{\overline{M}}_{11} & \overline{\overline{M}}_{12}\\ \overline{\overline{M}}_{21} & \overline{\overline{M}}_{22}\end{bmatrix}&=\left(\overline{\overline{t}}_1\otimes\overline{\overline{I}}+\frac{\eta_1}{2}\overline{\overline{e}}\otimes\overline{\overline{Y}}_1\right)\left(\overline{\overline{\phi}}_2\otimes\overline{\overline{I}}\right)\\ &\bullet\left(\overline{\overline{t}}_2\otimes\overline{\overline{I}}+\frac{\eta_2}{2}\overline{\overline{e}}\otimes\overline{\overline{Y}}_2\right)\left(\overline{\overline{\phi}}_3\otimes\overline{\overline{I}}\right)\\ &\bullet\left(\overline{\overline{t}}_3\otimes\overline{\overline{I}}+\frac{\eta_3}{2}\overline{\overline{e}}\otimes\overline{\overline{Y}}_3\right)\end{aligned} \tag{36}$$

The first term in parenthesis is associated with the first dielectric interface and sheet admittance. The second term represents the phase delay of the first dielectric spacer. Similar associations follow for the remaining terms. The operator $\otimes$ denotes the Kronecker tensor product, defined as

$$\overline{\overline{A}}_{n\times m}\otimes\overline{\overline{B}}_{p\times l}=\begin{bmatrix}a_{11}\overline{\overline{B}} & \cdots & a_{1m}\overline{\overline{B}}\\ \vdots & \ddots & \vdots\\ a_{n1}\overline{\overline{B}} & \cdots & a_{nm}\overline{\overline{B}}\end{bmatrix}_{np\times ml}. \tag{37}$$

The definitions of the various tensors in (36) are

$$\begin{aligned}&\overline{\overline{t}}_i=\frac{1}{T}\begin{bmatrix}1 & R\\ R & 1\end{bmatrix},\ R=\frac{\eta_{i+1}-\eta_i}{\eta_{i+1}+\eta_i},\ T=\frac{2\eta_{i+1}}{\eta_{i+1}+\eta_i}\\ &\overline{\overline{I}}=\begin{bmatrix}1 & 0\\ 0 & 1\end{bmatrix},\ \overline{\overline{e}}=\begin{bmatrix}1 & 1\\ -1 & -1\end{bmatrix},\ \overline{\overline{\phi}}_i=\begin{bmatrix}e^{j\phi_i} & 0\\ 0 & e^{-j\phi_i}\end{bmatrix}\end{aligned} \tag{38}$$

The total wave matrix of (36) is written in terms of the desired S-matrix in (34) to synthesize the necessary sheet admittances. Equating (36) and (34) results in the following design process [66]: First, the middle sheet admittance, $\bar{\bar{Y}}_2$ is solved for as

$$\bar{\bar{e}}\otimes\bar{\bar{Y}}_2=\frac{1}{a_2}\left(\left(\bar{\bar{e}}\otimes\bar{\bar{I}}\right)\bar{\bar{S}}_1\bar{\bar{S}}_2^{-1}\left(\bar{\bar{e}}\otimes\bar{\bar{I}}\right)-\left(\bar{\bar{e}}\bar{\bar{t}}_1\bar{\bar{\phi}}_2\bar{\bar{t}}_2\bar{\bar{\phi}}_3\bar{\bar{t}}_3\bar{\bar{e}}\right)\otimes\bar{\bar{I}}\right) \quad (39)$$

where $a_2$ satisfies $\frac{\eta_2}{2}\left(\bar{\bar{e}}\bar{\bar{t}}_1\bar{\bar{\phi}}_2\bar{\bar{e}}\bar{\bar{\phi}}_3\bar{\bar{t}}_3\bar{\bar{e}}\right)=a_2\bar{\bar{e}}$. Then, the outer sheet $\bar{\bar{Y}}_1$ is found in terms of $\bar{\bar{Y}}_2$ from

$$\begin{aligned}\bar{\bar{e}}\otimes\bar{\bar{Y}}_1=\frac{1}{a_1}\Bigg(&\bar{\bar{S}}_1\bar{\bar{S}}_2^{-1}\left(\bar{\bar{e}}\otimes\bar{\bar{I}}\right)-\left(\bar{\bar{t}}_1\bar{\bar{\phi}}_2\bar{\bar{t}}_2\bar{\bar{\phi}}_3\bar{\bar{t}}_3\bar{\bar{e}}\right)\otimes\bar{\bar{I}}\\ &-\frac{\eta_2}{2}\left(\bar{\bar{t}}_1\bar{\bar{\phi}}_2\bar{\bar{e}}\bar{\bar{\phi}}_3\bar{\bar{t}}_3\bar{\bar{e}}\right)\otimes\bar{\bar{Y}}_2\Bigg)\left(\bar{\bar{I}}\otimes\left(\bar{\bar{I}}+\frac{a_{12}}{a_1}\bar{\bar{Y}}_2\right)^{-1}\right)\end{aligned} \quad (40)$$

where $a_1$ and $a_{12}$ are given by $\frac{\eta_1}{2}\left(\bar{\bar{e}}\bar{\bar{\phi}}_2\bar{\bar{t}}_2\bar{\bar{\phi}}_3\bar{\bar{t}}_3\bar{\bar{e}}\right)=a_1\bar{\bar{e}}$ and $\frac{\eta_1\eta_2}{4}\left(\bar{\bar{e}}\bar{\bar{\phi}}_2\bar{\bar{e}}\bar{\bar{\phi}}_3\bar{\bar{t}}_3\bar{\bar{e}}\right)=a_{12}\bar{\bar{e}}$. And finally, the outer sheet $\bar{\bar{Y}}_3$ is found in terms of $\bar{\bar{Y}}_2$ from

$$\begin{aligned}\bar{\bar{e}}\otimes\bar{\bar{Y}}_3=\frac{1}{a_3}&\left(\bar{\bar{I}}\otimes\left(\bar{\bar{I}}+\frac{a_{23}}{a_3}\bar{\bar{Y}}_2\right)^{-1}\right)\Bigg(\left(\bar{\bar{e}}\otimes\bar{\bar{I}}\right)\bar{\bar{S}}_1\bar{\bar{S}}_2^{-1}\\ &-\left(\bar{\bar{e}}\bar{\bar{t}}_1\bar{\bar{\phi}}_2\bar{\bar{t}}_2\bar{\bar{\phi}}_3\bar{\bar{t}}_3\right)\otimes\bar{\bar{I}}-\frac{\eta_2}{2}\left(\bar{\bar{e}}\bar{\bar{t}}_1\bar{\bar{\phi}}_2\bar{\bar{e}}\bar{\bar{\phi}}_3\bar{\bar{t}}_3\right)\otimes\bar{\bar{Y}}_2\Bigg)\end{aligned} \quad (41)$$

where $a_3$ and $a_{23}$ are given by $\frac{\eta_3}{2}\left(\bar{\bar{e}}\bar{\bar{t}}_1\bar{\bar{\phi}}_2\bar{\bar{t}}_2\bar{\bar{\phi}}_3\bar{\bar{e}}\right)=a_3\bar{\bar{e}}$ and $\frac{\eta_2\eta_3}{4}\left(\bar{\bar{e}}\bar{\bar{t}}_1\bar{\bar{\phi}}_2\bar{\bar{e}}\bar{\bar{\phi}}_3\bar{\bar{e}}\right)=a_{23}\bar{\bar{e}}$. The preceding expressions provide an analytical approach to solve for the required sheet admittances to realize bianisotropic boundaries using three cascaded electric sheets. Several uses of this synthesis approach will be highlighted in section V (see Table II).

### *D. Synthesis using the Sheet Impedance Model: Integral Equation to Constituent Surface Parameters*

An integral equation can be constructed from the Sheet Impedance Model of (23) in terms of the unknown surface electric and magnetic current densities, $\vec{J}_s$ and $\vec{M}_s$ [34], [37], [85], [98]

$$\begin{aligned}\hat{z}\times\vec{E}_{av}^{inc}&=-\hat{z}\times\vec{E}_{av}^{sca}+\bar{\bar{Z}}_{ee}\bullet\vec{J}_s-\bar{\bar{K}}_{em}\bullet\vec{M}_s\\ \hat{z}\times\vec{H}_{av}^{inc}&=-\hat{z}\times\vec{H}_{av}^{sca}+\bar{\bar{K}}_{em}\bullet\vec{J}_s-\bar{\bar{Y}}_{mm}\bullet\vec{M}_s\end{aligned} \quad (42)$$

where $\vec{E}_{av}^{inc}$ and $\vec{H}_{av}^{inc}$ are the incident fields and $\vec{E}_{av}^{sca}$ and $\vec{H}_{av}^{sca}$ are the scattered fields found through spatial convolution of the appropriate current density with the corresponding Green's function. For example, in 2-dimensions (out of plane wavenumber is zero) and for a finite bianisotropic metasurface of width $w$ defined along the $y$-axis, these are

$$\begin{aligned}\vec{E}_z^{sca}(x,y)&=\frac{-\eta_0k_0}{4}\int_{-w/2}^{w/2}\vec{J}_z(y')H_0^{(2)}\left(k_0\left|y-y'\right|\right)dy'\\ \vec{H}^{sca}(x,y)&=\frac{-1}{4\eta_0k_0}\left(k_0^2+\frac{\partial^2}{\partial y^2}\right)\\ &\quad\int_{-w/2}^{w/2}\vec{M}_y(y')H_0^{(2)}\left(k_0\left|y-y'\right|\right)dy'\end{aligned} \quad (43)$$

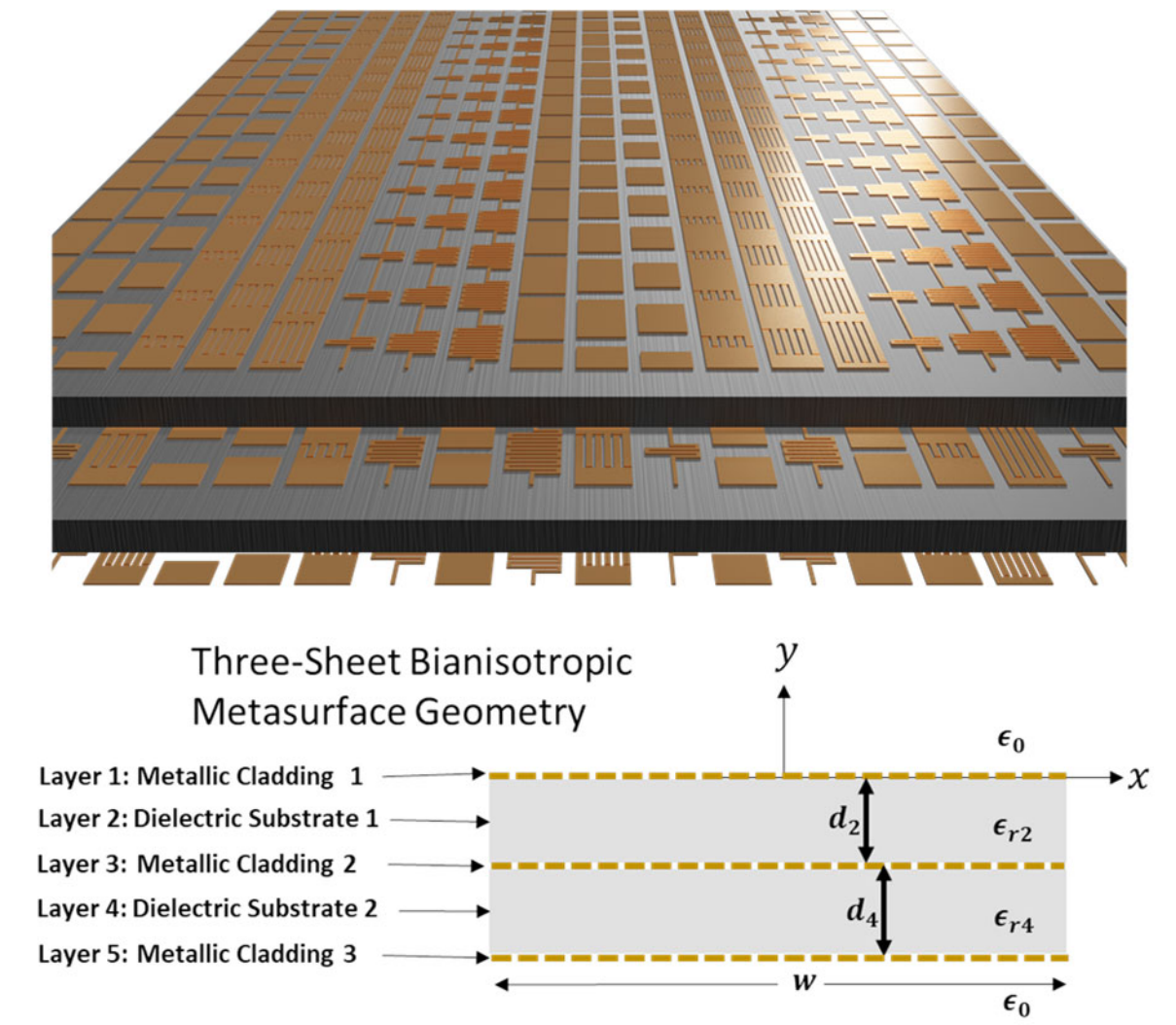

Fig. 3. A three-sheet metasurface consisting of three cascaded electric sheet admittances designed to realize a stipulated S-matrix.

where $H_0^{(2)}(\,)$ is the Hankel function of the second kind and of order zero. By applying the method of moments technique [99], the integral equation can be transformed into a linear matrix equation as [34]

$$\begin{bmatrix}\left[V_e^{inc}\right]\\ \left[V_h^{inc}\right]\end{bmatrix}=\begin{bmatrix}[Z_e]+[Z_{ee}] & -[K_{em}]\\ [K_{em}] & [Z_m]-[Y_{mm}]\end{bmatrix}\begin{bmatrix}[I_e]\\ [I_m]\end{bmatrix} \quad (44)$$

where $[V_e^{inc}]$ and $[V_h^{inc}]$ are the moment method voltage vectors for the incident electric and magnetic fields, $[Z_e]$ and $[Z_m]$ are the coupling matrices between basis and testing expansion functions placed along the bianisotropic sheet, and $[Z_{ee}]$, $[K_{em}]$, $[K_{em}]$, and $[Y_{mm}]$ are diagonal matrices of the constitutive surface parameters associated with the total fields in the last two terms of each line of (42)

$$\begin{aligned}\left[V_e^{tot}\right]&=[Z_{ee}][I_e]-[K_{em}][I_m]\\ \left[V_h^{tot}\right]&=-[Y_{mm}][I_m]+[K_{me}][I_e]\end{aligned} \quad (45)$$

Given the desired total fields on the left hand side of (45), the integral equation in (42) can be solved for the unknown current densities [37], [81].

Recall that a common way to realize a bianisotropic boundary is the three-sheet method (see section IV.C). The three-sheet realization of a bianisotropic boundary consists of a stack of three electric sheet impedances (layers 1, 3, and 5) each separated by dielectric spacers (layers 2 and 4), as shown in Fig. 3. For a 2-dimensional metasurface (sheet admittances vary only along one coordinate) illuminated with TE-polarized waves, (44) can be written for the three-sheet metasurface in Fig. 3 as [37], [57]

$$\begin{bmatrix} [V_e^{inc,1}] \\ [V_e^{inc,2}] \\ [V_e^{inc,3}] \\ [V_e^{inc,4}] \\ [V_e^{inc,5}] \end{bmatrix} = \left( \begin{bmatrix} [Z_e^{11}] & [Z_e^{12}] & [Z_e^{13}] & [Z_e^{14}] & [Z_e^{15}] \\ [Z_e^{21}] & [Z_e^{22}] & [Z_e^{23}] & [Z_e^{24}] & [Z_e^{25}] \\ [Z_e^{31}] & [Z_e^{32}] & [Z_e^{33}] & [Z_e^{34}] & [Z_e^{35}] \\ [Z_e^{41}] & [Z_e^{42}] & [Z_e^{43}] & [Z_e^{44}] & [Z_e^{45}] \\ [Z_e^{51}] & [Z_e^{52}] & [Z_e^{53}] & [Z_e^{54}] & [Z_e^{55}] \end{bmatrix} + \begin{bmatrix} [Z_{ee}^1] & 0 & 0 & 0 & 0 \\ 0 & [Z_{vol}^2] & 0 & 0 & 0 \\ 0 & 0 & [Z_{ee}^3] & 0 & 0 \\ 0 & 0 & 0 & [Z_{vol}^4] & 0 \\ 0 & 0 & 0 & 0 & [Z_{ee}^5] \end{bmatrix} \right) \begin{bmatrix} [I_e^1] \\ [I_e^2] \\ [I_e^3] \\ [I_e^4] \\ [I_e^5] \end{bmatrix} \tag{46}$$

where $[Z_{vol}^2] = diag[(j\omega\epsilon_0(\epsilon_{r2}-1))^{-1}]$ and $[Z_{vol}^4] = diag[(j\omega\epsilon_0(\epsilon_{r4}-1))^{-1}]$ and $diag[\ ]$ refers to the construction of a diagonal matrix with the argument appearing repeated along the diagonal. The superscripts indicate the various layer numbers. Thus, the matrices $[Z_e^{ij}]$ represent the coupling between basis (currents) on layer $j$ and testing functions (observations) on layer $i$. Defining the desired total fields on layers 1 (incident side of metasurface) and 5 (transmitted side of metasurface), the unknown sheet impedances $[Z_{ee}^1]$ and $[Z_{ee}^5]$ can be replaced by

$$\begin{aligned} \left[V_e^{tot,1}\right] &= \left[Z_{ee}^1\right]\left[I_e^1\right] \\ \left[V_e^{tot,5}\right] &= \left[Z_{ee}^5\right]\left[I_e^5\right] \end{aligned} \tag{47}$$

This leaves only $[Z_{ee}^3]$ left undetermined. To determine these unknown sheet impedances, an iterative technique that was originally introduced in [57] for dual band metasurfaces can be applied by choosing the second frequency a few hertz above the first. The process results in the description of the electric sheet impedances of layers 1, 3, and 5 necessary to achieve the desired field transformation defined in (47). The advantage of adopting the integral equation modelling technique over the approaches in III.B or III.C is that the integral equation modelling method accounts accurately for transverse coupling between elements within each layer and layer-to-layer coupling. It also accounts for the finite dimensions of the metasurface. Other approaches do not model the transverse coupling and solve the problem by including conducting baffles separating unit cells (see [50], [100] for example). The integral equation approach avoids the need for any of these unit cell separators. Several uses of the integral equation modelling technique will be presented in section V (see Table II).

### *E. Synthesis using the Susceptibility Model: Reflection and Transmission to Surface Susceptibilities*

One approach for direct synthesis of the tangential susceptibilities can be found in [90]–[92] and will be summarized here. Equation (25) can be written in matrix form as

$$\begin{bmatrix} \Delta H_y \\ \Delta H_x \\ \Delta E_y \\ \Delta E_x \end{bmatrix} = \begin{bmatrix} \tilde{\chi}_{ee}^{xx} & \tilde{\chi}_{ee}^{xy} & \tilde{\chi}_{em}^{xx} & \tilde{\chi}_{em}^{xy} \\ \tilde{\chi}_{ee}^{yx} & \tilde{\chi}_{ee}^{yy} & \tilde{\chi}_{em}^{yx} & \tilde{\chi}_{em}^{yy} \\ \tilde{\chi}_{me}^{xx} & \tilde{\chi}_{me}^{xy} & \tilde{\chi}_{mm}^{xx} & \tilde{\chi}_{mm}^{xy} \\ \tilde{\chi}_{me}^{yx} & \tilde{\chi}_{me}^{yy} & \tilde{\chi}_{mm}^{yx} & \tilde{\chi}_{mm}^{yy} \end{bmatrix} \begin{bmatrix} E_{x,av} \\ E_{y,av} \\ H_{x,av} \\ H_{y,av} \end{bmatrix} \tag{48}$$

In (48), the $'\sim'$ symbol denotes normalized susceptibilities which are related to the unnormalized susceptibilities as

$$\begin{bmatrix} \chi_{ee}^{xx} & \chi_{ee}^{xy} & \chi_{em}^{xx} & \chi_{em}^{xy} \\ \chi_{ee}^{yx} & \chi_{ee}^{yy} & \chi_{em}^{yx} & \chi_{em}^{yy} \\ \chi_{me}^{xx} & \chi_{me}^{xy} & \chi_{mm}^{xx} & \chi_{mm}^{xy} \\ \chi_{me}^{yx} & \chi_{me}^{yy} & \chi_{mm}^{yx} & \chi_{mm}^{yy} \end{bmatrix} = \begin{bmatrix} \frac{j}{\omega\varepsilon_0}\tilde{\chi}_{ee}^{xx} & \frac{j}{\omega\varepsilon_0}\tilde{\chi}_{ee}^{xy} & \frac{j}{k_0}\tilde{\chi}_{em}^{xx} & \frac{j}{k_0}\tilde{\chi}_{em}^{xy} \\ \frac{-j}{\omega\varepsilon_0}\tilde{\chi}_{ee}^{yx} & \frac{-j}{\omega\varepsilon_0}\tilde{\chi}_{ee}^{yy} & \frac{-j}{k_0}\tilde{\chi}_{em}^{yx} & \frac{-j}{k_0}\tilde{\chi}_{em}^{yy} \\ \frac{-j}{k_0}\tilde{\chi}_{me}^{xx} & \frac{-j}{k_0}\tilde{\chi}_{me}^{xy} & \frac{-j}{\omega\mu_0}\tilde{\chi}_{mm}^{xx} & \frac{-j}{\omega\mu_0}\tilde{\chi}_{mm}^{xy} \\ \frac{j}{k_0}\tilde{\chi}_{me}^{yx} & \frac{j}{k_0}\tilde{\chi}_{me}^{yy} & \frac{j}{\omega\mu_0}\tilde{\chi}_{mm}^{yx} & \frac{j}{\omega\mu_0}\tilde{\chi}_{mm}^{yy} \end{bmatrix} \tag{49}$$

When written in this form, (48) is the same as (22) apart from a factor of $j\omega$, and thus the following synthesis approach is similar to that in section III.B. The matrix equation in (48) represents 16 unknowns and 4 equations and thus is underdetermined. The 16 unknowns can either be solved for directly by defining 4 total field transformations the metasurface is to perform simultaneously or must be reduced to 4 unknowns to make the system determined. For a single field transformation (one involving one set of incident, reflected, and transmitted electric and magnetic fields), only 4 susceptibilities are required for the general case of both $x$ and $y$ polarization states (if only one polarization state is considered, then only 2 susceptibilities are required). Several ways to reduce (48) to determined forms can be found in [90]–[92].

The S-parameters can then be directly related to the susceptibilities [92]. First (48) can be written as

$$\overline{\overline{\Delta}} = \tilde{\overline{\overline{\chi}}} \bullet \overline{\overline{A}}_v \tag{50}$$

The matrices $\overline{\overline{\Delta}}$ and $\overline{\overline{A}}_v$ can be formed in terms of the desired S-parameters as

$$\overline{\overline{\Delta}} = \begin{bmatrix} -\frac{\overline{\overline{N}}_2}{\eta_1} + \frac{\overline{\overline{N}}_2\overline{\overline{S}}_{11}}{\eta_1} + \frac{\overline{\overline{N}}_2\overline{\overline{S}}_{21}}{\eta_2} & -\frac{\overline{\overline{N}}_2}{\eta_2} + \frac{\overline{\overline{N}}_2\overline{\overline{S}}_{12}}{\eta_1} + \frac{\overline{\overline{N}}_2\overline{\overline{S}}_{22}}{\eta_2} \\ -\overline{\overline{N}}_1\overline{\overline{N}}_2 - \overline{\overline{N}}_1\overline{\overline{N}}_2\overline{\overline{S}}_{11} + \overline{\overline{N}}_1\overline{\overline{N}}_2\overline{\overline{S}}_{21} & \overline{\overline{N}}_1\overline{\overline{N}}_2 - \overline{\overline{N}}_1\overline{\overline{N}}_2\overline{\overline{S}}_{12} + \overline{\overline{N}}_1\overline{\overline{N}}_2\overline{\overline{S}}_{22} \end{bmatrix}$$

$$\overline{\overline{A}}_v = \frac{1}{2}\begin{bmatrix} \overline{\overline{I}} + \overline{\overline{S}}_{11} + \overline{\overline{S}}_{21} & \overline{\overline{I}} + \overline{\overline{S}}_{12} + \overline{\overline{S}}_{22} \\ \frac{\overline{\overline{N}}_1}{\eta_1} - \frac{\overline{\overline{N}}_1\overline{\overline{S}}_{11}}{\eta_1} + \frac{\overline{\overline{N}}_1\overline{\overline{S}}_{21}}{\eta_2} & -\frac{\overline{\overline{N}}_1}{\eta_2} - \frac{\overline{\overline{N}}_1\overline{\overline{S}}_{12}}{\eta_1} + \frac{\overline{\overline{N}}_1\overline{\overline{S}}_{22}}{\eta_2} \end{bmatrix} \tag{51}$$

where

$$\overline{\overline{S}}_{ab} = \begin{bmatrix} S_{ab}^{xx} & S_{ab}^{xy} \\ S_{ab}^{yx} & S_{ab}^{yy} \end{bmatrix}, \overline{\overline{I}} = \begin{bmatrix} 1 & 0 \\ 0 & 1 \end{bmatrix}$$

$$\overline{\overline{N}}_1 = \begin{bmatrix} 0 & -1 \\ 1 & 0 \end{bmatrix}, \overline{\overline{N}}_2 = \begin{bmatrix} 1 & 0 \\ 0 & -1 \end{bmatrix}$$

Substitution of (51) into (50) allows the normalized susceptibilities $\tilde{\overline{\overline{\chi}}}$ to be obtained by matrix inversion. Examples of metasurfaces designed using this synthesis approach will be presented in section V.

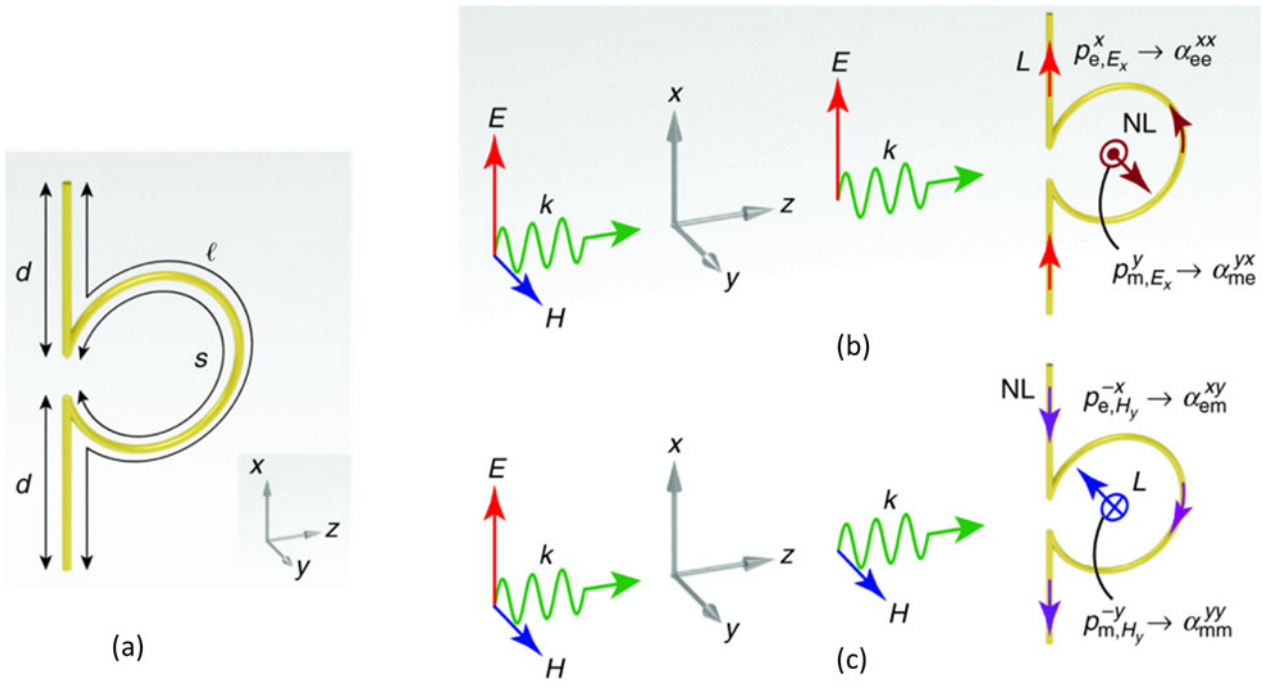

Fig. 4. Omega bianisotropic particle (reprinted with permission from [61]). *(a)* The Omega bianisotropic particle geometry. (b) induced dipole moments under electric excitation. (c) induced dipole moments under magnetic excitation.

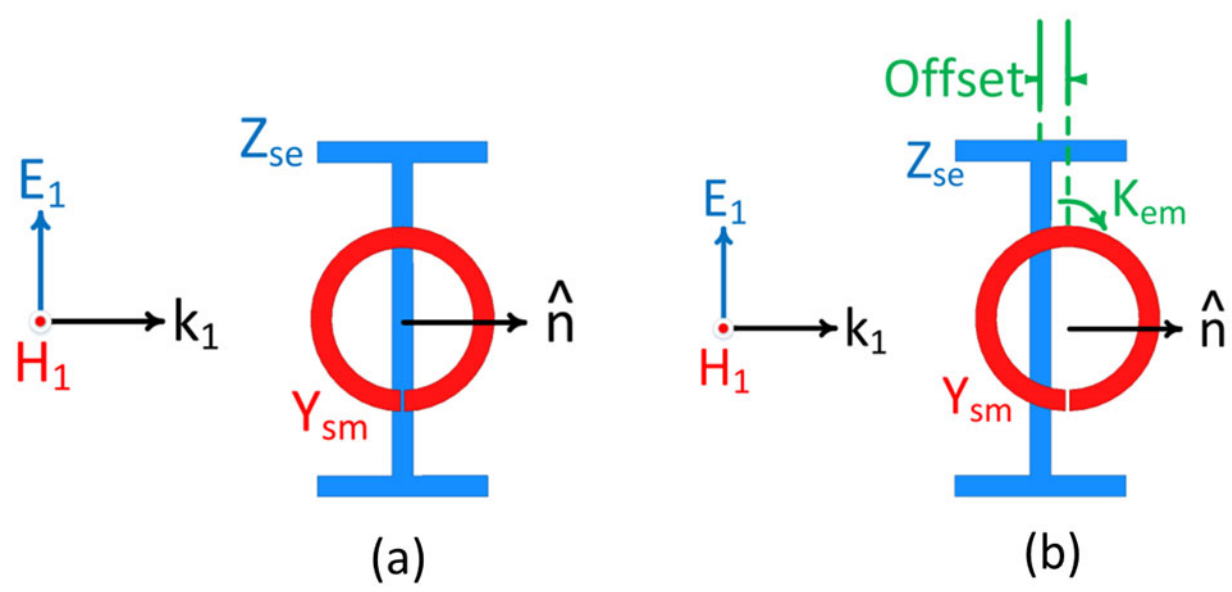

Fig. 5. Wire-loop Omega bianisotropic particle. The wire is isolated from the loop by an air-gap (reprinted with permission from [21]). *(a)* Symmetric wire-loop configuration only contains electric and magnetic responses and is not magnetoelectric. (b) By introducing an offset (asymmetry), the net magnetic flux through the loop created by the wire is non-zero leading to tunable magnetoelectric coupling.

## IV. Magnetoelectric Particle Design and Realization of Bianisotropic Boundaries

Design of a metasurface using bianisotropic boundary conditions and the associated synthesis techniques results in descriptions for either the surface susceptibility tensors $\bar{\bar{\chi}}_{ab}$, the constituent surface parameter tensors $\bar{\bar{Y}}, \bar{\bar{\gamma}}, \bar{\bar{\chi}}$ and $\bar{\bar{Z}}$, or the effective polarizability density tensors $\bar{\bar{\alpha}}$. Thus, appropriate magnetoelectric particles which exhibit the proper magnetoelectric response must be designed to realize these parameters. This section reviews some reciprocal and non-reciprocal magnetoelectric particle designs which can be used to realize bianisotropic metasurfaces. Each particle can be characterized by its polarizability tensors $\bar{\bar{\alpha}}_{ab}$ which relate the particle's dipole moment to its local excitation as in (9). To obtain these polarizabilities the particle is analyzed in isolation and thus the local field exciting the particle is simply the incident field only. Thus, each particles polarizabilities will be obtained by considering a plane wave excitation.

We will present four canonical magnetoelectric particles and their polarizability tensors. Following this, planar 3-sheet and 4-sheet realizations of magnetoelectric particles will be presented. Finally, implementations of all-dielectric magnetoelectric particles and planar all-dielectric realizations will be presented.

### *A. Reciprocal Magnetoelectric Particles*

The reciprocal class of magnetoelectric particles includes the Omega particle and the Chiral particle. Reciprocal magnetoelectric particles are characterized by the Onsager-Casimir symmetry relations [60], [61], [101]–[104]

$$\bar{\bar{\alpha}}_{ee} = \bar{\bar{\alpha}}_{ee}^{T}, \ \bar{\bar{\alpha}}_{mm} = \bar{\bar{\alpha}}_{mm}^{T}, \ \bar{\bar{\alpha}}_{me} = -\bar{\bar{\alpha}}_{em}^{T} \tag{52}$$

where the operator '$T$' denotes matrix transpose. The polarizabilities of each of the particles presented in this subsection will adhere to (52). We begin with the reciprocal Omega particle.

#### *1) Omega Particle*

The Omega magnetoelectric particle [61] consists of the combination of a resonant dipole conjoined to an in plane small loop antenna, as shown in Fig. 4a. As shown in Fig. 4b, when the particle is excited with an electric field parallel to the dipole axis, electric currents are induced in the dipole arms leading to non-zero $\alpha_{ee}^{xx}$ (the ratio of $E_{inc,x}$ to $p_{e,x}$). By continuity of current, the current also flows through the loop inducing a magnetic dipole moment in the $\hat{y}$-direction and leading to a non-zero $\alpha_{me}^{yx}$. Similarly in Fig. 4c, when the particle is excited by a $\hat{y}$-directed magnetic field, current is induced in the loop by magnetic induction. By Lenz's law, the induced magnetic dipole due to this current is in the opposite direction of the incident magnetic field leading to a non-zero $\alpha_{mm}^{yy}$. By continuity of current, current also flows in the dipole arms leading to non-zero $\alpha_{em}^{xy}$. Because the particle contains no non-reciprocal components, the magnetoelectric responses must be of equal magnitude, however, they are directed opposite one another. Hence, $\alpha_{me}^{yx} = -\alpha_{em}^{xy}$, indicating reciprocal Omega operation (see Table I). The Omega particle preserves polarization as the induced electric dipole moments are always parallel to the exciting electric field, and hence the Omega particle does not possess chiral properties. The polarizability tensor for the Omega particle appearing in Fig. 4 is given as

$$\begin{bmatrix} \bar{\bar{\alpha}}_{ee} & \bar{\bar{\alpha}}_{em} \\ \bar{\bar{\alpha}}_{me} & \bar{\bar{\alpha}}_{mm} \end{bmatrix} = \begin{bmatrix} \begin{bmatrix} \alpha_{ee}^{xx} & 0 & 0 \\ 0 & 0 & 0 \\ 0 & 0 & 0 \end{bmatrix} & \begin{bmatrix} 0 & \alpha_{em}^{xy} & 0 \\ 0 & 0 & 0 \\ 0 & 0 & 0 \end{bmatrix} \\ \begin{bmatrix} 0 & 0 & 0 \\ -\alpha_{em}^{xy} & 0 & 0 \\ 0 & 0 & 0 \end{bmatrix} & \begin{bmatrix} 0 & 0 & 0 \\ 0 & \alpha_{mm}^{yy} & 0 \\ 0 & 0 & 0 \end{bmatrix} \end{bmatrix} \tag{53}$$

The Omega particle's magnetoelectric terms are located on the off-diagonal matrix entries. The remaining terms of (53) can be obtained through coordinate rotations of the particle (see Table I). Analytical expressions for all particle polarizabilities $\alpha_{ab}^{uv}$ of the Omega particle in (53) can be found in [60].

Another realization of an Omega particle is the wire-loop topology shown in Fig. 5. The wire and loop are electrically isolated. The wire controls the electric response, and the loop controls the magnetic response. When the configuration is symmetric as in Fig. 5a, the net magnetic flux through the loop created by the wire is zero, and hence no magnetoelectric coupling. However, when the wire is offset with respect to the loop as in Fig. 5b, the net flux is non-zero and by magnetic induction, current flows in the loop generating a magnetic dipole. The magnetoelectric coupling is tunable through the degree of asymmetry introduced by the offset. The polarizability matrices are the same as (53) and analytical expressions for the polarizabilities can be found in [21].

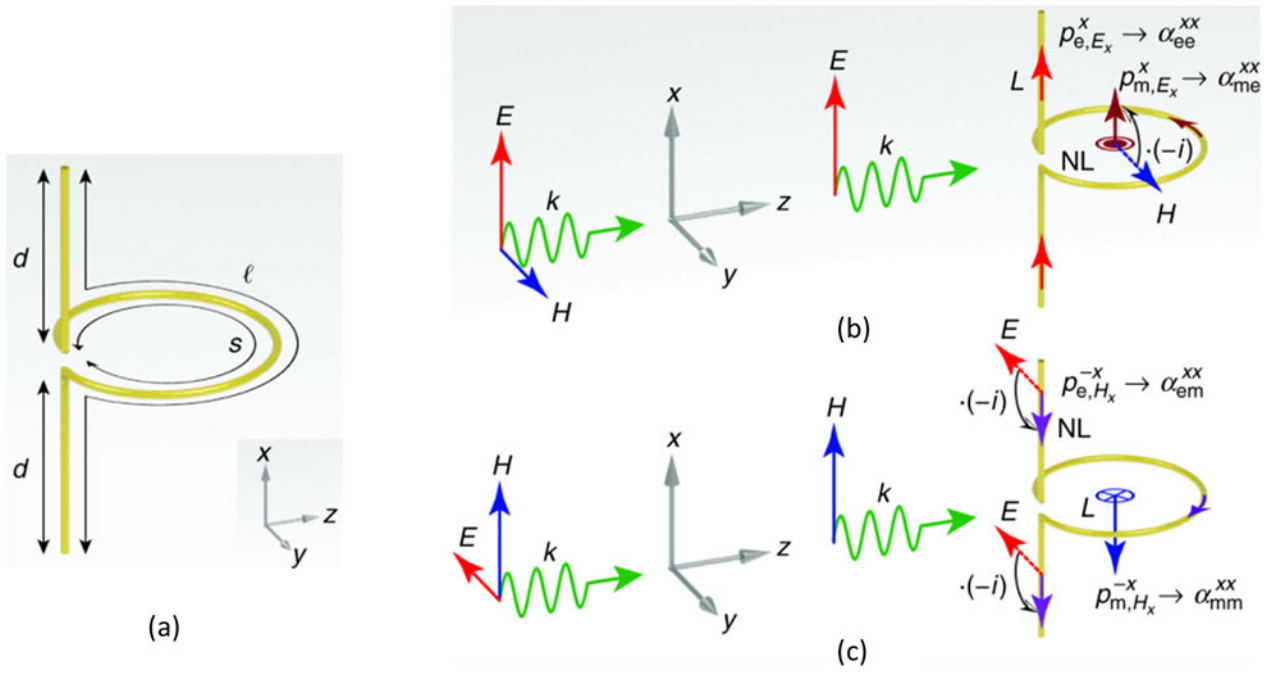

Fig. 6. Chiral bianisotropic particle (reprinted with permission from [61]). *(a)* The Chiral bianisotropic particle geometry. (b) induced dipole moments under electric excitation. (c) induced dipole moments under magnetic excitation.

### 2) *Chiral Particle*

The Chiral magnetoelectric particle [61] consists also of conjoined electric dipole and loop antennas, as in the case of the Omega particle. However, a 90° twist is added to the loop antenna to bring it out of plane with the dipole (see Fig. 6a). This simple change makes the particle chiral, as the particle now exhibits mirror asymmetry. As seen in Fig. 6b (6c), the induced magnetic (electric) dipole moment is now orthogonal to the incident magnetic (electric) field for the case of electric (magnetic) excitation leading to non-zero $\alpha_{em}^{xx}$ and $\alpha_{me}^{xx}$, and hence the Chiral particle rotates the polarization of the incident field upon excitation. Materials made from Chiral particles are therefore said to be Gyrotropic. The polarizability tensor for the Chiral particle shown in Fig. 6 is

$$\begin{bmatrix} \overline{\overline{\alpha}}_{ee} & \overline{\overline{\alpha}}_{em} \\ \overline{\overline{\alpha}}_{me} & \overline{\overline{\alpha}}_{mm} \end{bmatrix} = \begin{bmatrix} \begin{bmatrix} \alpha_{ee}^{xx} & 0 & 0 \\ 0 & 0 & 0 \\ 0 & 0 & 0 \end{bmatrix} & \begin{bmatrix} \alpha_{em}^{xx} & 0 & 0 \\ 0 & 0 & 0 \\ 0 & 0 & 0 \end{bmatrix} \\ \begin{bmatrix} -\alpha_{em}^{xx} & 0 & 0 \\ 0 & 0 & 0 \\ 0 & 0 & 0 \end{bmatrix} & \begin{bmatrix} \alpha_{mm}^{xx} & 0 & 0 \\ 0 & 0 & 0 \\ 0 & 0 & 0 \end{bmatrix} \end{bmatrix} \tag{54}$$

Note, the Chiral particle is described by the relationship $\alpha_{em}^{xx} = -\alpha_{me}^{xx}$ (see Table I). The Chiral particle's magnetoelectric terms are located along the main diagonal of the magnetoelectric polarizability matrices. The remaining terms of (54) can be obtained through coordinate rotations of the particle (see Table I). Analytical expressions for the particle polarizabilities $\alpha_{ab}^{uv}$ of the Chiral particle in (54) can be found in [60].

## B. *Non-Reciprocal Magnetoelectric Particles*

The non-reciprocal class of magnetoelectric particles includes the Tellegen-Omega particle and the Moving-Chiral particle. These particles are realized by including biased ferrite inclusions in the particles. Non-reciprocal magnetoelectric particles are characterized by the Onsager-Casimir relations [72], [101]–[104]

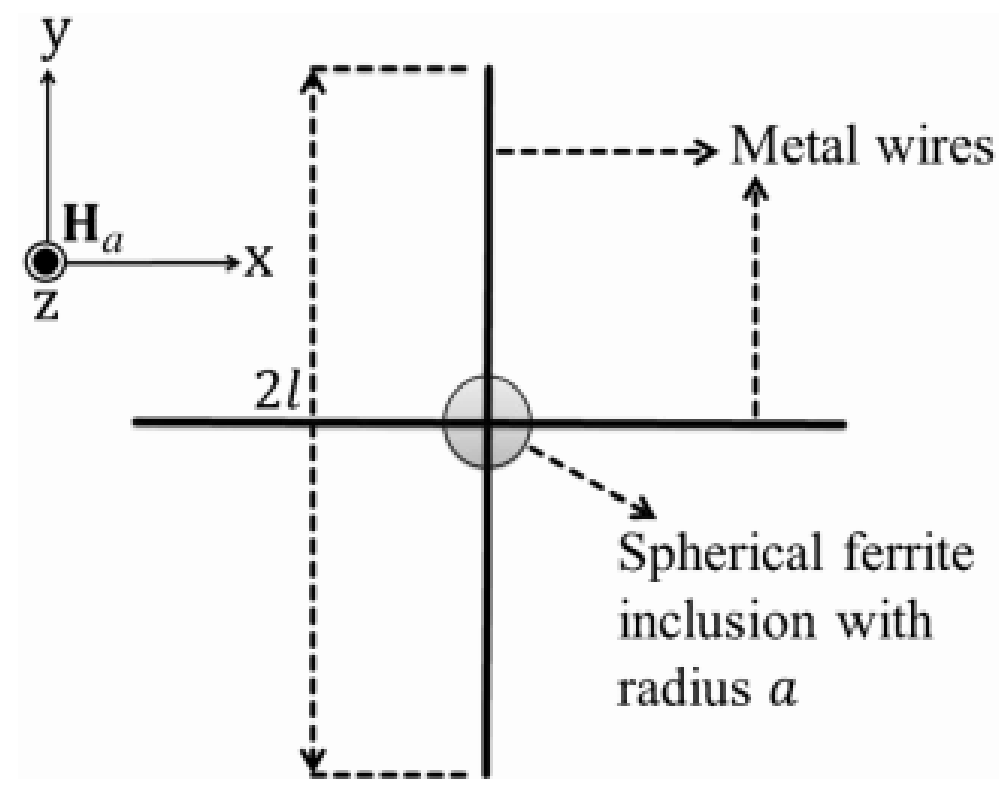

Fig. 7. Tellegen-Omega non-reciprocal magnetoelectric particle (reprinted with permission from [72]).

$$\begin{aligned} &\overline{\overline{\alpha}}_{ee}\left(\vec{H}_a\right) = \overline{\overline{\alpha}}_{ee}^{T}\left(-\vec{H}_a\right), \\ &\overline{\overline{\alpha}}_{mm}\left(-\vec{H}_a\right) = \overline{\overline{\alpha}}_{mm}^{T}\left(\vec{H}_a\right), \\ &\overline{\overline{\alpha}}_{me}\left(\vec{H}_a\right) = -\overline{\overline{\alpha}}_{em}^{T}\left(-\vec{H}_a\right) \end{aligned} \tag{55}$$

where $\vec{H}_a$ represents the external bias magnetic field. Thus, the particles are only reciprocal under bias field inversion and non-reciprocal otherwise. We begin with the non-reciprocal Tellegen-Omega particle.

### 1) *Tellegen-Omega Particles*

The Tellegen-Omega particle [72] is shown in Fig. 7. The particle geometry consists of a pair of orthogonal dipoles with a small spherical ferrite bead at the wire junction. The particle is located in the $xy$-plane. An external magnetic field, $\vec{H}_a$, directed along the $\hat{z}$-direction biases the ferrite bead to magnetization saturation. When an $\hat{x}$-directed electric field is incident upon the particle, electric current is induced in the dipole positioned along the $\hat{x}$-direction leading to a non-zero $\alpha_{ee}^{xx}$ polarizability. The induced current excites a magnetic field according to Ampere's Law. The $\hat{y}$-component of the magnetic field excites the ferrite bead inducing magnetic dipole moments in both the $\hat{x}$- and $\hat{y}$-directions leading to non-zero $\alpha_{me}^{xx}$ and $\alpha_{me}^{yx}$ polarizabilities. By magnetic induction, electric current is excited in the $\hat{y}$-directed wire leading to non-zero $\alpha_{ee}^{yx}$.

Now consider the particle being excited with an $\hat{x}$-directed high frequency magnetic field. Magnetic moments are excited in the ferrite sphere in both the $\hat{x}$- and $\hat{y}$-directions leading to non-zero $\alpha_{mm}^{xx}$ and $\alpha_{mm}^{yx}$. The magnetic moments in turn excite electric currents in both the wires by magnetic induction leading to non-zero $\alpha_{em}^{xx}$ and $\alpha_{em}^{yx}$. Due to the particle symmetry and bias field of the ferrite sphere, $\alpha_{me}^{xx} = \alpha_{em}^{xx}$ and $\alpha_{me}^{yx} = \alpha_{em}^{yx}$ and hence the particle is Tellegen ($\bar{\bar{\alpha}}_{em} = \bar{\bar{\alpha}}_{me}^{T}$). In this case, the polarizability tensors are

$$\begin{bmatrix} \bar{\bar{\alpha}}_{ee} & \bar{\bar{\alpha}}_{em} \\ \bar{\bar{\alpha}}_{me} & \bar{\bar{\alpha}}_{mm} \end{bmatrix} = \begin{bmatrix} \begin{bmatrix} \alpha_{ee}^{xx} & 0 & 0 \\ \alpha_{ee}^{yx} & 0 & 0 \\ 0 & 0 & 0 \end{bmatrix} & \begin{bmatrix} \alpha_{em}^{xx} & 0 & 0 \\ \alpha_{em}^{yx} & 0 & 0 \\ 0 & 0 & 0 \end{bmatrix} \\ \begin{bmatrix} \alpha_{em}^{xx} & 0 & 0 \\ \alpha_{em}^{yx} & 0 & 0 \\ 0 & 0 & 0 \end{bmatrix} & \begin{bmatrix} \alpha_{mm}^{xx} & 0 & 0 \\ \alpha_{mm}^{yx} & 0 & 0 \\ 0 & 0 & 0 \end{bmatrix} \end{bmatrix} \tag{56}$$

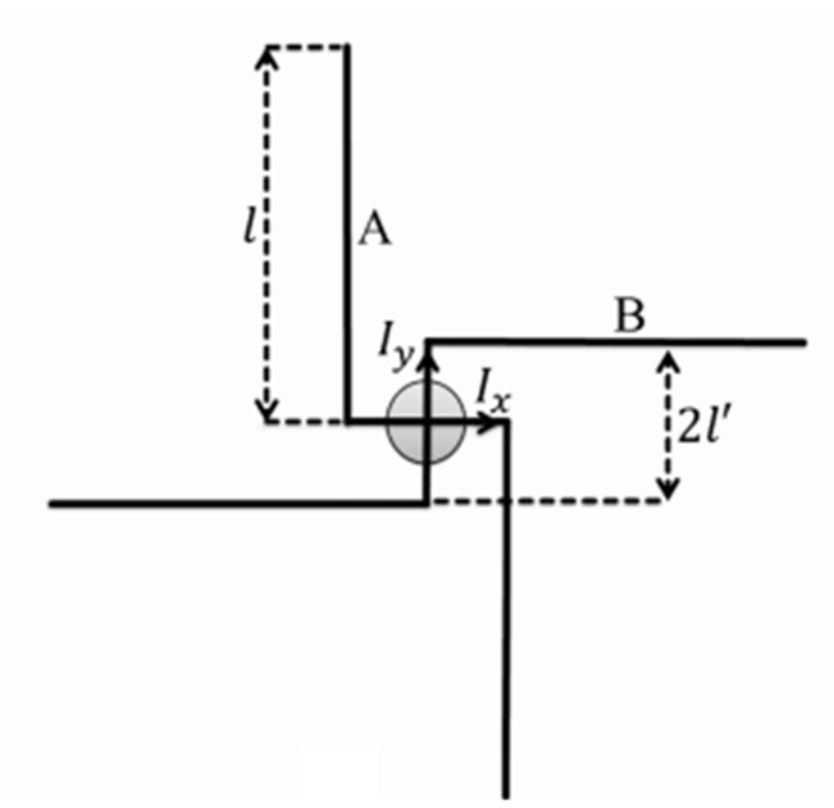

Fig. 8. Moving Chiral non-reciprocal magnetoelectric particle (reprinted with permission from [72]).

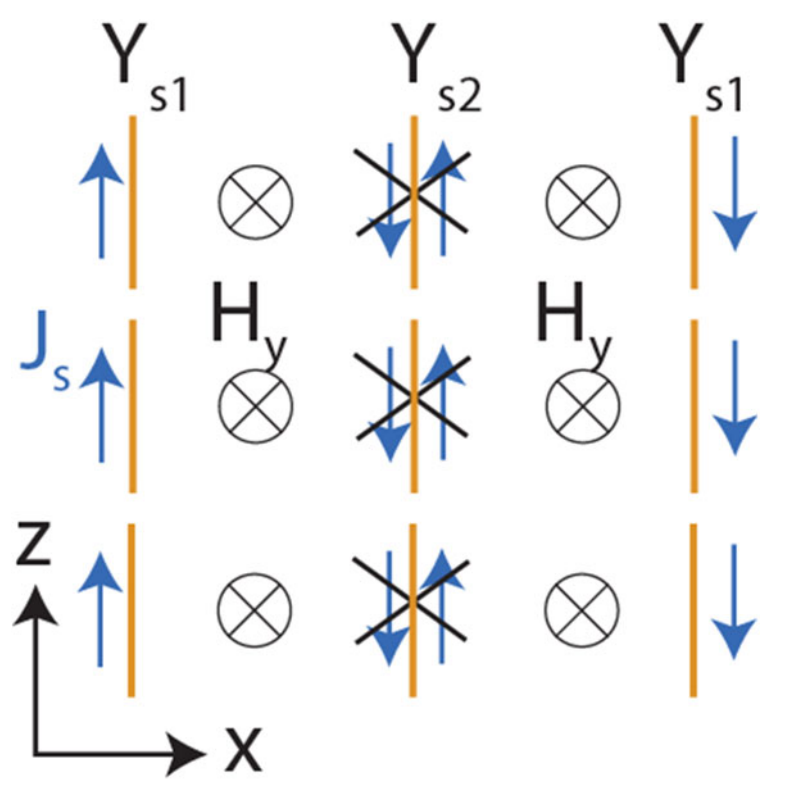

Fig. 9. A cascade of three sheet admittances can realize independent electric and magnetic responses.

The remaining terms of (56) can be obtained by considering excitations in the $\hat{y}$-direction (see Table I). Note, $\hat{z}$-directed excitations do not apply since the particle is uniaxial due to the applied bias. Analytical expressions for the particle polarizabilities $\alpha_{ab}^{uv}$ of the Tellegen-Omega particle in (56) can be found in [72].

### 2) *Moving-Chiral Particles*

The Moving-Chiral particle [72] is shown in Fig. 8. The particle geometry is similar to the Tellegen-Omega with a 90° twist added to the dipole arms. The twist makes the particle mirror-asymmetric leading to chirality [61], [105]. Similar to the Chiral particle, the induced dipole moments are now orthogonal to the exciting field. To see this, consider an $\hat{x}$-directed electric excitation which produces current in the longer wire segment of wire $B$. A non-zero $\alpha_{ee}^{xx}$ is observed. The currents in the shorter wire segment excites the ferrite inclusion, generating both an $\hat{x}$- and $\hat{y}$-directed magnetic dipole moments (only this time the moments are directed opposite to that of the Tellegen-Omega particle due to the twist) and hence non-zero $\alpha_{me}^{xx}$ and $\alpha_{me}^{yx}$. By magnetic induction again, electric currents are excited in wire $A$ leading to non-zero $\alpha_{ee}^{yx}$.

Next consider an $\hat{x}$-directed incident magnetic field. The incident magnetic field excites the ferrite sphere inducing both $\hat{x}$- and $\hat{y}$-directed magnetic dipole moments leading to non-zero $\alpha_{mm}^{xx}$ and $\alpha_{mm}^{yx}$. By magnetic induction current is induced in both wires leading to non-zero $\alpha_{em}^{xx}$ and $\alpha_{em}^{yx}$. Due to the twist, the magnetoelectric polarizabilities are directed in the opposite direction to one another. Hence, $\alpha_{em}^{xx} = -\alpha_{me}^{xx}$ (Chiral) and $\alpha_{em}^{yx} = -\alpha_{me}^{yx}$ (Moving). The polarizability tensors for the Moving-Chiral particle are

$$\begin{bmatrix} \bar{\bar{\alpha}}_{ee} & \bar{\bar{\alpha}}_{em} \\ \bar{\bar{\alpha}}_{me} & \bar{\bar{\alpha}}_{mm} \end{bmatrix} = \begin{bmatrix} \begin{bmatrix} \alpha_{ee}^{xx} & 0 & 0 \\ \alpha_{ee}^{yx} & 0 & 0 \\ 0 & 0 & 0 \end{bmatrix} & \begin{bmatrix} \alpha_{em}^{xx} & 0 & 0 \\ \alpha_{em}^{yx} & 0 & 0 \\ 0 & 0 & 0 \end{bmatrix} \\ \begin{bmatrix} -\alpha_{em}^{xx} & 0 & 0 \\ -\alpha_{em}^{yx} & 0 & 0 \\ 0 & 0 & 0 \end{bmatrix} & \begin{bmatrix} \alpha_{mm}^{xx} & 0 & 0 \\ \alpha_{mm}^{yx} & 0 & 0 \\ 0 & 0 & 0 \end{bmatrix} \end{bmatrix} \tag{57}$$

The remaining terms of (57) can be obtained by considering excitations in the $\hat{y}$-direction (see Table I). Analytical expressions for the particle polarizabilities $\alpha_{ab}^{uv}$ for the Moving-Chiral in (56) can be found in [72].

A summary of the transverse magnetoelectric polarizabilities of the presented canonical particles is provided in Table 1. The normal polarizabilities of the reciprocal particles are not shown. Note, the non-reciprocal particles are uniaxial due to the bias field and hence do not contain normal polarizabilities. From the table, it is easy to verify (52) and (55).

**Table I. Summary of Transverse Magnetoelectric Polarizabilities of Canonical Particles**

| | $\alpha_{em}^{xx}$ | $\alpha_{em}^{yy}$ | $\alpha_{em}^{xy}$ | $\alpha_{em}^{yx}$ | $\alpha_{me}^{xx}$ | $\alpha_{me}^{yy}$ | $\alpha_{me}^{xy}$ | $\alpha_{me}^{yx}$ |
|---|---|---|---|---|---|---|---|---|
| Omega | 0 | 0 | $-\alpha$ | $\alpha$ | 0 | 0 | $-\alpha$ | $\alpha$ |
| Chiral | $\alpha$ | $\alpha$ | 0 | 0 | $-\alpha$ | $-\alpha$ | 0 | 0 |
| Pure Tellegen | $\alpha$ | $\alpha$ | 0 | 0 | $\alpha$ | $\alpha$ | 0 | 0 |
| Pure Moving | 0 | 0 | $-\alpha$ | $\alpha$ | 0 | 0 | $\alpha$ | $-\alpha$ |
| Tellegen-Omega | $\alpha$ | $\alpha$ | $-\alpha$ | $\alpha$ | $\alpha$ | $\alpha$ | $-\alpha$ | $\alpha$ |
| Moving - Chiral | $\alpha$ | $\alpha$ | $-\alpha$ | $\alpha$ | $-\alpha$ | $-\alpha$ | $\alpha$ | $-\alpha$ |

## C. *Three-Sheet and Four-Sheet Implementations of Bianisotropic Particles*

Bianisotropic responses can also be realized as a series of cascaded sheet admittances [15], [65], [66], [106]. For an intuitive understanding of how this works, let us first consider a quasi-static magnetic field interacting with an *isotropic* metasurface consisting of a symmetric cascade of three sheets (see Fig. 9). A $\hat{y}$-directed magnetic field generates circulating electric currents on the outer sheets (Ys1), thus creating an equivalent magnetic current. However, the magnetic field does not interact with the middle sheet (Ys2) since the induced currents are canceled due to symmetry. Conversely, all three sheets will interact with a $\hat{z}$-directed electric field. Therefore, the outer sheets can be designed to realize a desired magnetic response, and the middle sheet adjusted to independently control the electric response. This enables independent control over both the electric and magnetic response. This is similar to how symmetry in the symmetric wire-loop particle enables independent control of the electric and magnetic responses.

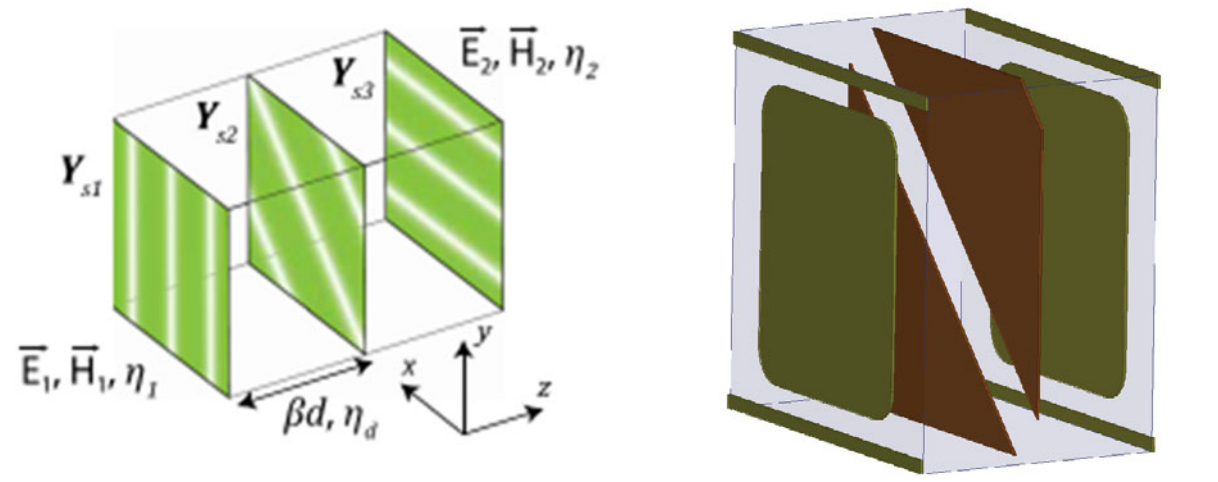

(a) (b)

Fig. 10. (Reprinted with permission from [15]) (a) Metasurface consisting of patterned plasmonic sheets (anisotropic sheet admittances) separated by optically thin dielectric layers. The metasurface can be designed to exhibit electric, magnetic, and chiral constituent surface parameters. (b) An example of a designed bianisotropic metasurface that acts as an asymmetric circular polarizer at mm-wave frequencies. This unit cell converts an incident right-handed circular polarization to left-handed circular polarization, but completely reflects incident left-handed circular polarization.

To realize an isotropic omega response, the cascade of three sheets should be asymmetric (the outer sheets are not equal). To realize full bianisotropic surface parameters, the three sheets should also be *anisotropic* [15]. This will cause the electric response to couple to the magnetic response since the cancellation shown in Fig. 9 will not fully occur, similar to the asymmetric wire loop combination. A transfer (ABCD) or wave matrix approach can be employed to develop a relation between the cascaded sheet admittances (representing the patterned sheets) of the metasurface and its S-parameters. This transfer matrix approach relates the total field in regions 1 and 2 (see Fig. 1) through the ABCD matrix

$$\begin{bmatrix} \vec{E}_1 \\ \vec{H}_1 \end{bmatrix} = \begin{bmatrix} \overline{\overline{A}} & \overline{\overline{B}} \\ \overline{\overline{C}} & \overline{\overline{D}} \end{bmatrix} \begin{bmatrix} \vec{E}_2 \\ \vec{H}_2 \end{bmatrix} \tag{58}$$

where $\bar{\bar{A}}$, $\bar{\bar{B}}$, $\bar{\bar{C}}$, and $\bar{\bar{D}}$ are each 2×2 matrices relating the $\hat{x}$ and $\hat{y}$ field components. For example, the transfer matrix for the metasurface consisting of three cascaded patterned sheets (sheet admittances) shown in Fig. 10 can be written as

$$\begin{bmatrix} \overline{\overline{A}} & \overline{\overline{B}} \\ \overline{\overline{C}} & \overline{\overline{D}} \end{bmatrix} = \begin{bmatrix} \overline{\overline{I}} & \overline{\overline{0}} \\ \overline{\overline{nY}}_{s1} & \overline{\overline{I}} \end{bmatrix} \begin{bmatrix} \cos(\beta d)\overline{\overline{I}} & -j\sin(\beta d)\eta_d\overline{\overline{n}} \\ j\sin(\beta d)\eta_d^{-1}\overline{\overline{n}} & \cos(\beta d)\overline{\overline{I}} \end{bmatrix}$$
$$\bullet \begin{bmatrix} \overline{\overline{I}} & \overline{\overline{0}} \\ \overline{\overline{nY}}_{s2} & \overline{\overline{I}} \end{bmatrix} \begin{bmatrix} \cos(\beta d)\overline{\overline{I}} & -j\sin(\beta d)\eta_d\overline{\overline{n}} \\ j\sin(\beta d)\eta_d^{-1}\overline{\overline{n}} & \cos(\beta d)\overline{\overline{I}} \end{bmatrix} \begin{bmatrix} \overline{\overline{I}} & \overline{\overline{0}} \\ \overline{\overline{nY}}_{s3} & \overline{\overline{I}} \end{bmatrix} \tag{59}$$

Here, $\eta_d = \sqrt{\mu_d/\varepsilon_d}$ is the wave impedance of the inter-sheet dielectric layers, $\beta d = \omega\sqrt{\mu_d\varepsilon_d}d$ is the electrical thickness of the dielectric layers (i.e. propagation delay), and $\overline{\overline{Y}}_{sn}$ is the admittance of the $n^{\text{th}}$ sheet (see Fig. 10). Once again, $\eta_1, \eta_2$ are the wave impedances on the incident (region 1) and transmitted (region 2) side of the metasurface, respectively. The scattering matrix of the metasurface can then be related to the sheet admittances through the ABCD matrix

$$\begin{bmatrix} \overline{\overline{S}}_{11} & \overline{\overline{S}}_{12} \\ \overline{\overline{S}}_{21} & \overline{\overline{S}}_{22} \end{bmatrix} = \begin{bmatrix} -\overline{\overline{I}} & \dfrac{\overline{\overline{Bn}}}{\eta_2} + \overline{\overline{A}} \\ \dfrac{\overline{\overline{n}}}{\eta_1} & \dfrac{\overline{\overline{Dn}}}{\eta_2} + \overline{\overline{C}} \end{bmatrix}^{-1} \begin{bmatrix} \overline{\overline{I}} & \dfrac{\overline{\overline{Bn}}}{\eta_2} - \overline{\overline{A}} \\ \dfrac{\overline{\overline{n}}}{\eta_1} & \dfrac{\overline{\overline{Dn}}}{\eta_2} - \overline{\overline{C}} \end{bmatrix} \tag{60}$$

By equating the scattering parameters of (31) and (60), one can relate the constituent surface parameters of a bianisotropic metasurface to the sheet admittances comprising it. Therefore, the sheets can be systematically designed to achieve arbitrary bianisotropic surface parameters, or desired transmission and reflection characteristics, limited only by reciprocity and passivity.

In fact, 4 sheets are needed to realize the full $4 \times 4$ scattering matrix in (30). There are 32 entries in the scattering matrix but if one applies lossless and reciprocal constraints, there are only 10 distinct entries of the scattering matrix. These 10 distinct entries can be realized using 4 sheets, since each sheet can provide 3 entries [107]. Therefore, three sheets are insufficient if one is to realize a full scattering matrix. The same ABCD matrix cascading approach can be extended to the four-sheet case, and equality between (31) and the four-sheet version of (59) (add an additional sheet and dielectric spacer) and (60) can be made. Adding a fourth sheet can also provide a wider bandwidth, as has been shown in [14], [15], [66].

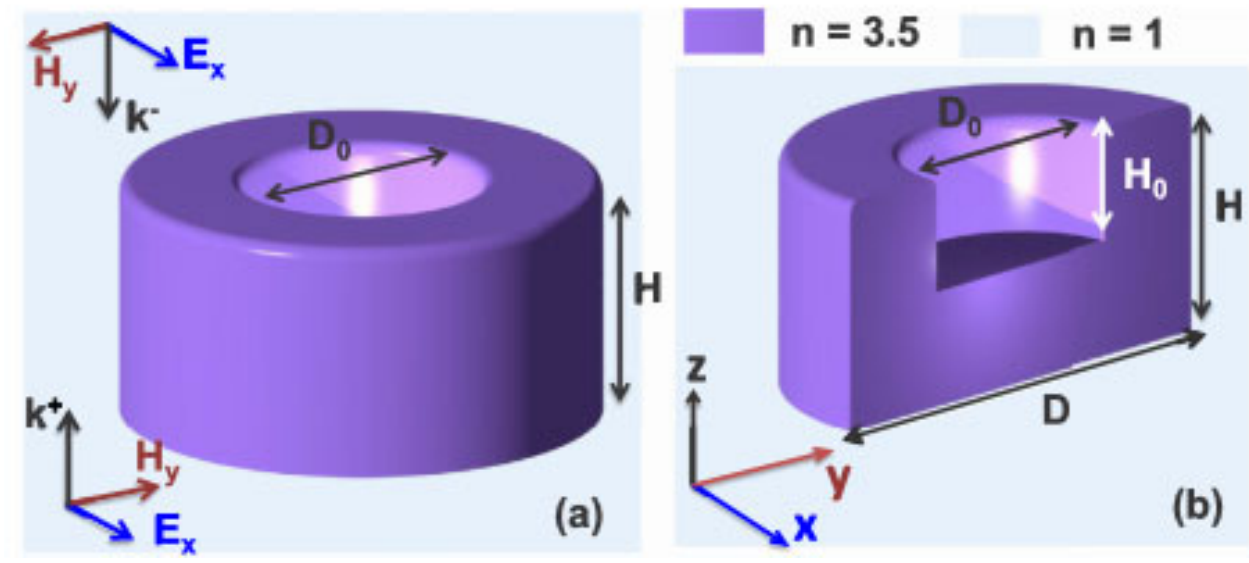

Fig. 11. All-dielectric reciprocal Omega-type bianisotropic nanoparticles (reprinted with permission from [77]).

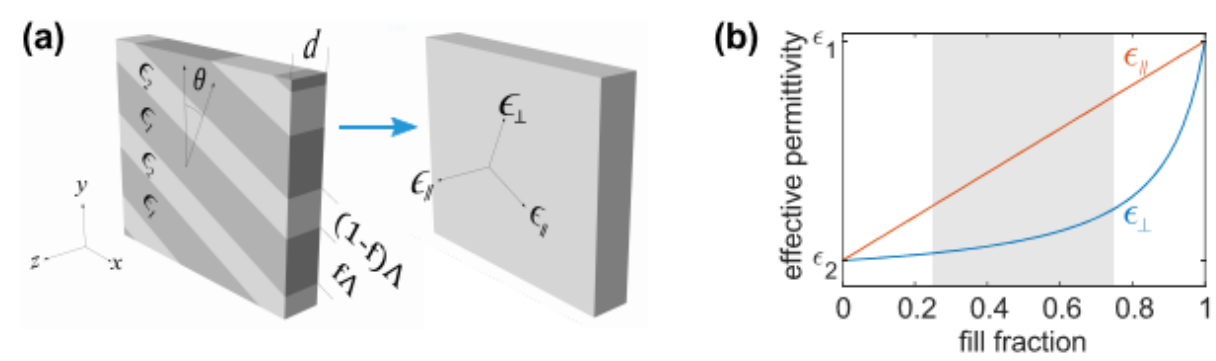

Fig. 12. All-dielectric stacks of high contrast gratings (reprinted with permission from [67]). (a) Each grating is homogenized into an anisotropic layer using effective medium theory. By controlling the fill fraction within each layer, the anisotropic effective permittivity can be controlled. Stacks of the effective anisotropic layers provide bianisotropic responses analogous to the three electric sheet implementations of bianisotropic boundaries.

*D. All-Dielectric Bianisotropic Particles*

An all-dielectric reciprocal Omega-type bianisotropic nanoparticle is shown in Fig. 11 [77]. The cylindrical dielectric puck of radius $D$ and height $H$ has a hole drilled into it of depth $D_0$ and height $H_0$ breaking the symmetry of the nanoparticle. It is well known that cylindrical nanoparticles can exhibit electric and magnetic dipolar resonances which can be de-scribed through effective electric and magnetic dipole moments [108]. By introducing the partially drilled hole and breaking the symmetry of the particle, a magnetoelectric response is created. The polarizabilities of the reciprocal Omega-type bianisotropic particle when illuminated by the plane waves indicated in Fig. 11 are given as [77]

$$\begin{aligned}\frac{p_x^{\pm}}{\varepsilon_0} &= \alpha_{ee}E_x^{inc} \pm \alpha_{em}Z_0H_y^{inc} \\ Z_0m_y^{\pm} &= \alpha_{me}E_x^{inc} \pm \alpha_{mm}Z_0H_y^{inc}\end{aligned} \tag{61}$$

In (61), $p_x^{\pm}$ and $m_y^{\pm}$ are the $\hat{x}$- and $\hat{y}$-directed electric and magnetic dipole moments of the nanoparticle when illuminated by the $\hat{x}$- and $\hat{y}$-directed incident electric, $E_x^{inc}$, and magnetic, $H_y^{inc}$, fields, respectively. Note, when the $+$ is chosen in the $\pm$ symbol, the illumination is from below, whereas when the $-$ sign is chosen, the illumination is from above. Because of the broken symmetry, $p_x^+ \neq p_x^-$ and $m_y^+ \neq m_y^-$ which can be explained as due to electro-magnetic/magnetoelectric coupling and hence non-zero $\alpha_{em}$ and $\alpha_{me}$. Values for the polarizabilities in (61) are provided in [77].

An all-dielectric analog of the three-sheet and four-sheet implementations of section IV.C are presented in [67] and shown in Fig. 12. Each layer of the multilayer stack consists of a high-contrast, subwavelength dielectric grating rotated by some angle $\theta$ with respect to a global cartesian coordinate system. By using effective medium theory, each layer can be homogenized into an anisotropic layer following from

$$\begin{aligned}\frac{1}{\varepsilon_{\perp}} &= \frac{f}{\varepsilon_1} + \frac{1-f}{\varepsilon_2} \\ \varepsilon_{\parallel} &= f\varepsilon_1 + (1-f)\varepsilon_2\end{aligned} \tag{62}$$

where $\epsilon_{\perp}$ and $\epsilon_{\parallel}$ are the effective permittivity of the grating in the direction perpendicular and parallel to the direction of dielectric contrast, $\epsilon_1$ and $\epsilon_2$ are the permittivity of the two materials used to fabricate the grating, and $f$ is the fill fraction of the grating. Stacking the effective homogenized anisotropic layers produces a bianisotropic response analogous to the stack of anisotropic electric sheets case in Fig. 10. Typically, each grating provides a narrower range of constitutive surface parameters, therefore more layers (thicker stacks) are required in designs.

## V. Capabilities and Recent Applications of Bianisotropic Boundaries

The examples, taken from scientific works in open literature, of this section are a collective representation of the state of the art in metasurface design using bianisotropic boundary conditions. The synthesis approach and realization used in each example, can be traced back to the previous sections and are summarized in Table II.

**Table II. Summary of Boundary Conditions, Synthesis Approach, and Realization Technique**

| Example (Section V) | GSTC Model (Section II) | Synthesis Approach (Section III) | Realization (Section IV) |
|---|---|---|---|
| A | C, B | B, A | C, A |
| B | C, C | C, B | C, A |
| C | A | B | C |
| D | D | E | A |
| E | C | D | C |
| F | C | B | A |
| G | C | B | C |
| H | C | C | C |
| I | C | D | C |
| J | C | D | $x$ |
| K | C | B | $x$ |
| L | C | C | $x$ |
| M | C | D | C |
| N | D | * | $x$ |
| O | C | B | $x$ |
| P | A | * | D |
| Q | B | A | A |

$x$ = example not realized in referenced publication
* = synthesis method not reviewed in section III

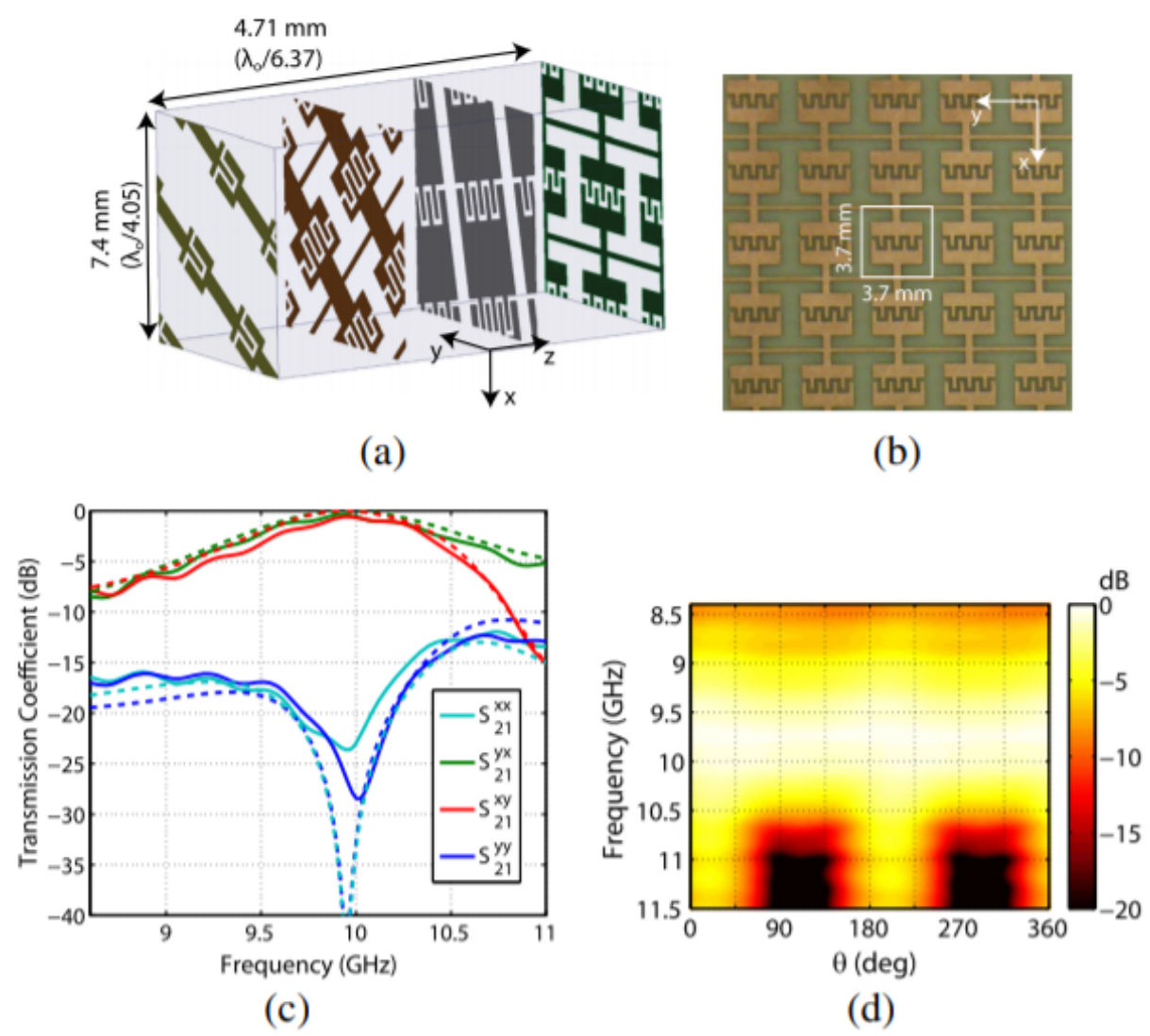

Fig. 13. (reprinted with permission from [15]). Metasurface exhibiting polarization rotation near 10 GHz. *(a)* Schematic of the unit cell. For clarity, the $z$ axis is scaled by a factor of 3 so that all four sheets are visible. *(b)* Bottom sheet ($\overline{\overline{Y}}_{s4}$) of the fabricated polarization rotator. *(c)* Transmission coefficient for an incident plane wave traveling in the $+z$ direction. Measured data are denoted by solid lines, whereas simulated are denoted by dashed lines. For clarity, the measured data are frequency shifted by +0.20 GHz in the plot. *(d)* Measured cross-polarized transmission coefficient as a function of frequency and incident linear polarization. The angle $\theta$ refers to the angle between the $x$ and $y$ axes of the incident linear polarization. It can be seen that the cross-polarized transmission coefficient is near 0dB, independent of $\theta$.

### A. Polarization Control

The first application of bianisotropic metasurfaces was in polarization control. Several bianisotropic metasurfaces for polarization control have been reported in literature [13]–[20]. As an illustrative example, consider a reflectionless metasurface that rotates an arbitrary linearly polarized plane wave by 90° upon transmission [15]. Thus, the desired scattering matrix is

$$\overline{\overline{S}}_{11} = \overline{\overline{0}}, \quad \overline{\overline{S}}_{21} = e^{j\phi}\begin{bmatrix} 0 & -1 \\ 1 & 0 \end{bmatrix} \tag{63}$$

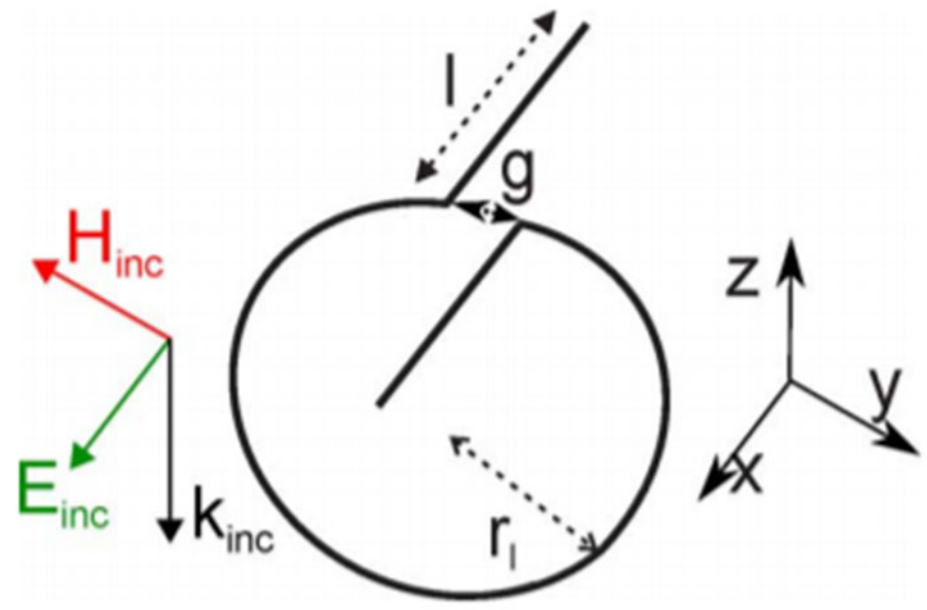

Fig. 14. Geometry of single Chiral Particle for reciprocal twist polarizer (reprinted with permission from [20]).

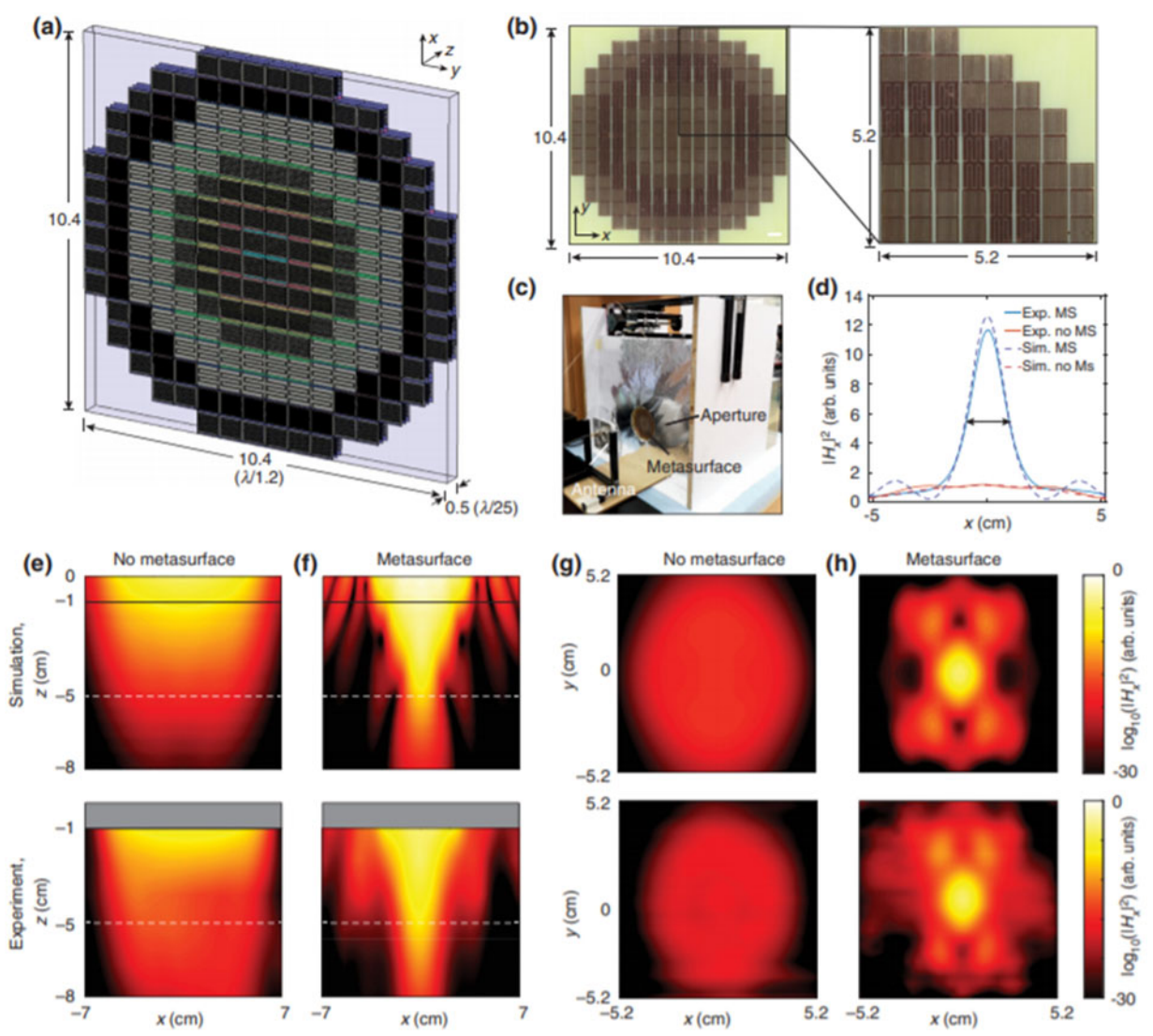

Fig. 15. (Reprinted with permission from [22]). Antireflection focusing from air into 5-cm (λ/2.5) depth in water. (a) A perspective view of the metasurface using eight unit cells distributed in a circular ring form. Cells occupying the same background color are identical. (b) A photograph of the fabricated metasurface: a quarter of the metasurface is enlarged on the right. (c) A photograph of the experimental setup. (d) The normalized magnetic field intensity along the focusing depth with and without the metasurface (MS) from simulation (Sim.) and experiment (Exp.). The arrow indicates the full width at half maximum (FWHM) of 0.16λ. (e),(f) The simulated (top) and measured (bottom) magnetic field intensity at the x-z plane (e) without and (f) with the metasurface. (g),(h) The simulated (top) and measured (bottom) magnetic field intensity at the 5-cm-deep focal plane (x-y plane) (g) without and (h) with the metasurface. The water is simulated with a complex relative permittivity ε2 = 78 − 11.7j.

By inserting (63) into (32), the constituent surface parameters are obtained as

$$\begin{bmatrix} \overline{\overline{Y}} & \overline{\overline{\chi}} \\ \overline{\overline{\gamma}} & \overline{\overline{Z}} \end{bmatrix} = \begin{bmatrix} -2j\eta_0^{-1}\tan\phi & 0 & -2\sec\phi & 0 \\ 0 & -2j\eta_0^{-1}\tan\phi & 0 & -2\sec\phi \\ 2\sec\phi & 0 & -2j\eta_0\tan\phi & 0 \\ 0 & 2\sec\phi & 0 & -2j\eta_0\tan\phi \end{bmatrix} \tag{64}$$

Comparison of (64) with Table I shows the metasurface is isotropic and chiral. The metasurface was realized using the techniques of section IV.C and is shown in Fig. 13.

Next consider the same polarizer designed using the particle polarizability model [20] (see section II.B). Since the metasurface is isotropic and chiral, the same metasurface should be able to be realized using the chiral particle of section IV.A.2.

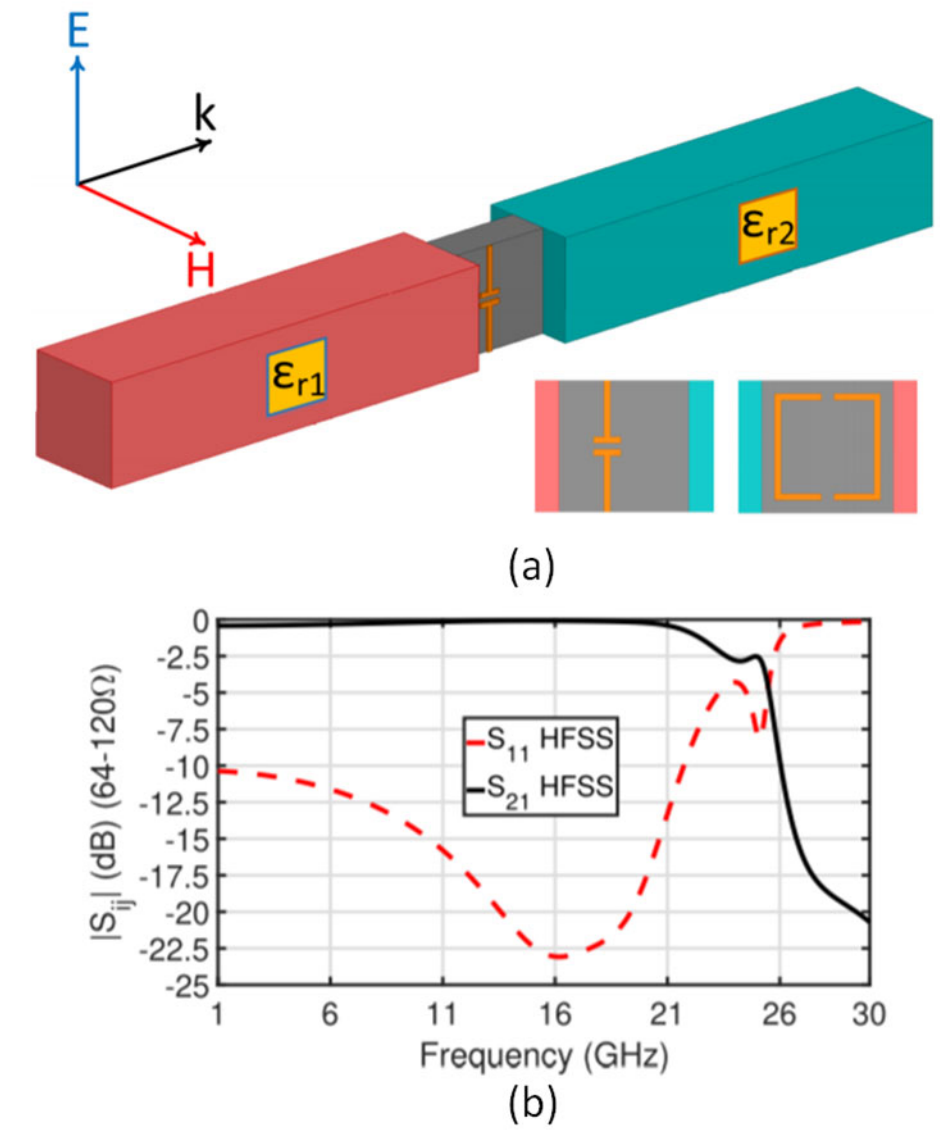

Fig. 16. (Reprinted with permission from [21]). (a) Finite wire and split-ring bianisotropic Huygens' unit cell printed using 1 oz copper (36 μm) on a 0.635 mm Rogers RO3010 substrate placed in a 1.58 mm air gap between two dielectric half-spaces with relative permittivities $\epsilon_{r1}$= 35 and $\epsilon_{r2}$ = 10. The wire is offset by −203 μm compared to the symmetric case. (b) Magnitude of the reflection and transmission of the bianisotropic Huygens' metasurface unit cell with copper and dielectric loss.

Starting from the polarizability model and noting (63) can be written as

$$\vec{E}_r = 0,\ \vec{E}_t = A\overline{\overline{J}}_t \bullet \vec{E}_{inc} \tag{65}$$

where the factor $A = e^{j\phi}$, the particle's polarizabilities can be found by setting the first of (28) and the co-polarized component of the second to zero while setting the cross-polarized component the second to $A$. The solution results in the description of a Chiral particle

$$\hat{\alpha}_{ee}^{co} = \frac{S}{j\omega\eta_0}\hat{\alpha}_{mm}^{co},\ \hat{\alpha}_{mm}^{co} = \frac{\eta_0 S}{j\omega},\ \hat{\alpha}_{me}^{co} = -\hat{\alpha}_{em}^{co} = -A\frac{S}{j\omega} \tag{66}$$

and is shown in Fig. 14. Note, the superscript '$co$' refers to the main diagonal entries of the polarizability tensors.

### *B. Impedance Matching, Antireflection, and Absorption*

Another useful application of bianisotropic metasurfaces is seamless impedance matching and antireflection [21], [22], or absorption [109]. Consider the work presented in [22]. A metasurface which focuses an incident plane wave from air into a medium with a high dielectric constant, in this case water with a relative permittivity of $78 - j11.7$ (simultaneous impedance matching and wavefront manipulation) is shown in Fig. 15. The metasurface was designed using the wave matrix approach of section III.C and realized using the three-sheet method of section IV.C. The detailed results of the design process can be found in [22]. As the figure shows, the metasurface simultaneously performs impedance matching and focusing into the water medium. By impedance matching with the metasurface and thus avoiding the reflection at the interface, the measured transmission efficiency has been increased by over 20% over the case of no matching metasurface.

As another example, consider the work presented in [21]. A metasurface was designed as a wideband impedance matching layer between two dielectric half spaces of differing permittivities of $\epsilon_{r1} = 35$ and $\epsilon_{r2} = 10$. In Fig. 16a, a unit cell of the infinite periodic metasurface is shown. The wire-loop unit cell consists of a Rogers RO3010 substrate with a loaded wire printed on one side and a loaded loop printed on the reverse side. The center of the wire is offset with respect to the center of the loop to create the magnetoelectric response. The reflection and transmission coefficients of the matching layer are shown in Fig. 16b. It is observed that the matching layer achieves a minimum reflection coefficient of -10dB over the wideband of 1-22 GHz.

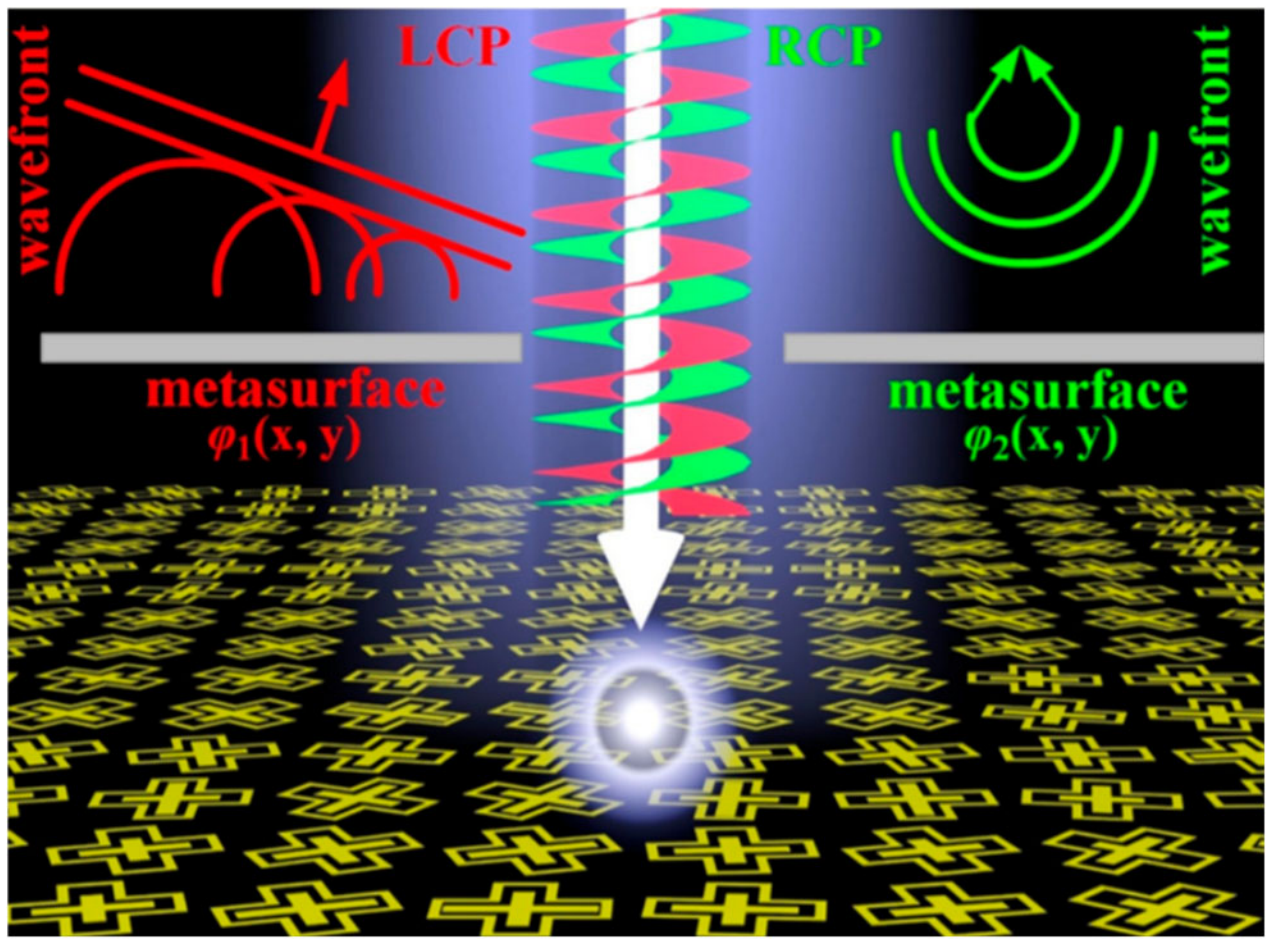

Fig. 17. (Reprinted with permission from [110]). Scattering of a wave impinging on the interface under an arbitrary angle $(\theta_a)$, with conventional Fresnel transmission and reflection for the case of the bare interface (left) and with reflectionless (Brewster) transmission when a properly designed metasurface is placed at the interface (right).

### C. *Multifunctional Metasurfaces*

Bianisotropic metasurfaces can also be designed for multi-functional control of the wavefront [110]–[113] or polarization state [19], [69], [112], [114]. In [69], multifunctional polarization converters are made from cascaded subwavelength gratings (see Fig. 12). One example is a dual-band, dual-function metasurface which functions as a left-handed symmetric circular polarizer for one band and a left-handed asymmetric circular polarizer at the other higher band. In another work [19], an LP-to-CP polarizer that operates in two bands is reported for SatCom applications. In the lower band, LP is converted to LHCP, while in the upper band, LP is converted to RHCP.

An example of multifunctional wavefront control can be found in [110]. There, a bianisotropic metasurface operating at 10.5 GHz was designed to function as a beam deflector when illuminated by an LHCP wave, and a reflective focusing lens when illuminated by an RHCP wave (see Fig. 17). The metasurface elements were designed such that the polarization-dependent responses were decoupled with minimal interference between responses, by incorporating higher order resonances and patch/loop structures. The elements consist of two identical layers of these elements backed by a ground plane. The interplay between the two layers forms a Fabry-Perot resonance and thus enhances the phase accumulation. Given that the structure is asymmetric, it can be modeled as a bianisotropic boundary. Measured results of the multifunctional metasurface, shown in Fig. 17, can be found in [110].

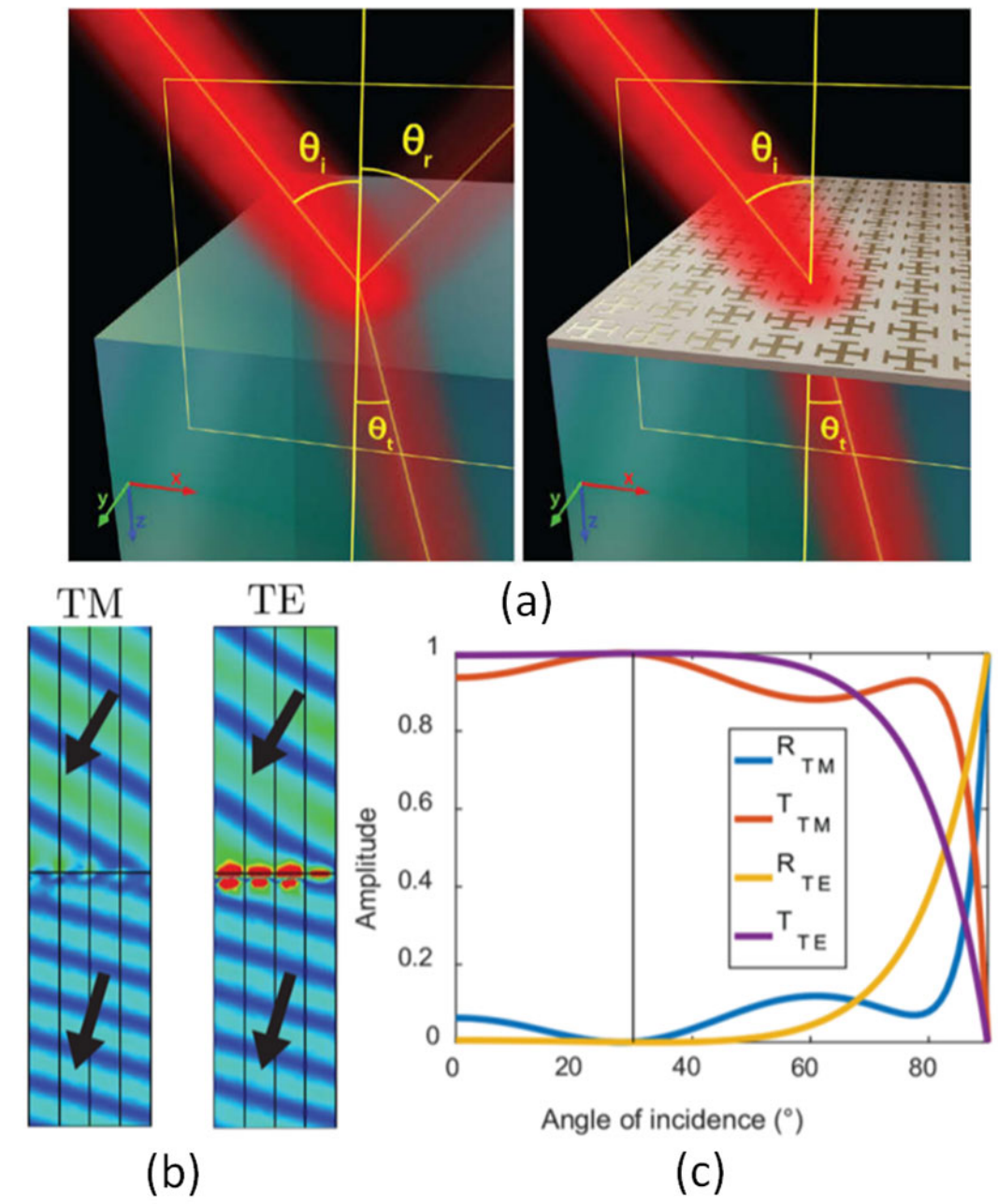

Fig. 18. (Reprinted with permission from [26], © The Optical Society). (a) Scattering of a wave impinging on the interface under an arbitrary angle $(\theta_a)$, with conventional Fresnel transmission and reflection for the case of the bare interface (left) and with reflectionless (Brewster) transmission when a properly designed metasurface is placed at the interface (right). (b) Full wave simulated electric field amplitude distribution and (c) angular response of the reflectance and transmittance for polarization-independent $xz$-plane Brewster metasurfaces with the general parameters $(\epsilon_{r,a}, \epsilon_{r,b}) = (1,3)$ for generalized Brewster angle at $\theta_a$=30°.

In [113], multifunctional reflectors are designed which utilize the propagating spatial harmonics of a periodic metasurface to create multifunctional metasurfaces. These metasurfaces can simultaneously control reflection from and into several directions in space. Examples of three channel retroreflectors and three channel power splitters which can send specified amounts of power into the three diffraction order directions are demonstrated.

### D. *Generalized Brewster Effect*

Bianisotropic metasurfaces can also be used to generalize the Brewster effect [26]. Under the generalized Brewster effect, incident waves are totally transmitted with no reflections for both polarizations and for arbitrary incidence angles (see Fig. 18a). In [26], the authors find total transmission with no reflections from metasurfaces with only electric and magnetic susceptibility requires complex susceptibilities and hence require particles exhibiting loss and gain. However, by including magnetoelectric coupling, the metasurface can realize the generalized Brewster effect from purely passive and lossless metasurfaces. An example of a bianisotropic metasurface designed to exhibit the Brewster effect at $\theta_a = 30°$ is shown in Fig. 18. The full-wave simulated electric field amplitude distribution is shown in Fig. 18b and the angular dependance is shown in Fig. 18c. As can be seen, the metasurface exhibits generalized Brewster angle for both polarizations at $\theta_a = 30°$. Due to the inclusion of bianisotropy, the metasurface was made lossless and passive.

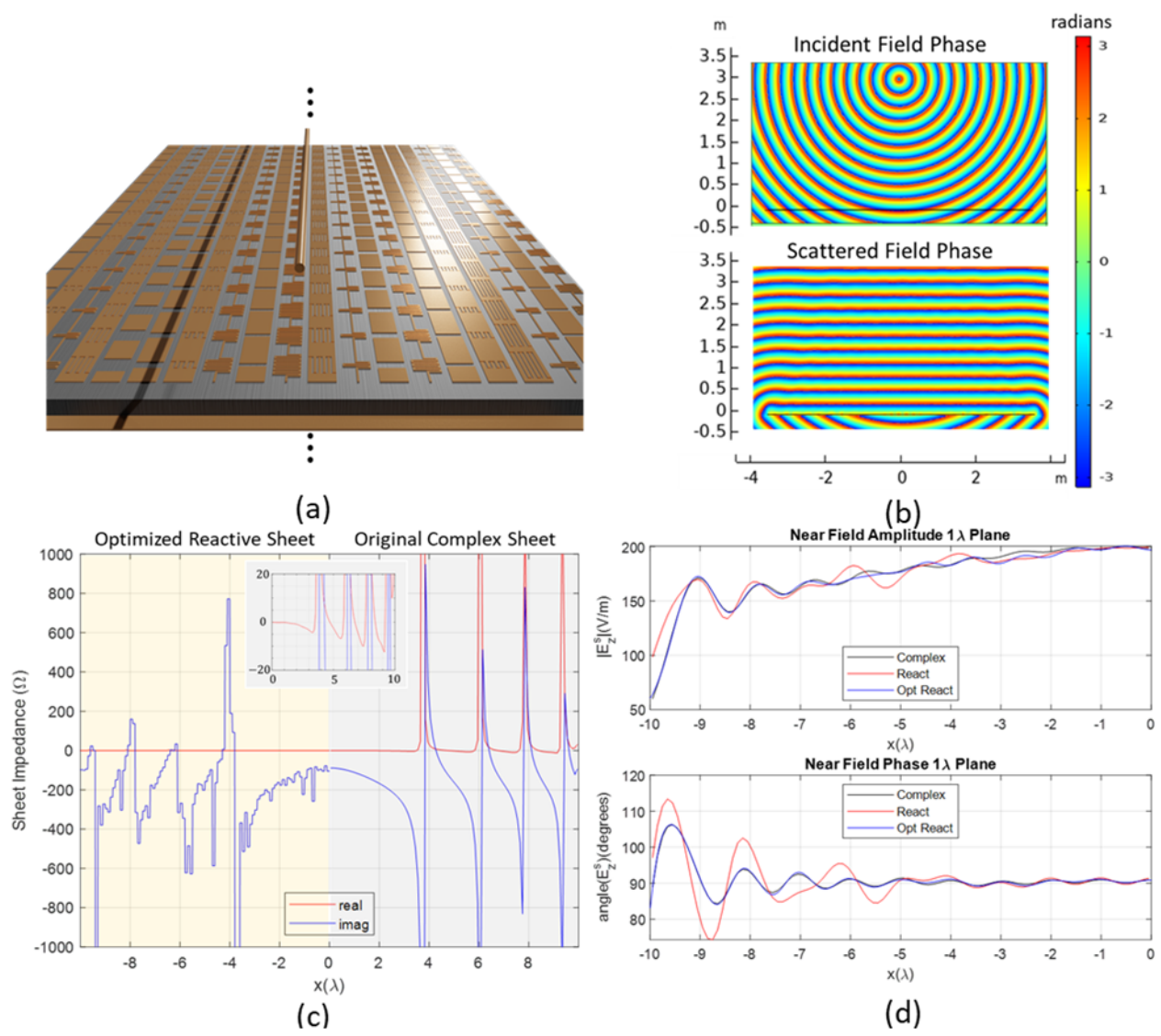

Fig. 19. (Reprinted with permission from [32]). (a) The perfectly reflecting metasurface. The geometry is finite in the transverse directions and infinite and invariant in the axial direction (2D electromagnetics problem). (b) Phase plots of the incident and scattered fields. (c) Comparison of original complex-valued sheet impedance (right half) and optimized reactive sheet impedance (left half). (d) Comparison of near fields observed at a plane $1\lambda$ above the metasurface as scattered by the complex-valued sheet ('Complex'), the complex-valued sheet with resistances discarded ('React') and the optimized reactive sheet ('Opt React').

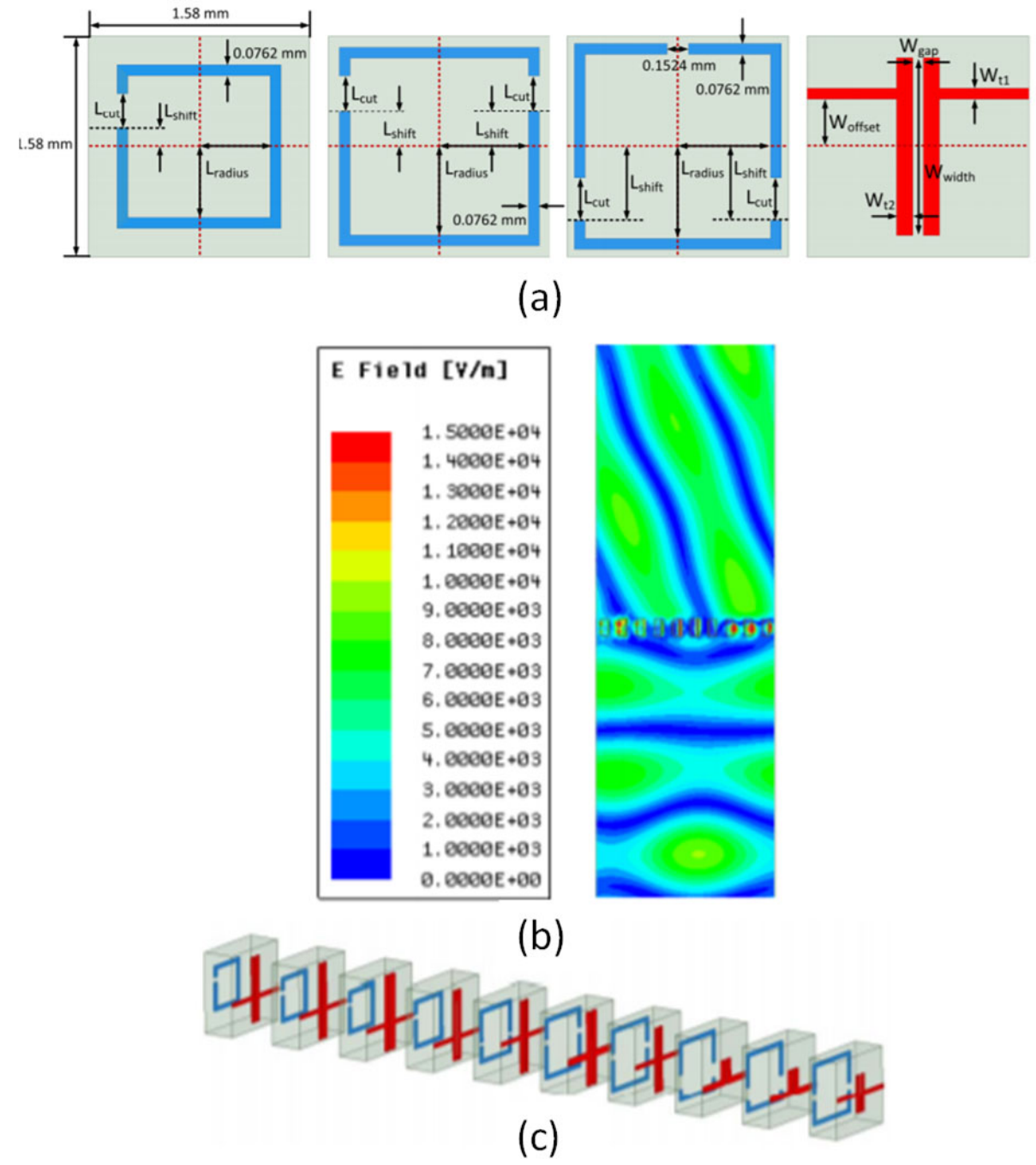

Fig. 20. (Reprinted with permission from [24]). (a) Proposed omega-bianisotropic wire-loop geometries for (leftmost) single-cut loop, (second from left) double-cut loop, (third from left) triple-cut loop, (fourth from left) shifted wire. (b) TE refraction design full-wave simulation, Electric field distribution of one period via periodic simulation in HFSS at 20 GHz. (c) Physical period of the TE refraction design.

### E. *Perfect Reflection*

An application of metasurfaces that has also received attention recently is the concept of perfect reflection. Here, we define perfect reflection as the transformation of an incident wavefront into a desired reflected wavefront without the generation of additional undesired radiation. Such perfect transformation typically requires complex sheet impedances [37]. Since the real part of the sheet impedance represents loss and/or gain, which results in inefficient designs or the need for active components, there is a desire to perform the transformation with a purely reactive sheet. Authors have approached this problem in a variety of ways [37], [38], [41]–[43], [115], [116].

For example, in [32], [37], the authors solve the problem by beginning with the local metasurface design (requiring loss and/or gain) with a complex sheet impedance above a grounded dielectric substrate and find the scattered field amplitude and phase in the radiative near field using the integral equation modelling technique of section III.D (see Fig. 19c). Setting a targeted field distribution in the radiative near field as the optimization goal, they discard the real part of the initial complex-valued sheet impedance and optimize the remaining reactances such that the targeted field distribution is achieved using only the reactances. In other words, the amplitude and phase of the radiative near field is shaped with only a single fully reactive electric layer by introducing surface waves which add to the total field on the metasurface in a way that leads to a passive and lossless metasurface. Observation of the fields along a plane in the radiative near field avoids the difficulty of reconstructing evanescent components through optimization. Thus, by optimizing the fields at one wavelength away from the surface, only the radiated fields are considered. The resulting sheet reactances are non-intuitive (see Fig. 19c) and produce the same radiative near fields (see Fig. 19d) and far field patterns as the design involving complex sheets.

### F. *Perfect Wide-Angle Refraction*

In addition to perfect reflection, perfect wide-angle refraction is also possible [23]–[25], [41], [48]. Consider the work in [24] where a wire-loop unit cell topology is used to achieve reflectionless wide-angle refraction of a normally incident plane wave to a refraction angle of 71.8° with respect to the surface normal. By specifying the TE-polarized desired incident, reflected, and transmitted fields in both regions 1 (below the metasurface) and 2 (above the metasurface),

$$\begin{aligned}
E_{t,1} &= E_{x,1}(y,z) = E_{0,1}e^{-jk_0\cos\theta_{in}z}e^{-jk_0\sin\theta_{in}y}\\
H_{t,1} &= H_{y,1}(y,z) = \frac{1}{Z_{0,1}}E_{0,1}e^{-jk_0\cos\theta_{in}z}e^{-jk_0\sin\theta_{in}y}\\
E_{t,2} &= E_{x,2}(y,z) = E_{0,2}e^{-jk_0\cos\theta_{out}z}e^{-jk_0\sin\theta_{out}y}\\
H_{t,2} &= H_{y,2}(y,z) = \frac{1}{Z_{0,2}}E_{0,2}e^{-jk_0\cos\theta_{out}z}e^{-jk_0\sin\theta_{out}y}
\end{aligned} \tag{67}$$

where

$$Z_{0,1} = \frac{\eta}{\cos\theta_{in}},\; Z_{0,2} = \frac{\eta}{\cos\theta_{out}},\; \left|E_{0,2}\right| = \sqrt{\frac{Z_{0,2}}{Z_{0,1}}}\left|E_{0,1}\right|$$

the IBC (22) can be used to derive the constituent surface parameters necessary for wide- angle refraction as

$$
\begin{aligned}
Z_{se} &= -j\left[\frac{1}{2}\operatorname{Im}\left\{\frac{E_{1,x}+E_{2,x}}{H_{2,y}-H_{1,y}}\right\}\right] - j\left[K_{em}\operatorname{Im}\left\{\frac{E_{2,x}-E_{1,x}}{H_{2,y}-H_{1,y}}\right\}\right] \\
Y_{sm} &= -j\left[\frac{1}{2}\operatorname{Im}\left\{\frac{H_{1,y}+H_{2,y}}{E_{2,x}-E_{1,x}}\right\}\right] + j\left[K_{em}\operatorname{Im}\left\{\frac{H_{2,y}-H_{1,y}}{E_{2,x}-E_{1,x}}\right\}\right] \\
K_{em} &= \frac{1}{2}\frac{\operatorname{Re}\left\{E_{2,x}H_{1,y}^{*}-E_{1,x}H_{2,y}^{*}\right\}}{\operatorname{Re}\left\{\left(E_{1,x}-E_{2,x}\right)\left(H_{2,y}-H_{1,y}\right)^{*}\right\}}
\end{aligned}
\tag{68}
$$

where Im[ ] denotes the imaginary part operator. To realize these constituent surface parameters, either a three-sheet implementation or an offset wire-loop unit cell can be used. When the wire is offset with respect to the loop or the loading of the loop is offset with respect to the loop, as shown in Fig 20a, the parameters of (68) can be realized. A metasurface was built from these unit cells (shown in Fig. 20c) and simulated using HFSS. The results in Fig. 20b show the metasurface is performing the wide-angle refraction. In [24], the authors also design a wide-angle refracting metasurface for TM polarization and obtain measurements from fabricated samples.

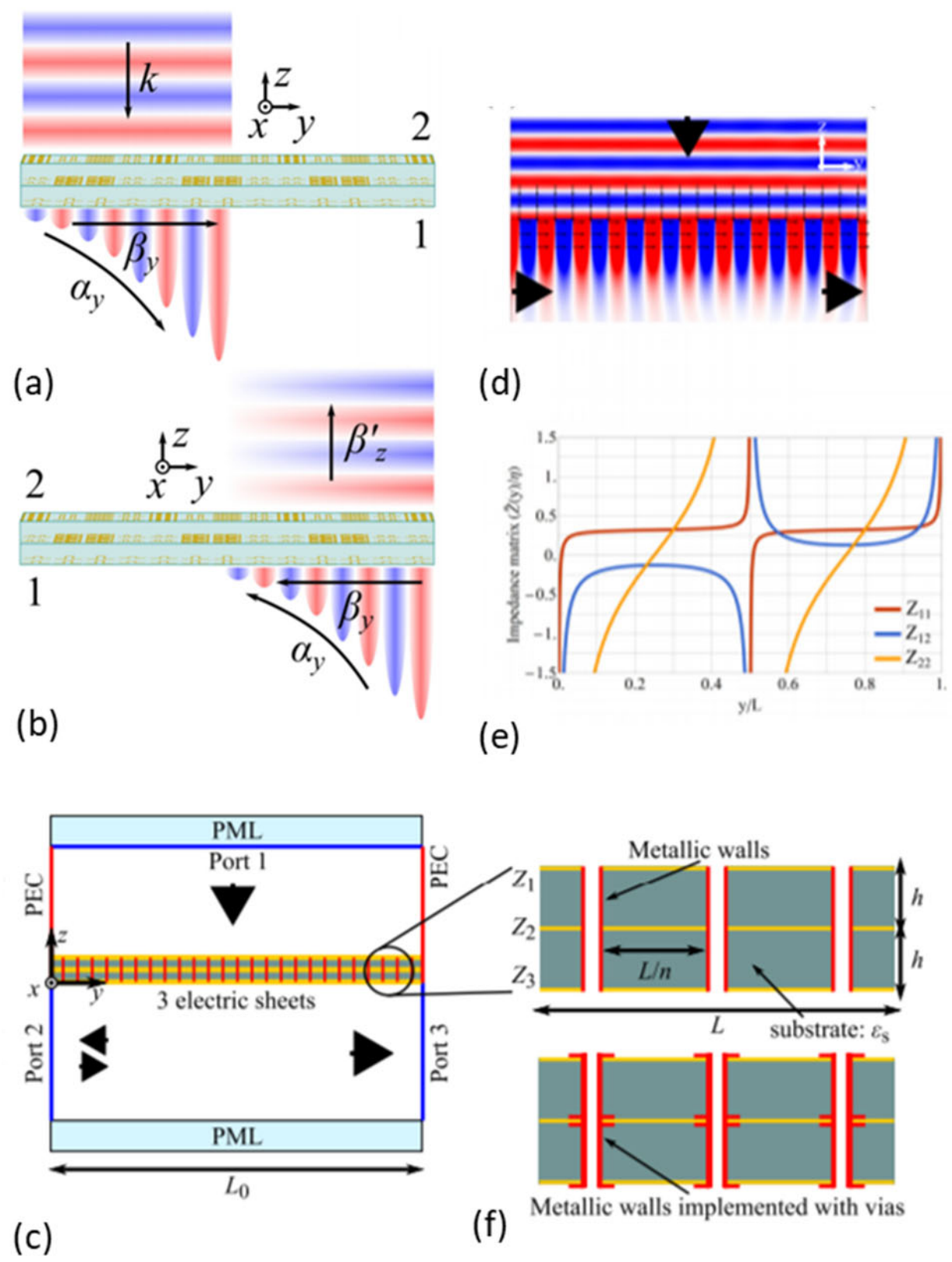

Fig. 21. (Reprinted with permission from [50]). (a) Schematics of a metasurface converting a normally incident plane wave into a transmitted surface wave with ethe propagation constant $\beta_y$ and the growth rate $\alpha_y$. (b) Schematics of a metasurface converting a surface wave into an inhomogeneous plane wave propagating in the normal direction with the propagation constant $\beta'_z$. (c) Schematic of the COMSOL models used for simulating the conversion with asymmetric three-layer structure. Port 1 launches the normally incident plane wave. Port 2 either launches or accepts the surface wave. Port 3 only accepts the excited surface wave. (d) Snapshot of the magnetic field for a metasurface with 10 periods, the growth rate is $\alpha_y = 0.001k$ with Port 2 on. The arrows depict the directions of power flow density. The metasurface is represented by an omega-bianisotropic combined sheet and propagation constant of the surface wave equals $\beta_y = 1.05k$. (e) Imaginary part of the impedance matrix as functions of the $y$-coordinate. (f) Zooming of the three-layer metasurface with metallic walls (implemented with vias in upper) separating individual unit cells, $n$ is the number of unit cells per super cell, $Z_i$ $(i = 1,2,3)$ is the electric surface impedance of the corresponding sheet.

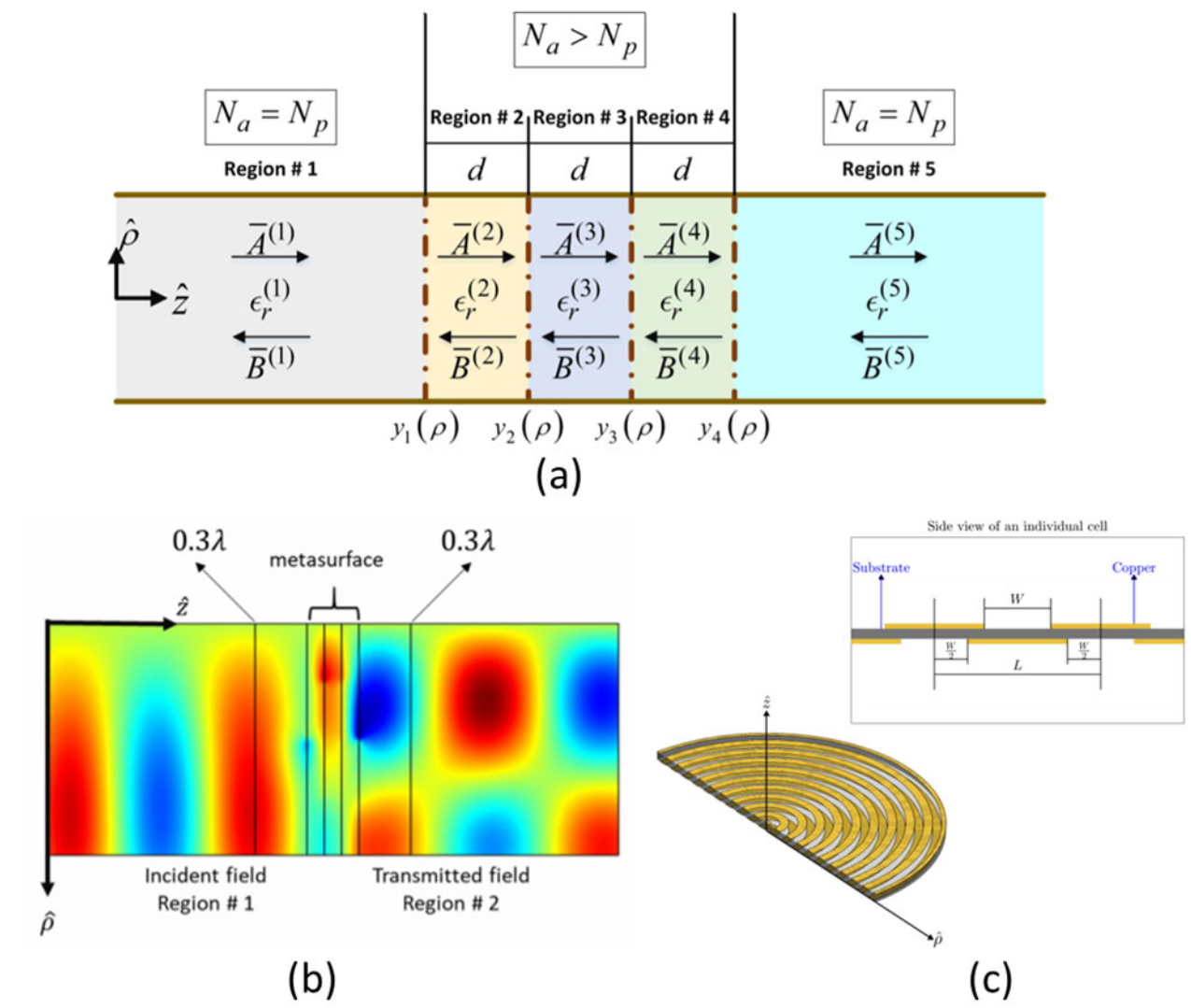

Fig. 22. (Reprinted with permission from [56]). (a) A metasurface consisting of cascaded electric sheets placed perpendicular to the propagation axis within an over-moded cylindrical waveguide. The metasurface comprises four electric sheets described by inhomogeneous admittance profiles $y_n(\rho)$. The sheets are separated by dielectric spacers of thickness $d$. The metasurface divides the waveguide into two outer regions (Region 1 & 5), and three inner regions (Region 2 and 4). (b) Simulated performance of the metasurface-based single mode converter with ideal electric admittance sheets using the 2D, axially symmetric fullwave solver COMSOL Multiphysics. A 2D surface plot of the real part of the electric field for the metasurface-based single mode converter. (c) Realization of the capacitive sheets. Metallic rings are printed on both sides of a thin substrate. The bottom rings are shifted by a half a cell with respect to the top rings.

### *G. Perfect Leaky Wave to Surface Wave Transformations*

Another perfect transformation enabled by bianisotropic metasurfaces is the perfect conversion of a surface wave to a leaky wave [49] or the near-perfect conversion of the converse, a propagating wave into a surface wave [50]. In [51], plane wave to surface wave couplers, which can perform both functions (either plane wave to surface wave or surface wave to leaky wave), are reported. Here, we review the work of [50]. The Omega-bianisotropic metasurface converts an incident plane wave into a surface wave as depicted in Fig. 21a. The metasurface is excited from above by a normally incident plane wave (Port 1 in Fig. 21c). An input surface wave is launched from Port 2 in Fig. 21c. The combined input surface wave and converted surface wave from the incident plane wave are absorbed by Port 4 in Fig. 21c. The input surface wave is necessary to obtain a reactive and symmetric impedance matrix representing the metasurface. The design process results in the impedance matrix elements shown in Fig. 21e. The metasurface is realized as a stack of three sheets (see section IV.C) with each cell separated by metallic vias to combat transverse coupling not modelled in the design (see Fig. 21f). The result of the 2D simulation is shown in Fig. 21d. The conversion efficiency, defined as the difference of the output power from Port 3 ($P_3$) and the input power from Port 2 ($P_2$) divided by the power delivered by the incident plane wave from Port 1 ($P_1$) or $(P_3 - P_2)/P_1$, is between 90-95% depending on the length of the metasurface.

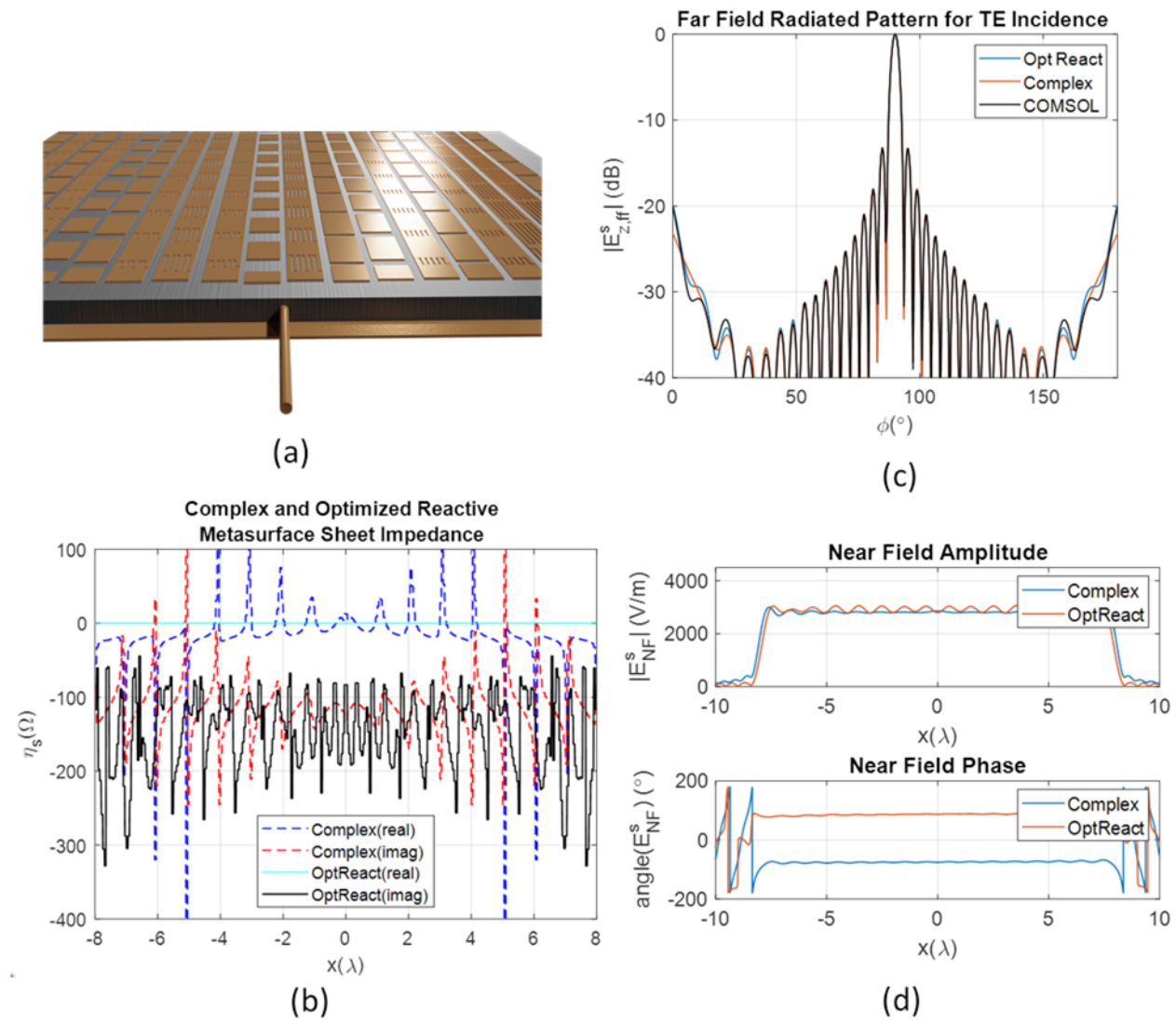

Fig. 23. (Reprinted with permission from [52]). (a) A metasurface antenna consisting of a patterned metallic cladding supported by a grounded dielectric substrate fed by an infinite electric line source placed within the substrate. The geometry is finite in the transverse directions and infinite in the axial direction and hence the electromagnetics problem is 2-dimensional. (b) The initial complex-valued sheet impedance and the optimized reactive sheet impedance. (c) Far fields of the complex-valued sheet design, the optimized reactive sheet design, and a full wave verification of the optimized reactive sheet design in COMSOL. (d) Backprojected far fields to the metasurface plane to show aperture fields.

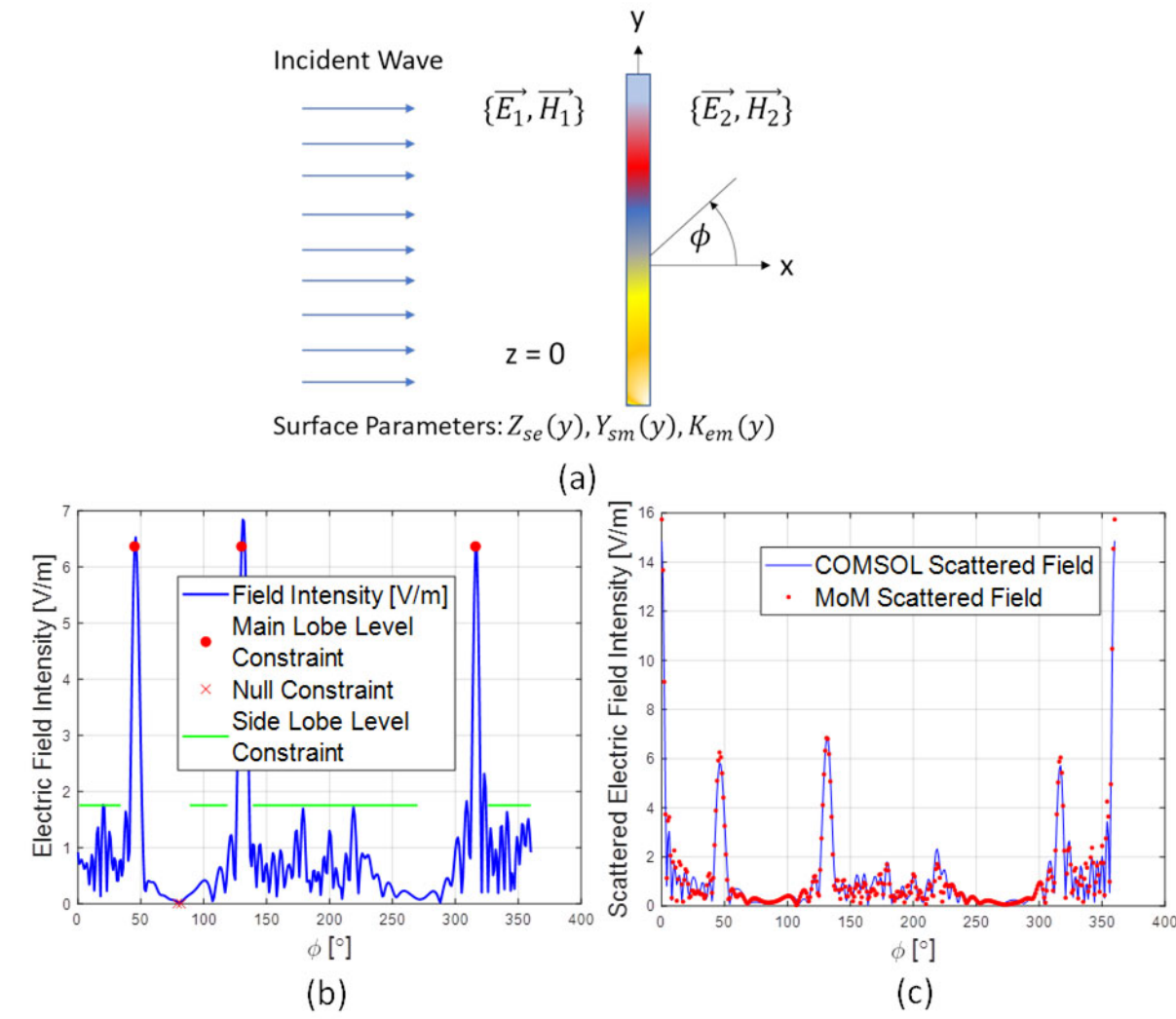

Fig. 24. (Reprinted with permission from [34]). (a) Depiction of a general electromagnetic metasurface (EMMS). Fields on each side of the EMMS ($\vec{E}_1, \vec{H}_1$ and $\vec{E}_2, \vec{H}_2$) induce electric and magnetic surface currents ($\vec{J}_s, \vec{M}_s$) according to the surface parameters ($\bar{\bar{Z}}_{se}, \bar{\bar{Y}}_{sm}, \bar{\bar{K}}_{em}$, and $\bar{\bar{K}}_{me}$). (b) Multi-criteria optimization of far-field parameters with an omega-type bianisotropic EMMS. (c) Scattered electric field intensities comparison for multi-criteria optimization results with COMSOL and MATLAB-based MoM solver.

### H. Perfect Mode Converters

Bianisotropic metasurfaces can also be used for perfect conversion between two different sets of waveguide modes [56]. In this work, the authors create bianisotropic metasurfaces constructed from a cascade of four admittance sheets (see section IV.C) to perfectly convert a set of $TM_{0n}$ modes to a desired set of $TM_{0n}$ reflected/transmitted modes within an over-moded cylindrical waveguide (see Fig. 22a). In the full-wave results shown in Fig. 22b, the metasurface is designed to convert the $TM_{01}$ mode to the $TM_{02}$ mode with a -45° transmission phase. The metasurfaces are designed using a combination of modal network theory accelerated by Discrete Hankel Transforms, and optimization. The approach allows rapid synthesis based exclusively on matrix operations. The metasurface's electric sheet admittances are realized as arrays of conductive cylindrical rings (see Fig. 22c). The technique can also be applied to the synthesis of aperture antennas [55].

### I. Perfect Antennas

Another application of metasurfaces is in enhanced antenna design. For example, in [52], [117], [118], a metasurface antenna is designed to achieve perfect (100%) aperture efficiency. In [52], the metasurface antenna consists of a patterned metallic cladding supported by a grounded dielectric substrate and fed by an infinite electric line source placed within the substrate, as seen in Fig. 23a. The metasurface is modeled using a reduced version (the first three rows and columns) of the matrix equation in (46) to account for 1 sheet impedance layer, 1 dielectric spacer, and a ground plane (an impedance sheet of zero impedance). Using the first of (47), the matrix equation can be directly solved since the desired total field is equal to the summation of the known incident cylindrical wave field generated from the line source placed within the substrate and the desired scattered aperture field of uniform amplitude and phase. The solution results in a complex-valued sheet design labeled 'Complex' in Fig. 23. The radiative near and far fields of the complex-valued sheet show the desired performance (Fig. 23c and 23d). Using the retained reactances (with resistances discarded) of the complex-valued sheet as a seed solution, and the amplitude of the far field pattern as an optimization goal, gradient descent optimization accelerated by the Adjoint Method [119] is applied to convert the complex-valued sheet into a purely reactive one by introducing surface waves which facilitate passivity and losslessness. The results of the optimization are labeled as 'OptReact' in Fig. 23. As can be seen, the performance of the optimized reactive sheet is identical to the complex-valued design. Also shown in Fig. 23c is the full wave simulation results from COMSOL Multiphysics of the optimized reactive sheet for validation. The metasurface generates a uniform aperture field from a passive and lossless metasurface, and hence exhibits perfect aperture efficiency in a compact form factor.

### J. Beamforming

With the introduction of surface waves and evanescent field engineering through optimization, metasurfaces are capable of beamforming in a passive and lossless manner [31], [33], [34], [36], [38]. Epstein was the first to introduce the concept of adding surface waves to achieve passivity in [38]. In the work of [34], the authors design Omega-type bianisotropic beamforming metasurfaces using integral equation modelling techniques and numerical optimization. The metasurface shown in Fig. 24a is modeled using (42) and (43) and converted into a matrix equation (44) following from the method of moments in 2-dimensions [99]. The matrix equation is solved by optimization (the alternating direction method of multipliers). The optimization goals are formulated around the desired far field beams calculated from the vector potentials (see Fig. 24b). The optimization results in the specifications for the constitutive surface parameters ($\bar{\bar{Z}}_{se}, \bar{\bar{Y}}_{sm}, \bar{\bar{K}}_{em}$, and $\bar{\bar{K}}_{me}$). These parameters are inserted back into (42) to form a boundary condition which can be enforced in simulation for analysis. The results of the COMSOL simulation are shown compared to the MATLAB-based method of moments results in Fig. 24c. These metasurfaces can potentially by realized using the three-sheet method.

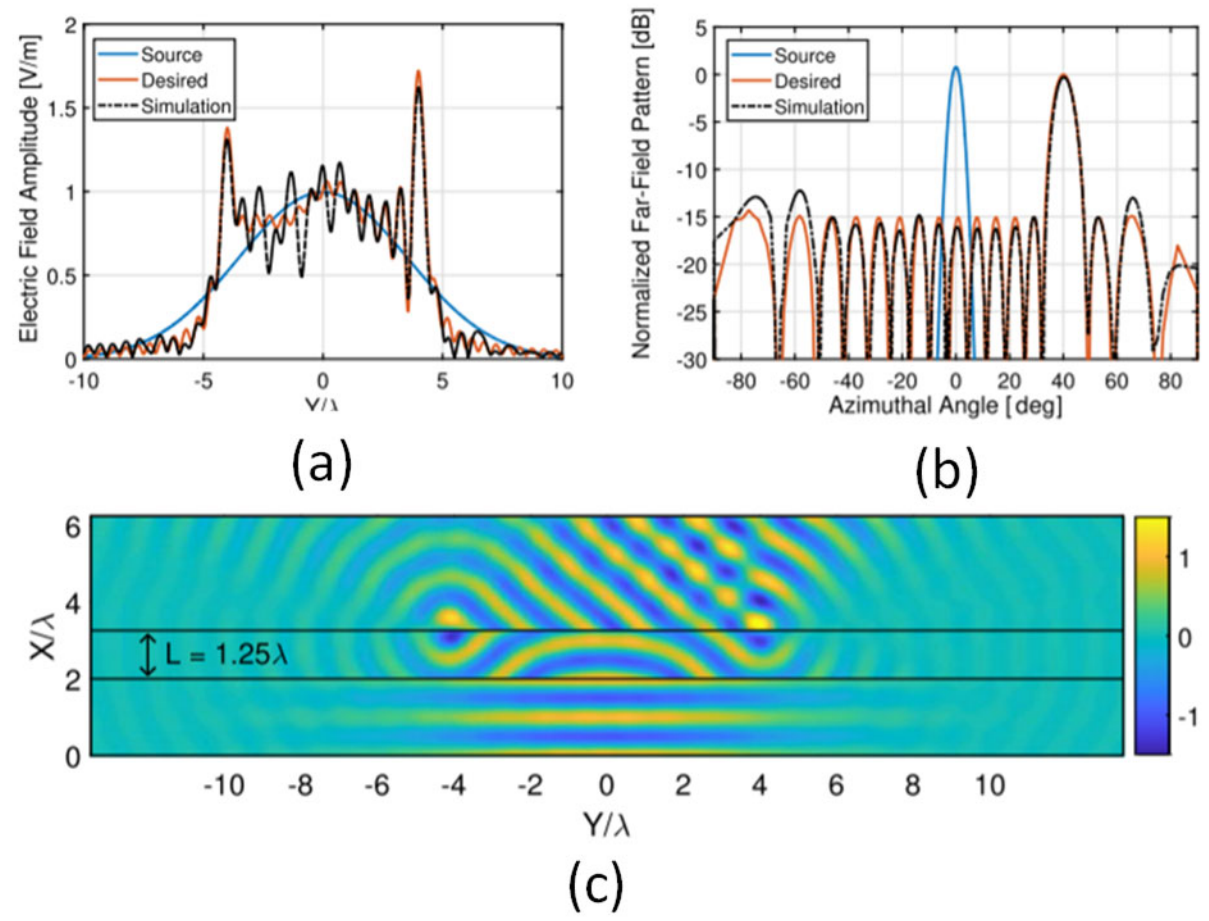

Fig. 25. (Reprinted with permission from [114]). A compound metaoptic reshapes an incident Gaussian beam to produce a Dolph-Chebyshev far-field pattern pointed towards 40 deg. The metaoptic performance is shown in (a) as the transmitted electric field amplitude λ=3 from the metaoptic, (b) as the far-field radiation pattern, and (c) as the real part of the simulated electric field.

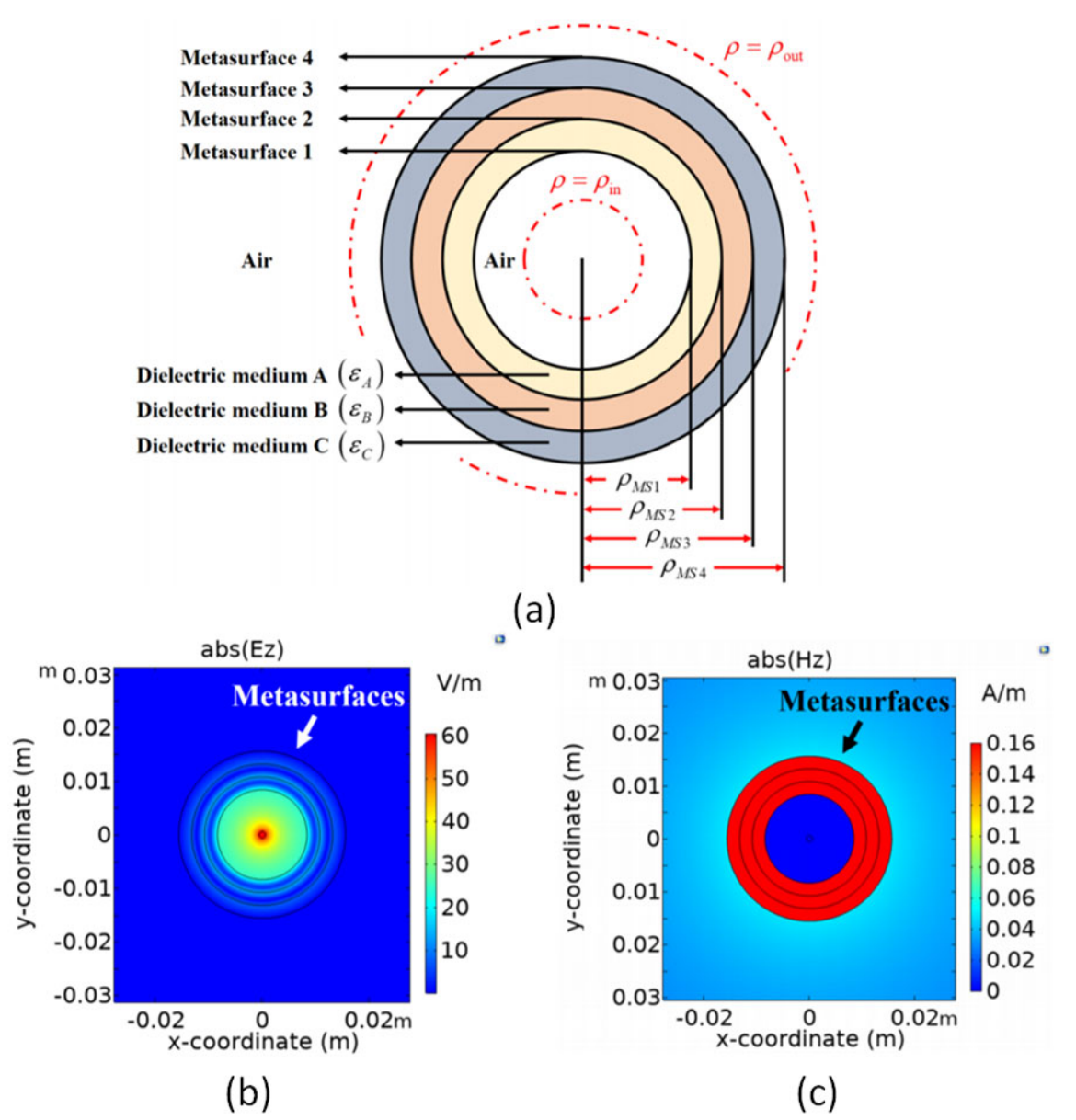

Fig. 26. (Reprinted with permission from [107]). (a) Schematic of cascaded conformal cylindrical metasurface. (b-c) Field plots of the polarization converter when excited by an electric surface current density near the center. (b) Amplitude of $E_z$. (c) Amplitude of $H_z$.

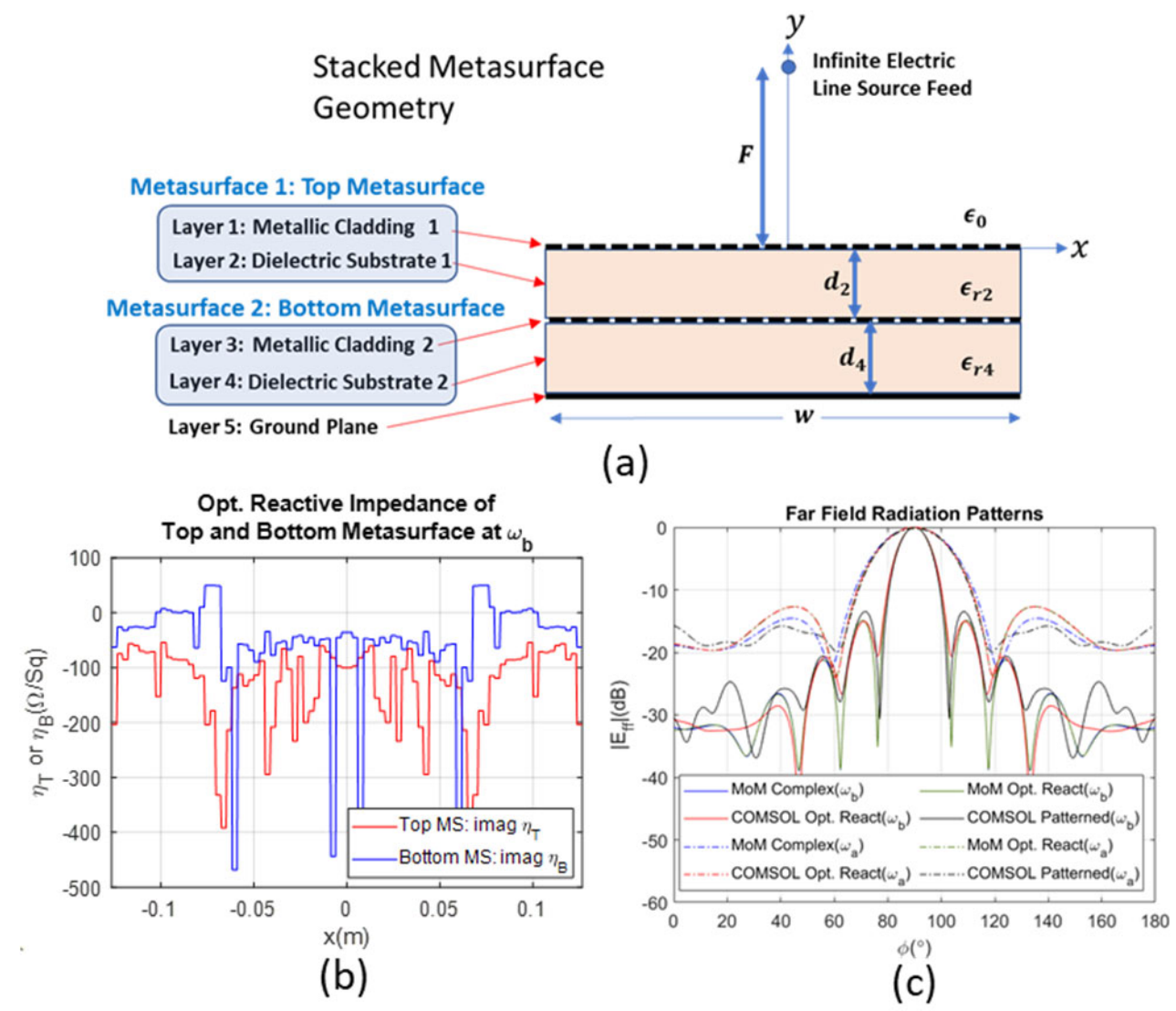

Fig. 27. (Reprinted with permission from [57]). (a) Dual band stacked metasurfaces geometry. The geometry is two-dimensional. The metasurface contains five layers: two metasurface layers (layers 1 and 3) and a ground plane (layer 5) separated by two dielectric spacers (layers 2 and 4). An infinite electric line source placed $F$ meters above the aperture feeds the metasurface. (b) Optimized reactive sheet impedance for top and bottom metasurface at $\omega_b$. (c) Far field patterns of dual band stacked metasurface.

### *K. Metasurface Pairs*

A pair of lossless and passive metasurfaces separated by a wavelength scale distance can perform arbitrary wavefront shaping (amplitude and phase) as well as beamforming [35], [40], [120]. In the work of [120], a pair of bianisotropic metasurfaces modeled using the IBC in (22) are synthesized using a modified Gerchberg-Saxton phase retrieval algorithm to reshape an incident Gaussian beam into a Dolph-Chebyshev far-field pattern pointing toward 40°. The metasurfaces are separated by 1.25 wavelengths. The results of the beamforming synthesis are shown in Fig. 25. Each metasurface in the pair is realized through the three-sheet technique of section IV.C.

### *L. Conformal Metasurfaces*

Metasurfaces conformal to different shaped surfaces have also seen interest in recent scientific works [107], [121]. Consider the work in [107]. By formulating the wave matrix synthesis approach of section III.C in terms of cylindrical modes, the same approach can be used to synthesize cascaded cylindrical metasurfaces. The authors consider a cascade of four electric admittance sheets each separated by dielectric spacers as seen in Fig. 26a. The cylindrical metasurface is designed such that the $TE_z$ modes created by the electric line source placed at the center of the geometry (Fig. 26b) are completely converted to $TM_z$ modes at the output (Fig. 26c). The authors also report polarization splitters which split half of the incident power to $TE_z$ waves and half to $TM_z$ waves.

### *M. Multiband Metasurfaces*

Metasurfaces can be stacked to enable operation at multiple bands. For example, in [57]–[59], an algorithm to synthesize dual-band, stacked metasurfaces is presented. The reflective, stacked metasurface configuration of two metasurfaces (each metasurface consists of a patterned metallic cladding and a dielectric spacer) stacked one upon the other is shown in Fig 27a. The stacked metasurface is modeled using (46) except the impedance of the 5th layer, $Z_{ee}^{5}$, is set equal to zero to represent the ground plane. An iterative algorithm introduced in [57] is used to synthesize the dual band metasurface to collimate the incident cylindrical wave at two different frequencies (for example, at both $f_a = 2.4$ GHz and $f_b = 5.1$ GHz for the case in Fig. 27). The synthesis approach results in complex-valued sheet impedances for both cladding layers. The reactances (resistances discarded) of these complex-valued sheets are used as a seed for a gradient descent optimization, accelerated by the Woodbury Matrix Identity [122], to convert the complex-valued sheet into a purely reactive sheet which scatters the same far fields as the complex-valued sheet design. The results of the optimization are shown in Fig. 27b. The far-field patterns scattered by the stacked metasurface for both the complex-valued sheets and the purely reactive sheets are shown in Fig. 27c. Also included in the figure are the full wave verifications in COMSOL Multiphysics for both the homogenized ideal sheet case and for the realized patterned metallic cladding.

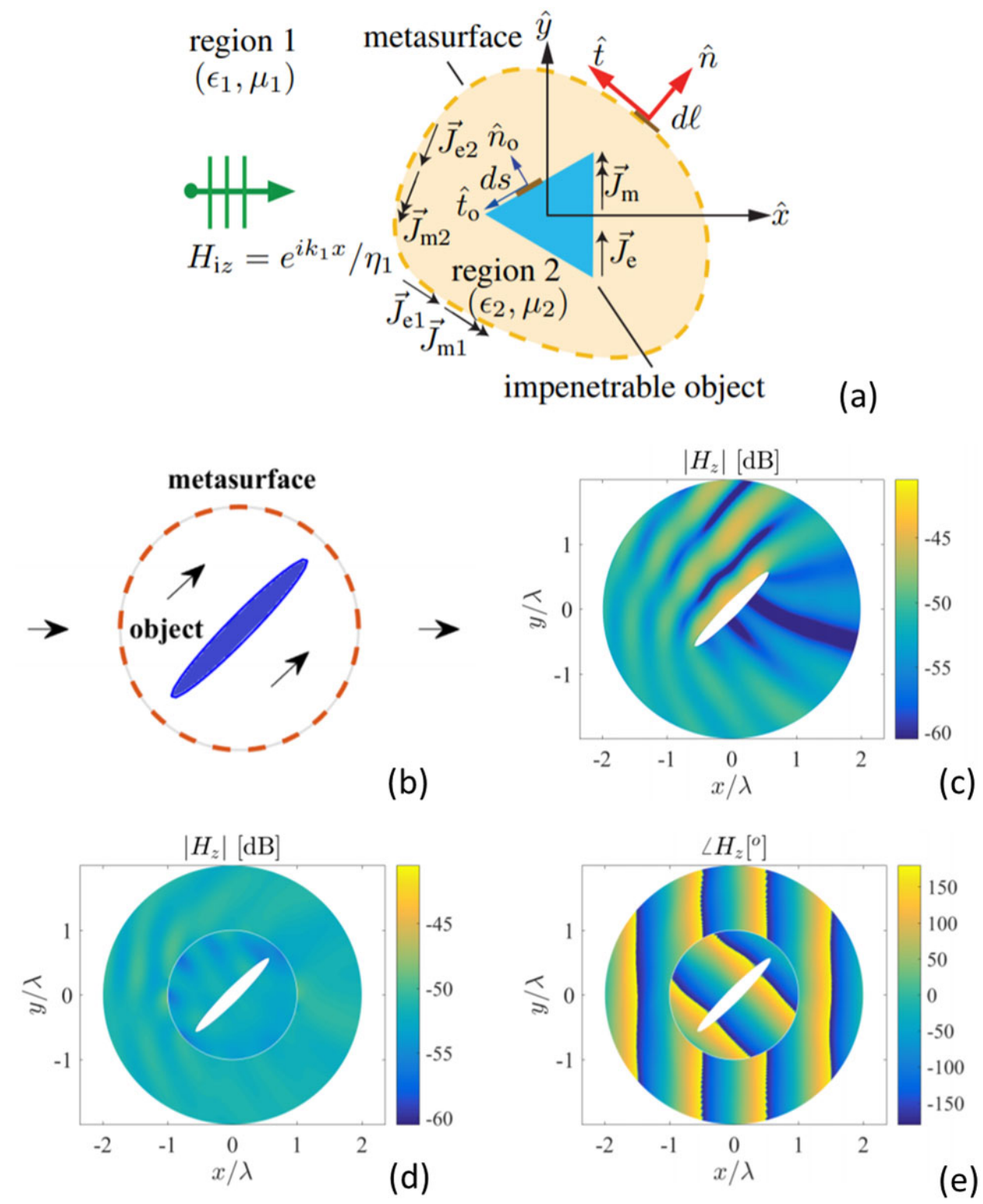

Fig. 28. (Reprinted with permission from [86]). (a) Scattering by an impenetrable cylindrical object inside a metasurface cavity using equivalent electric and magnetic currents. (b) A tilted thin elliptic cylinder inside a circular metasurface cavity. Arrows show the propagation directions of plane waves inside and outside of the metasurface cavity. (c) Magnitude of the total magnetic field when there is no metasurface. (d) Magnitude and (e) phase of the total magnetic field inside and outside of the metasurface cavity when there is metasurface.

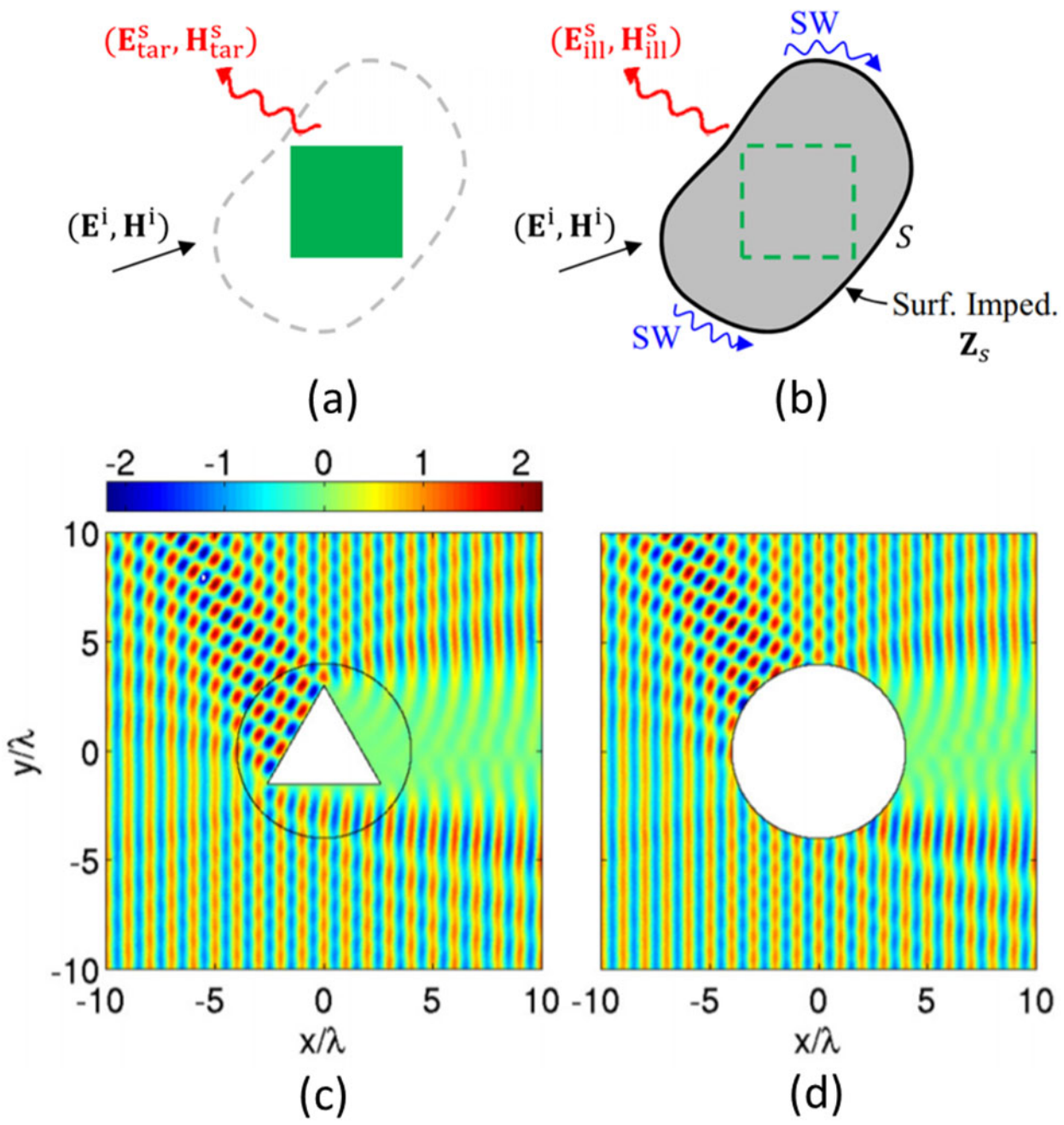

Fig. 29. (Reprinted with permission from [85]). (a-b) Design of an illusion device. (a) A target configuration with scattered fields ($\vec{E}_{tar}^{s}$, $\vec{H}_{tar}^{s}$). (b) An illusion device having an impenetrable bounding surface S with scattered fields ($\vec{E}_{ill}^{s}$, $\vec{H}_{ill}^{s}$) on and outside S. (c-d) Distributions of $E_z$(x, y) in V/m for a triangular PEC cylinder and its illusion device. The target is a cylinder of an equilateral triangular cross section with a side of 3√3λ. (c) An E-field snapshot for the target cylinder. The black circle indicates the imaginary contour of the illusion device to be designed. (d) An E-field snapshot for the cylindrical illusion device of a = 4λ.

## N. *Perfect Cloaking*

The field of cloaking has excited researchers and the public in general. The first cloaks were metamaterial based and designed using a transformation optics approach [123]. Metasurfaces have enabled mantle cloaks which can lay conformal to surfaces allowing the cloaking of objects hidden within thin metasurface coverings [83] through polarization cancellation. Early mantle cloaks relied on electric surface impedances only. By introducing bianisotropic metasurfaces, better cloaks were created. These cloaks were made from penetrable bianisotropic metasurfaces and were termed perfect in literature [87]. These perfect cloaks were designed using an integral equation formulation similar to (42)-(44). Following from [88], by combining the GSTC with integral equations written in both region 1 (outside the metasurface cloak) and region 2 (inside the metasurface cloak), a system of four equations (two from the GSTC and two from the IE) in four unknowns (tangential electric and magnetic fields on both sides of the metasurface) is created (see Fig. 28a). The system of integral equations is solved together to obtain the tangential fields on each side of the metasurface. From these fields, the surface susceptibilities can be obtained. In [88], the authors noted that perfect cloaking requires active/lossy metasurfaces. However, they employed metasurfaces with only electric and magnetic polarizabilities. In a subsequent publication, the same authors show that inclusion of magnetoelectric coupling can lead to perfect, passive, and lossless cloaks [87]. The cloaking of an elongated elliptical cylinder is shown in Fig. 28b. The total fields in region 1 are stipulated to be the same as the incident illuminating plane wave field. The object has minimum radar cross section at φ = 45° and 225° observation angles. Thus, the metasurface is synthesized so that the plane wave inside the cavity propagates along $\vec{k}_2 = k_2(\hat{x}cos\alpha + \hat{y}sin\alpha)$ where α = 45°. The arrows in Fig. 28b show the propagation directions of plane waves inside and outside of the metasurface cavity. Fig. 28c shows the magnitude of the total magnetic field when there is no metasurface. By including the bianisotropic metasurface, the object is perfectly cloaked as seen in Fig. 28d and 28e.

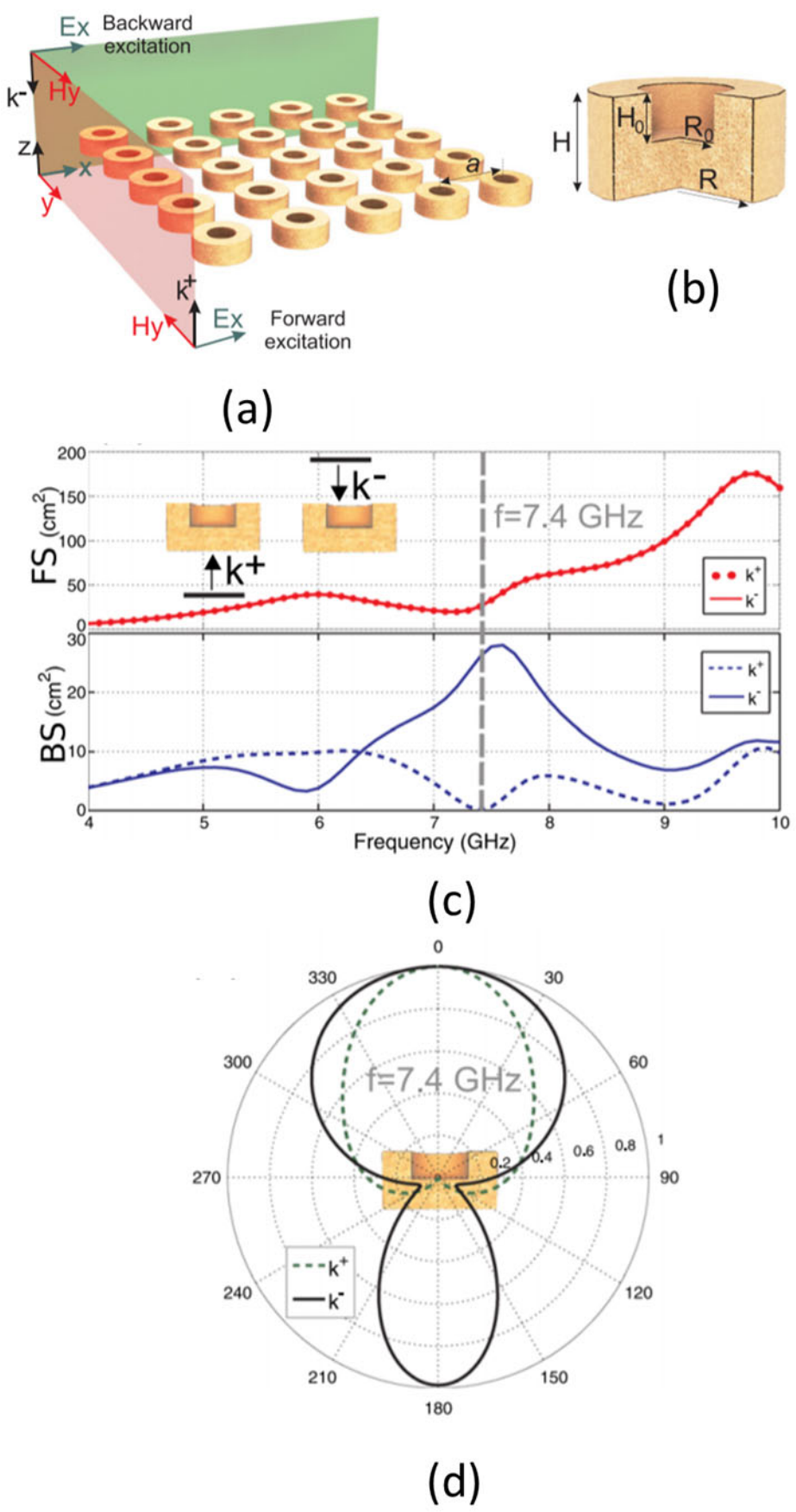

Fig. 30. (Reprinted with permission from [76]). (a) Schematic of a bianisotropic high-index dielectric metasurface. (b) A cutaway drawing of a single bianisotropic particle. (c) Simulated forward scattering (FS) and backward scattering (BS) of a bianisotropic particle. (d) Numerically simulated radiation patterns of a bianisotropic particle at the frequency of 7.4 GHz.

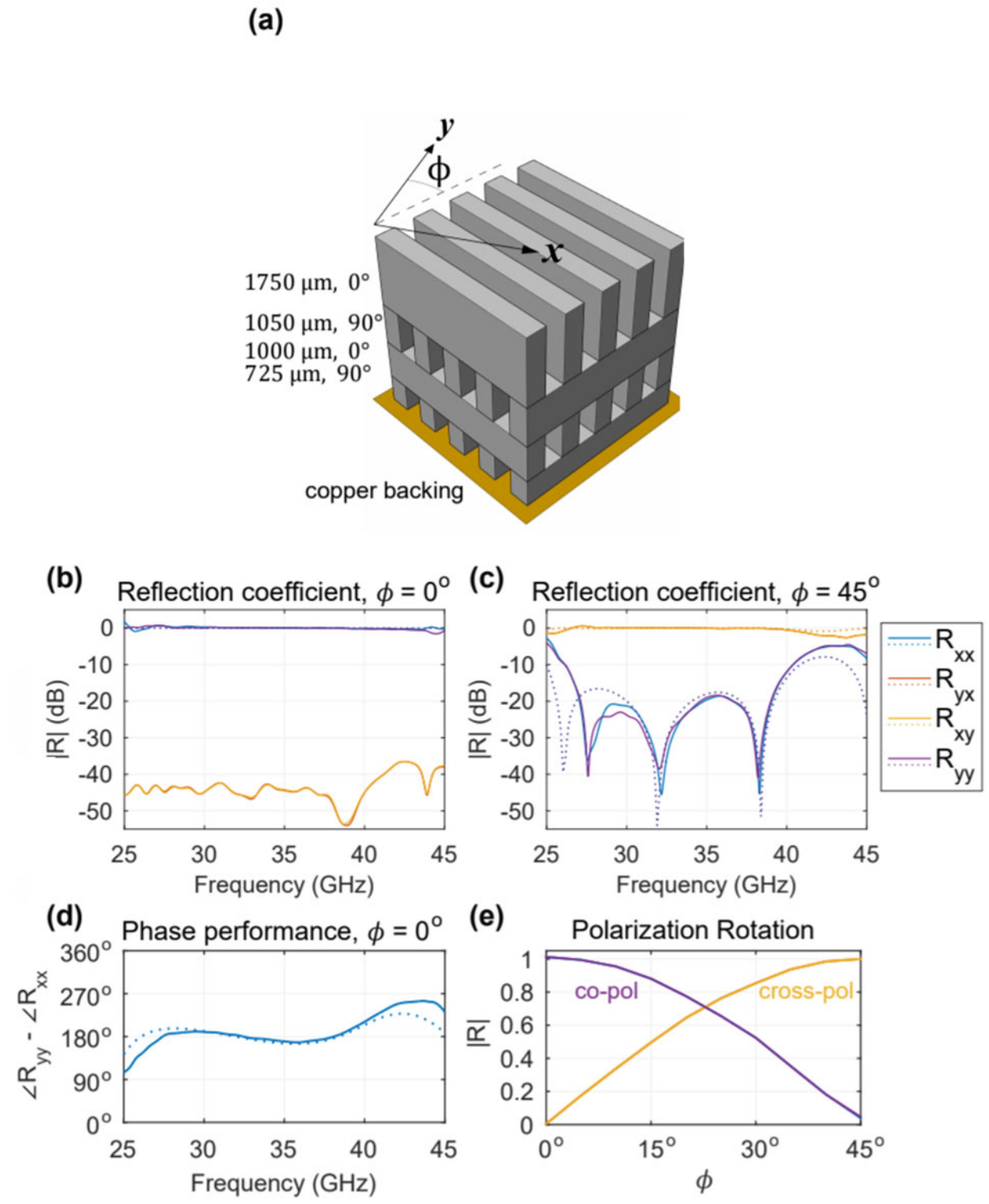

Fig. 31. (Reprinted with permission from [67]). (a) Half-wave plate design and measurement configuration. Excitation is at normal incidence with linear polarization along x and y. The waveplate's fast optic axis is rotated by an angle φ to the y axis. (b) Measured (solid) and analytically calculated (dotted) reflection coefficients with φ = 0◦. (c) Measured (solid) and calculated (dotted) reflection coefficients with φ = 45◦. (d) Measured (solid) and calculated (dotted) phase performance with φ = 0◦. Optimal is ∠Ryy − ∠Rxx = 180◦. (e) Polarization rotation at 33 GHz as a function of waveplate angle φ.

## *O. Electromagnetic Illusions*

A closely related electromagnetic phenomenon to cloaking is electromagnetic illusion. A device capable of electromagnetic illusion scatters the same field as a different object. In [85], cylindrical bianisotropic metasurfaces are designed to produce electromagnetic illusions. The concept is illustrated in Fig. 29a and 28b. In Fig. 29a, the target fields are acquired by recording the field scattered from a targeted object. Then in Fig. 29b, an impenetrable bianisotropic metasurface is designed using the IBC such that it scatters the target field when illuminated by the same incident field. The bianisotropic metasurface is made passive and lossless by including a number of evanescent surface waves in the design which travel around the perimeter of the metasurface. In Fig. 29c and 29d, an example of an impenetrable cylindrical electromagnetic illusion metasurface which scatters the same fields as a triangular PEC object is shown. The authors also provide additional examples of both PEC and dielectric objects.

## *P. All Dielectric Bianisotropic Metastructures*

Bianisotropic metasurfaces made from all-dielectric magnetoelectric particles can avoid losses associated with plasmonic metals when operated at optical or infrared frequencies. To this end, an all dielectric bianisotropic metasurface was fabricated and measured in [76] using the dielectric magnetoelectric particles presented in section IV.D. The metasurface is shown in Fig. 30a and the particle in 30b. Numerical simulation results of the particle are shown in Fig. 30c. As can be seen, the particle exhibits the same forward scattering since the particle is reciprocal, however, its backscattering differs due to the magnetoelectric coupling induced by the broken symmetry of the particle. This is further evident in Fig. 30d, where the far field patterns are shown for a single particle. In the forward direction, the electric and magnetic dipoles spectrally overlap producing a Huygens type source with unidirectional radiation, whereas when excited in the backward direction, a different pattern emerges giving rise to the scattering asymmetry indicative of bianisotropic operation. In [76], the authors show measured scattering parameters for a fabricated array of bianisotropic particles. The measured scattering parameters agree well with the theoretical operation.

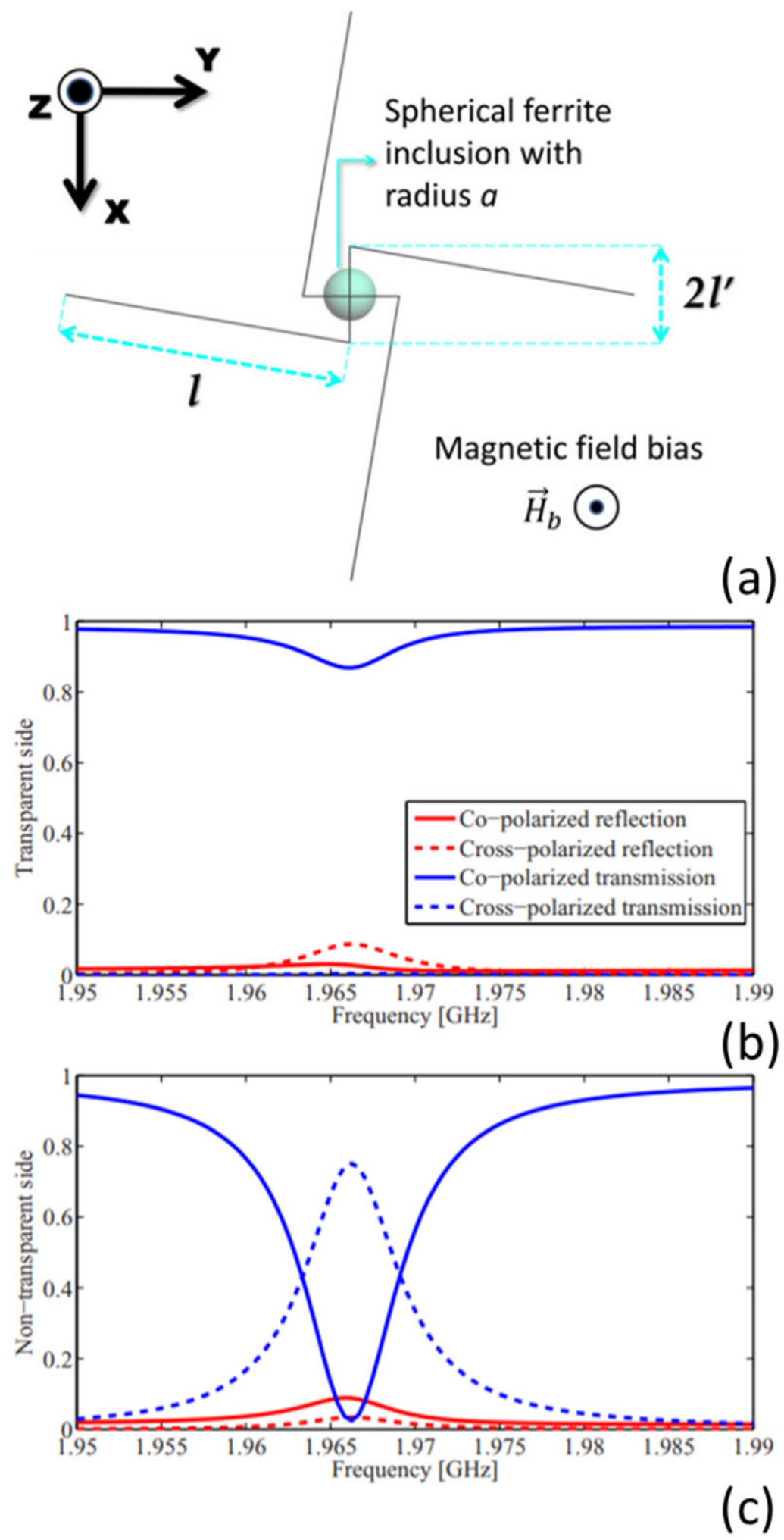

Fig. 32. (Reprinted with permission from [124]). (a) Geometry of a Moving-Chiral particle. The external magnetic field bias is along the $\hat{z}$-axis. (b-c) Simulated reflection and transmission (in terms of intensity) for the sheet when the incident wave propagates along the (b) transparent side and (c) non-transparent side.

Using the all-dielectric particles constructed from layers of anisotropic dielectric gratings also presented in section IV.D, a Half-wave plate was demonstrated in [67] and is shown in Fig. 31. The half wave plate is made from four stacked staked anisotropic gratings and shows excellent agreement between simulated and measured values. Different scattering matrices can be realized using this technique. This structure may find use in applications where low loss is required.

### *Q. Nonreciprocal Bianisotropic Metasurfaces*

An application of non-reciprocal metasurfaces using the Moving-Chiral particles from section IV.A is one-way transparent sheets [124], [125]. A one-way transparent sheet is a metasurface which is transparent when illuminated from one side (the transparent side) and has controllable properties when illuminated from the opposite side (the non-transparent side). For example, in [124], the authors design a one-way transparent sheet metasurface which completely transmits from the transparent side and acts as a polarization rotator metasurface (rotates the incident polarization by 90° upon transmission) when illuminated from the non-transparent side. To design the metasurface, the synthesis technique based on the polarizability model (see section III.A) is used. By writing the magnetoelectric effective polarizabilities with the coupling coefficients responsible for reciprocal and non-reciprocal coupling processes separated

$$\begin{aligned}\overline{\overline{\hat{\alpha}}}_{em} &= \left(\hat{\chi} - j\hat{\kappa}\right)\overline{\overline{I}}_t + \left(\hat{V} - j\hat{\Omega}\right)\overline{\overline{J}}_t \\ \overline{\overline{\hat{\alpha}}}_{me} &= \left(\hat{\chi} + j\hat{\kappa}\right)\overline{\overline{I}}_t + \left(-\hat{V} + j\hat{\Omega}\right)\overline{\overline{J}}_t\end{aligned} \tag{69}$$

The conditions on the polarizabilities needed for one-way sheet design can be obtained. Note, in (69), $\hat{\kappa}, \hat{\Omega}, \hat{V}$, and $\hat{\chi}$ are the chiral, omega, moving, and Tellegen coefficients. Substituting (69) into (28) and noting the one-way sheet is characterized by

$$\vec{E}_r = 0,\ \ \vec{E}_t = \hat{z} \times \vec{E}_{inc} \tag{70}$$

the required polarizabilities for one-way sheet which rotates polarization by 90° when illuminated by the non-transparent side can be obtained. It is found that Moving-Chiral particles are required (see Fig. 32a). The reflection and transmission properties of an infinite sheet composed of Moving-Chiral particles is shown in Fig. 32b and c, respectively. The results show that the incident wave is completely transmitted when illumined from the transparent side (one-way sheet) and has its polarization rotated by 90° when illuminated by the non-transparent side (polarization twist).

## VI. Future Prospects

The next generation of bianisotropic metasurfaces will involve dynamic and active bianisotropic properties which can overcome the design limitations imposed by linearity, passivity, and reciprocity. Non-linear effects [126], [127] can be incorporated into the magnetoelectric particles for new functionality unexplored to date. Time-modulated meta-atoms can also be included in bianisotropic metasurfaces to allow new design dimensions [128]–[132]. Both of these approaches lead to control over not only the spatial spectrum but also the temporal spectrum as well. New forms of the GSTC applicable in the time-domain can also be envisaged. In [133], the authors develop an extension of the GSTC to the time domain. Other extensions of the GSTC have also been formulated. In [134], the GSTC was extended to include spatial dispersion. In the future, one can envision fast reconfigurable meta-atoms coupled with fast optimization algorithms allowing for software controlled bianisotropic metasurfaces. Toward this end, the work in [135] develops a reconfigurable metasurface for cloaking of objects hidden inside a bump. If these metasurfaces can be made multiband and conformal, cloaks capable of hiding large objects or creating dynamic optical illusions can be envisioned. This could enable three-dimensional dynamic multi-color holograms for next generation display and communications systems. Non-reciprocal metasurfaces [101], [136], metasurfaces with parametric gain, and Multi-Input-Multi-Output (MIMO) metasurfaces [137], [138] are also emerging and promise new capabilities. In [139], a time-modulated metasurface that locally mimics a rotating aniso-tropic metasurface is introduced that performs frequency up/down conversion and provides parametric gain for reflections of circularly polarized incident waves. All of these research directions will require new computational and numerical design, analysis, and optimization algorithms as well as fabrication approaches. Undoubtedly, the next generation of electromagnetic and optical devices will involve bianisotropic metasurfaces.

## VII. Conclusion

In conclusion, bianisotropic boundary conditions used in metasurface design have been reviewed in a complete tutorial-like manner. The most common bianisotropic boundary models used in metasurface design (Susceptibility, Impedance, and Polarizability Models) have been derived. Several synthesis methods based on these models have been presented. Several bianisotropic particles which exhibit magnetoelectric coupling have also been reviewed. These include the reciprocal Omega, reciprocal Chiral, non-reciprocal Tellegen-Omega, and non-reciprocal Moving-Chiral particles, three-sheet and four-sheet realizations of bianisotropic particles, and all dielectric realizations. Finally, numerous metasurfaces, taken from literature, which utilize bianisotropic boundaries, synthesis methods, and particle designs described were highlighted in the final section of this review article. The survey of examples taken from recent scientific works represents the state of the art in bianisotropic metasurface design. The article also included a short section on future prospects. This review article serves as a one-stop collection on recent advances in the theory, realization, capabilities, and applications of bianisotropic boundary conditions in metasurface design.